\PassOptionsToPackage{unicode}{hyperref}
\PassOptionsToPackage{hyphens}{url}
\PassOptionsToPackage{dvipsnames,svgnames,x11names}{xcolor}
\documentclass[
  12pt]{article}

\usepackage{amsmath,amssymb}
\usepackage{iftex}
\ifPDFTeX
  \usepackage[T1]{fontenc}
  \usepackage[utf8]{inputenc}
  \usepackage{textcomp} 
\else 
  \usepackage{unicode-math}
  \defaultfontfeatures{Scale=MatchLowercase}
  \defaultfontfeatures[\rmfamily]{Ligatures=TeX,Scale=1}
\fi
\usepackage{lmodern}
\ifPDFTeX\else  
\fi
\IfFileExists{upquote.sty}{\usepackage{upquote}}{}
\IfFileExists{microtype.sty}{
  \usepackage[]{microtype}
  \UseMicrotypeSet[protrusion]{basicmath} 
}{}
\makeatletter
\@ifundefined{KOMAClassName}{
  \IfFileExists{parskip.sty}{%
    \usepackage{parskip}
  }{
    \setlength{\parindent}{0pt}
    \setlength{\parskip}{6pt plus 2pt minus 1pt}}
}{
  \KOMAoptions{parskip=half}}
\makeatother
\usepackage{xcolor}
\makeatletter
\ifx\paragraph\undefined\else
  \let\oldparagraph\paragraph
  \renewcommand{\paragraph}{
    \@ifstar
      \xxxParagraphStar
      \xxxParagraphNoStar
  }
  \newcommand{\xxxParagraphStar}[1]{\oldparagraph*{#1}\mbox{}}
  \newcommand{\xxxParagraphNoStar}[1]{\oldparagraph{#1}\mbox{}}
\fi
\ifx\subparagraph\undefined\else
  \let\oldsubparagraph\subparagraph
  \renewcommand{\subparagraph}{
    \@ifstar
      \xxxSubParagraphStar
      \xxxSubParagraphNoStar
  }
  \newcommand{\xxxSubParagraphStar}[1]{\oldsubparagraph*{#1}\mbox{}}
  \newcommand{\xxxSubParagraphNoStar}[1]{\oldsubparagraph{#1}\mbox{}}
\fi
\makeatother

\usepackage{longtable,booktabs,array}
\usepackage{calc} 
\usepackage{etoolbox}
\makeatletter
\patchcmd\longtable{\par}{\if@noskipsec\mbox{}\fi\par}{}{}
\makeatother
\IfFileExists{footnotehyper.sty}{\usepackage{footnotehyper}}{\usepackage{footnote}}
\makesavenoteenv{longtable}
\usepackage{graphicx}
\makeatletter
\def\maxwidth{\ifdim\Gin@nat@width>\linewidth\linewidth\else\Gin@nat@width\fi}
\def\maxheight{\ifdim\Gin@nat@height>\textheight\textheight\else\Gin@nat@height\fi}
\makeatother
\setkeys{Gin}{width=\maxwidth,height=\maxheight,keepaspectratio}
\makeatletter
\def\fps@figure{htbp}
\makeatother

\makeatletter
\@ifpackageloaded{caption}{}{\usepackage{caption}}
\AtBeginDocument{%
\ifdefined\contentsname
  \renewcommand*\contentsname{Table of contents}
\else
  \newcommand\contentsname{Table of contents}
\fi
\ifdefined\listfigurename
  \renewcommand*\listfigurename{List of Figures}
\else
  \newcommand\listfigurename{List of Figures}
\fi
\ifdefined\listtablename
  \renewcommand*\listtablename{List of Tables}
\else
  \newcommand\listtablename{List of Tables}
\fi
\ifdefined\figurename
  \renewcommand*\figurename{Figure}
\else
  \newcommand\figurename{Figure}
\fi
\ifdefined\tablename
  \renewcommand*\tablename{Table}
\else
  \newcommand\tablename{Table}
\fi
}
\@ifpackageloaded{float}{}{\usepackage{float}}
\floatstyle{ruled}
\@ifundefined{c@chapter}{\newfloat{codelisting}{h}{lop}}{\newfloat{codelisting}{h}{lop}[chapter]}
\floatname{codelisting}{Listing}

\makeatother
\makeatletter
\@ifpackageloaded{caption}{}{\usepackage{caption}}
\@ifpackageloaded{subcaption}{}{\usepackage{subcaption}}
\makeatother

\ifLuaTeX
  \usepackage{selnolig}  
\fi
\usepackage[]{natbib}
\usepackage{bookmark}

\IfFileExists{xurl.sty}{\usepackage{xurl}}{} 
\hypersetup{
  pdftitle={Title},
  pdfauthor={Author 1; Author 2},
  pdfkeywords={3 to 6 keywords, that do not appear in the title},
  colorlinks=true,
  linkcolor={blue},
  filecolor={Maroon},
  citecolor={Blue},
  urlcolor={Blue},
  pdfcreator={LaTeX via pandoc}}

\usepackage{graphicx}
\usepackage{amsfonts, bbm, bm, mathtools, mathrsfs, amsthm}
\usepackage{xr-hyper} 
\usepackage{hyperref}
\usepackage{url}
\usepackage{soul}
\usepackage{cancel}
\usepackage{natbib}

\usepackage{array}
\usepackage{caption}
\usepackage{rotating} 
\usepackage{hhline}
\usepackage{calc}
\usepackage{tabularx}
\usepackage{textcomp}

\newcolumntype{C}[1]{>{\centering\arraybackslash}p{#1}} 
\usepackage[normalem]{ulem}
\newcommand{\stkout}[1]{\ifmmode\text{\sout{\ensuremath{#1}}}\else\sout{#1}\fi}

\usepackage{setspace}
\usepackage{graphicx}         
\usepackage{caption}          
\usepackage{subcaption}       
\usepackage{booktabs}         
\usepackage{multirow}         
\usepackage{array}            
\usepackage{float}            
\usepackage{wrapfig}          
\usepackage{siunitx}
\usepackage[normalem]{ulem}
\usepackage{arydshln}         
\newcommand{\thickline}{\noalign{\hrule height 0.8pt} } 
\usepackage{pdflscape}
\usepackage{titletoc}
\usepackage{tikz}
\usepackage{tikz-qtree}
\usetikzlibrary{trees}
\usetikzlibrary{automata,positioning}

\usepackage{algorithm}        
\usepackage{algpseudocode}    

\allowdisplaybreaks

\DeclareMathOperator{\doo}{do}

\newtheorem{theorem}{Theorem}

\newtheorem{corollary}[theorem]{Corollary}

\newtheorem{lemma}[theorem]{Lemma}

\newtheorem{remark}[theorem]{Remark}

\usepackage{xr}

\usepackage{mathtools}
\usepackage{enumerate}
\usepackage{parskip}
\newcommand{\uniform}{\text{Uniform}}
\newcommand{\bin}{\text{Binomial}}
\newcommand{\expit}{\text{expit}}

\newcommand{\razi}[1]{\textcolor{olive}}
\newcommand{\anna}[1]{\textcolor{Salmon}}

\newcommand{\I}{\mathbb{I}}

\newcommand{\E}{\mathbb{E}}
\newcommand{\R}{\mathbb{R}}
\newcommand{\N}{\mathcal{N}}

\DeclareMathOperator*{\argmin}{\arg\!\min}

\newcommand{\hilbert}{\hbox{${\rm I\kern-.2em H}$}}

\newcommand{\openn}{\hbox{${\rm I\kern-.2em N}$}}
\newcommand{\opend}{\hbox{${\rm I\kern-.2em D}$}}
\newcommand{\smallO}{o_P}

\newcommand{\anon}{1}

\begin{document}

\def\spacingset#1{\renewcommand{\baselinestretch}%
{#1}\small\normalsize} \spacingset{1}


\if1\anon
{
  \title{\bf Causal Inference with the Napkin Graph}
  \author{Anna Guo$^1$, Lin Liu$^2$, David Benkeser$^1$, Razieh Nabi$^1$\\[0.35cm]
    $^1$Department of Biostatistics and Bioinformatics, Emory University\\[0.25cm]
    $^2$Institute of Natural Sciences, MOE--LSC; School of Mathematical Sciences,\\
    CMA--Shanghai; SJTU--Yale Joint Center for Biostatistics and Data Science,\\
    Shanghai Jiao Tong University}
  \maketitle
} \fi

\if0\anon
{
  \bigskip
  \bigskip
  \bigskip
  \begin{center}
    {\LARGE\bf Causal Inference with the ``Napkin Graph''}
\end{center}
  \medskip
} \fi

\bigskip
\begin{abstract}
Unmeasured confounding can render identification strategies based on adjustment functionals invalid. We study the ``Napkin'' graph, a causal structure that encapsulates features of M-bias, instrumental variables, and classical back-door and front-door settings, yet identifies the average treatment effect through a nonstandard ratio of two g-formulas. We develop influence-function-based estimators for this functional, including doubly-robust one-step and targeted minimum loss-based estimators that remain asymptotically linear under slower-than-parametric nuisance estimation using machine learning. A distinguishing feature of the Napkin graph is that it imposes a \textit{generalized independence restriction}, known as a \textit{Verma constraint}, rather than ordinary conditional independence restrictions, on the observed data distribution. We develop semiparametric efficiency theory for causal effects under a moment restriction corresponding to this Verma constraint,  characterizing the orthocomplement of the tangent space, deriving the class of influence functions, and obtaining the semiparametric efficiency bound. More broadly, our analysis provides a framework for semiparametric inference in causal models defined by Verma constraints and demonstrates how such restrictions may yield efficiency gains. Simulations confirm the estimators' theoretical properties and demonstrate substantial efficiency gains. A real-data application using the Finnish Life Course Study estimates the effect of educational attainment on income. An accompanying \textsf{R} package, \href{https://github.com/annaguo-bios/napkincausal}{\texttt{napkincausal}}, implements our methods.
\end{abstract}

\noindent%
\vfill

\newpage
\spacingset{1.8} 

\section{Introduction}
\label{sec:intro}

Pearl's \textit{do-calculus} \citep{tian2002general, pearl2009causality} provides a  graphical framework for determining whether and how causal effects such as the average treatment effect (ATE) can be identified from observational data, by leveraging assumptions about both observed and unobserved variables encoded in a hidden variable directed acyclic graph (DAG). Subsequent work has established the \textit{completeness} of the do-calculus \citep{shpitser06id, huang06do}, showing that any identifiable causal effect can, in principle, be derived using its rules. Variations of a sound and complete identification algorithm have been proposed by \cite{richardson2023nested} and \cite{bhattacharya2022semiparametric}.

While do-calculus provides a complete identification theory, several subclasses of hidden-variable DAGs admit simpler graphical criteria. The \textit{back-door} criterion identifies the~ATE by adjusting for covariates that block all non-causal paths between treatment and outcome \citep{robins1986new, pearl2009causality}. The \textit{front-door} criterion applies when mediators are not subject to unmeasured confounding of treatment–mediator or mediator–outcome relations and~fully mediate the effect of treatment on the outcome \citep{pearl2009causality}. The more general \textit{primal fixability} criterion requires that the treatment and its children lie in distinct latent components (c-component/district) \citep{tian2002general, richardson03markov,bhattacharya2022semiparametric, guo2024average}. These criteria are easy to verify and lend themselves~to practical estimation strategies as recent work has developed estimators with favorable asymptotic properties even under machine learning nuisance estimation \citep{fulcher2020robust, bhattacharya2022semiparametric, jung2021estimating, jung2024unified, guo2023flexible, guo2024average}. 

Despite these advances, some causal graphs fall outside these convenient classes. A prominent example is the \textit{Napkin graph} \citep{pearl2018book} shown in Figure~\ref{fig:graphs}(a) with $X$ as treatment, $Y$ as outcome, unmeasured confounders $U_1, U_2$, and observed pre-treatment variables $W, Z$. Several features illustrate why standard identification criteria fail. First, the path $X \leftarrow U_1 \rightarrow W \leftarrow U_2 \rightarrow Y$ forms a collider at $W$, a hallmark of M-bias \citep{pearl2009causality, ding2015adjust}; conditioning on $W$ (or on the downstream $Z$) opens this path, invalidating the back-door criterion. Second, no observed mediators lie between $X$ and $Y$, ruling out the use of the front-door criterion. Third, $X$ and $Y$ belong to the same latent component, so primal fixability fails, concluding that $P(Y^x, Z, W)$ is not identified \citep{tian2002general}. Here $Y^x$ denotes the potential outcome under $X = x$.

\begin{figure}[t] 
	\begin{center}
    \scalebox{0.65}{
    \begin{tikzpicture}[>=stealth, node distance=1.5cm]
        \tikzstyle{format} = [thick, circle, minimum size=1.0mm, inner sep=2pt]
        \tikzstyle{square} = [draw, thick, minimum size=4.5mm, inner sep=2pt]

    \begin{scope}[xshift=0cm, yshift=0cm]
		\path[->, thick]
		
		node[] (w) {$W$}
        node[right of=w, xshift=0.5cm] (z) {$Z$}
        node[above of=z, yshift=-0.75cm] (u1) {$U_1$}
        node[above of=u1, xshift=0.5cm, yshift=-0.75cm] (u2) {$U_2$}
		node[right of=z, xshift=0.5cm] (x) {$X$}
        node[right of=x, xshift=0.5cm] (y) {$Y$}
		 
        (w) edge[blue] (z)
		(z) edge[blue] (x) 
		(x) edge[blue] (y) 
		(u1) edge[red] (w)
        (u1) edge[red] (x)
        (u2) edge[red] (w)
        (u2) edge[red] (y)
        
        node[below of=z, xshift=0.85cm, yshift=0.5cm] (a_t) {(a)} ;
	\end{scope}
	\begin{scope}[xshift=8cm, yshift=0cm]
		\path[->, thick]
		
		node[] (w) {$W$}
        node[right of=w, xshift=0.5cm] (z) {$Z$}
        node[above of=z, yshift=-0.75cm] (u1) {$U_1$}
        node[above of=u1, xshift=0.5cm, yshift=-0.75cm] (u2) {$U_2$}
		node[right of=z, xshift=0.5cm] (x) {$X$}
        node[above of=x, xshift=1cm, yshift=-0.5cm] (c) {$C$}
        node[right of=x, xshift=0.5cm] (y) {$Y$}
    
        (c) edge[blue] (w)
        (c) edge[blue] (z)
        (c) edge[blue] (x)
        (c) edge[blue] (y)

        (u1) edge[red] (w)
        (u1) edge[red] (x)
        (u2) edge[red] (w)
        (u2) edge[red] (y)
   
        (w) edge[blue] (z)
		(z) edge[blue] (x) 
		(x) edge[blue] (y) 

        node[below of=z, xshift=1.2cm, yshift=0.5cm] (b_t) {(b)} ; 
	\end{scope}
    \begin{scope}[xshift=16.cm, yshift=0cm]
		\path[->, thick]
		
		node[] (w) {$W$}
        node[square, right of=w, xshift=0.5cm] (z) {$Z$}
        node[above of=z, yshift=-0.75cm] (u1) {$U_1$}
        node[above of=u1, xshift=0.5cm, yshift=-0.75cm] (u2) {$U_2$}
		node[right of=z, xshift=0.5cm] (x) {$X$}
        node[right of=x, xshift=0.5cm] (y) {$Y$}

		(z) edge[blue] (x) 
		(x) edge[blue] (y) 

        (u1) edge[red] (w)
        (u1) edge[red] (x)
        (u2) edge[red] (w)
        (u2) edge[red] (y)
        
        node[below of=z, xshift=0.85cm, yshift=0.5cm] (c_t) {(c)} ; 
	\end{scope}
	\end{tikzpicture}}
	\caption{(a) The Napkin DAG; (b) A generalized Napkin DAG with measured confounders $C$; (c) The conditional DAG resulting from intervention on $Z$ in panel (a).} 
	\label{fig:graphs}
	\end{center}
\end{figure}

Even though $P(Y^x, Z, W)$ is not identifiable in the Napkin model, do-calculus proves that the marginal distribution $P(Y^x)$ is identifiable. In particular, the counterfactual mean $\E(Y^x)$ can be expressed as a ratio of two g-formulas: the numerator characterizes the effect of $Z$ on $(X,Y)$ jointly, while the denominator captures its effect on $X$ alone. This ratio representation makes the Napkin graph a canonical example of causal effect identification in hidden-variable DAGs beyond standard back-door, front-door, and primal-fixability criteria. 

The ratio-of-g-formulas representation also invites comparison with the instrumental variable (IV) literature. While $Z$ qualifies as an IV and can be used for ATE identification under additional assumptions, the ATE in the Napkin model is in fact identifiable without assumptions beyond those encoded by the graph itself. In particular, $Z$ satisfies the classical IV conditions of exclusion, independence, and relevance conditional on the measured covariate $W$ \citep{abadie2003semiparametric}, permitting standard IV analyses under additional assumptions. For example, when $Z$ is binary, monotonicity identifies the \textit{local average treatment effect} \citep{angrist1996identification,baiocchi2014instrumental}, while alternative assumptions such as no additive interaction conditions \citep{wang2018bounded,hartwig2023average} identify the ATE. For continuous instruments, extensions of these assumptions are typically required \citep{dong2025marginal,chen2025identification}, or alternatively stronger monotonicity assumptions are imposed to identify \textit{local IV curves} \citep{kennedy2019robust}. In contrast, under the Napkin model, $P(Y^x)$ (and hence the ATE) is identifiable, without requiring the additional assumptions typically invoked in the IV literature and regardless of whether $Z$ is discrete or continuous. 

ATE estimation under the Napkin graph has previously been considered by \citet{helske2021estimation}, who termed $Z$ a trapdoor variable and proposed parametric plug-in estimators, and by \citet{bhattacharya19cinu}, who developed inverse-probability-weighted estimators. Both approaches establish identifiability and yield consistent estimators, which constitute important progress. Nevertheless, the prior proposed estimators do not protect against nuisance model misspecification bias, in contrast to doubly robust or debiased estimation, making valid inference challenging and motivating the development of new estimation procedures grounded in semiparametric theory. 

Aside from its graphical features, the Napkin graph is statistically distinctive because the observed data distribution is not characterized by ordinary independence restrictions. Instead, it satisfies a \textit{generalized independence restriction}, known as a \textit{Verma constraint} \citep{verma90equiv,robins1986new}. Semiparametric efficiency theory has been extensively developed for models defined by ordinary independencies \citep{tsiatis2007semiparametric}, but analogous theory for models defined by Verma constraints remains largely undeveloped. Consequently, the impact of such constraints on tangent space, efficiency bounds, and estimation efficiency remains poorly understood.

This paper makes two main methodological contributions.
\textit{First,} we address limitations of existing estimators \citep{helske2021estimation, bhattacharya19cinu} by developing nonparametric influence-function-based inference for the ATE under the Napkin graph. Because the identification functional exhibits distinct differentiability properties depending on whether the trapdoor variable $Z$ is discrete or continuous, a unified semiparametric treatment is not immediate. We derive an equivalent representation of the causal estimand based on averaging over an arbitrary valid density on $Z$, preserving the target parameter while enabling semiparametric inference. Building on these representations, we develop one-step and targeted minimum loss estimators that remain asymptotically linear under flexible machine-learning estimation of nuisance functions and enjoy robustness to nuisance-model misspecification. 
\textit{Second,} we develop semiparametric efficiency theory for a model incorporating the Verma constraint implied by the Napkin graph. By characterizing the associated tangent space, its orthocomplement, and the resulting class of influence functions, we derive the semiparametric efficiency bound. These results yield semiparametric efficient estimators when the trapdoor variable $Z$ is discrete and locally efficient estimators within a broad class of semiparametric submodels when $Z$ is continuous. More broadly, our analysis demonstrates how Verma constraints can be incorporated into semiparametric efficiency theory, a relatively under-explored area in the causal inference literature, despite recent contributions by \citet{liu2021efficient} and \citet{von2025evaluation}. Since such constraints arise naturally in nested Markov models associated with hidden-variable DAGs \citep{richardson2023nested}, our new results constitute a concrete step toward a broader efficiency theory for causal models defined by latent-variable graphical structures. Relative to fully nonparametric estimators, the resulting procedures can achieve substantial efficiency gains, with variance reductions approaching a factor of three in simple simulations. 

The paper proceeds as follows. Section~\ref{sec:id} presents the Napkin model and the identification assumptions. Section~\ref{sec:est} derives estimators based on the corresponding nonparametric influence function and establishes their asymptotic properties. Section~\ref{sec:efficiency} develops semiparametric efficiency theory and characterizes the tangent space and the efficiency bound. Section~\ref{sec:sims} reports results from simulation studies, followed by Section~\ref{sec:real-data} presenting an analysis on the Finnish Life Course study on the effect of education on income. Section~\ref{sec:conc} concludes. All proofs are deferred to the Appendix.

\section{Target parameter and identification}
\label{sec:id}

Let $O=\{W,Z,X,Y\}$ denote the observed variables and $U=\{U_1,U_2\}$ the unobserved variables. We assume $P(O,U)$ factorizes according to the Napkin DAG in Figure~\ref{fig:graphs}(a), with $P(O)$ obtained by marginalizing $P(O,U)$ over $U$. We use calligraphic notation to denote supports, e.g., $\mathcal{Z}$ for $Z$. We write $P$ for distributions and $p$ for densities, assuming continuous variables admit Lebesgue densities (though this is not required). 

We consider a binary treatment $X$ and define the ATE as $\E(Y^1-Y^0)$, noting that results extend directly to categorical treatments. Since the identification and estimation of $\E(Y^1)$ and $\E(Y^0)$ are analogous, we focus on the generic target parameter $\E(Y^{x_0})$ for $x_0 \in \{0,1\}$. Under the Napkin model, identification relies on: 
(i) \textit{Consistency:} if $Z=z$, then $X^z=X$, and in addition if $X=x$ then $Y^{z, x}=Y$; 
(ii) \textit{Conditional ignorability:} $X^z \perp Y^{z,x}$ and $\{Y^{z,x}, X^z\} \perp Z \mid W$;
(iii) \textit{No direct effect of $Z$ on $Y$:} $Y^{z,x}=Y^{x}$; and
(iv) \textit{Positivity:} $p(X=1, Z=z\mid W=w)>0$ for all $z$ and $w$ where $p(W=w)>0$. Specifically, assumption~(iii) corresponds to the Verma (or generalized independence/nested Markovian) constraint; see later Section~\ref{sec:verma} for a more detailed explanation.

\begin{lemma}\label{lem:id}
Under assumptions (i)--(iv) and for any $z^*\in\mathcal{Z}$, $\E(Y^{x_0})$ is identified via the following functional, denoted by $\psi_{x_0}(P; z^*)$:
\vspace{-0.25cm}
\begin{align}
\psi_{x_0}(P; z^*) = \frac{\kappa_{x_0,1}(P;z^*)}{\kappa_{x_0,2}(P;z^*)} \ ,
\label{eq:ID_z*}
\end{align}

\vspace{-0.5cm}
where $\kappa_{x_0,1}(P;z) = \int y \, p(y\, | \, x_0,z, w) \, p(x_0\, | \, z, w) \, p(w) \, dw$ and $\kappa_{x_0,2}(P;z) = \int p(x_0\, | \, z, w) \, p(w) \, dw$. 
Given a pre-specified density over $Z$, denoted $\tilde{p}(z)$, $\E(Y^{x_0})$ can also be identified via: 
\vspace{-0.25cm}
\begin{align} 
\psi_{x_0}(P;\tilde{p}_z) &= \frac{ \int \kappa_{x_0,1}(P;z) \, \tilde{p}(z) \, dz }{\int  \kappa_{x_0,2}(P;z)  \, \tilde{p}(z) \, dz} \ .
\label{eq:ID_tilde_pZ} 
\end{align}
\end{lemma}

\vspace{-0.6cm}
See a proof in Appendix~\ref{app:proofs_ID}. 

Identification via \eqref{eq:ID_z*} follows directly from do-calculus. According to \eqref{eq:ID_z*}, $\E(Y^{x_0})$ is identified as the ratio of two g-formulas: $\kappa_{x_0,1}(P; z^*)$ identifies the mean of the composite outcome $Y \I(X = x_0)$ under the intervention $Z = z^*$, while $\kappa_{x_0,2}(P; z^*)$ identifies the mean of $\I(X = x_0)$ under the same intervention. The invariance of $\psi_{x_0}(P; z^*)$ to $z^*$ is a manifestation of the Verma constraint encoded in the Napkin graph, which can also be seen directly through assumption~(iii), which encodes ``no direct effect of $Z$ on $Y$ except through $X$''. Although identification does not depend on the choice of $z^*$, efficiency of estimators may, as will be discussed in Section~\ref{sec:efficiency}. 

The representation in \eqref{eq:ID_tilde_pZ} follows from this invariance. Although algebraically equivalent to \eqref{eq:ID_z*}, it replaces point evaluation at a fixed value $z^*$ with an average over $z$. Thus \eqref{eq:ID_tilde_pZ} yields a regular parameterization for continuous $Z$, enabling the influence-function analyses developed in subsequent sections. The indexing density $\tilde p_z$ is an auxiliary distribution over $\mathcal Z$ and need not coincide with the true marginal density of $Z$. We require $\tilde p_z$ to be valid in the sense that it assigns positive mass only to values of $z$ for which $p(z\, | \, w)>0$ for all $w$ satisfying $p(w)>0$, and for which $\kappa_{x_0,2}(P;z)>0$. Different choices of $\tilde p_z$ yield the same target parameter but may lead to different efficiency properties, see Section \ref{sec:efficiency}.

The Napkin graph can be extended to incorporate measured confounders, as illustrated in Figure~\ref{fig:graphs}(b). For clarity, we focus on the original formulation in the main text and defer the identification and estimation procedures for this extension to Appendix~\ref{app:extension_confounders}.

Next, we develop estimation and inference for $\E(Y^{x_0})$ using the representation in \eqref{eq:ID_tilde_pZ} under a fully nonparametric model. This formulation subsumes the discrete case, since for discrete $Z$ one may choose $\tilde p(z)$ to be a probability mass function; in particular, placing all mass at a fixed value $z^* \in\mathcal Z$ (i.e., $\tilde p(z^*)=1$) recovers the representation in \eqref{eq:ID_z*}.

\section{Estimation and inference}
\label{sec:est}

We consider $n$ i.i.d. copies of $O=(Y,X,Z,W)$ drawn from a distribution $P$ that is Markov relative to the Napkin graph in Figure~\ref{fig:graphs}(a). Estimation of $\psi_{x_0}(P;\tilde{p}_z)$ requires several nuisance functions. We define
the outcome regression $\mu(x,z,w) = \E(Y\, | \, X=x,Z=z,W=w)$,
the propensity score $\pi(x\, | \, z,w) = P(X=x\, | \, Z=z,W=w)$,
the conditional density $f_Z(z\, | \, w) = p(Z=z\, | \, W=w)$, and
the marginal distribution $p_W(w) = p(W=w)$.
These nuisances are collected in $Q=\{\mu,\pi,f_Z,p_W\}$. The functional $\psi_{x_0}(P;\tilde{p}_z)$ and its components $\kappa_{x_0,1}(P;z)$ and $\kappa_{x_0,2}(P;z)$ depend on $P$ only through $Q$. To emphasize this, we instead write $\psi_{x_0}(Q;\tilde{p}_z)$, $\kappa_{x_0,1}(Q;z)$, and $\kappa_{x_0,2}(Q;z)$ throughout the remainder of the paper. 

For notational simplicity, we define $\kappa_{x_0, 1}(Q;\tilde{p}_z) = \int \kappa_{x_0, 1}(Q;z) \, \tilde p(z) \, dz$ and $\kappa_{x_0, 2}(Q;\tilde{p}_z) = \int \kappa_{x_0, 2}(Q;z) \, \tilde p(z) \, dz$. Then the identification functional in \eqref{eq:ID_tilde_pZ} can be expressed as 
\vspace{-0.25cm}
\begin{align*}
    \psi_{x_0}(P;\tilde{p}_z) = \frac{\kappa_{x_0, 1}(Q;\tilde{p}_z)}{\kappa_{x_0, 2}(Q;\tilde{p}_z)} \ .
\end{align*}

\vspace{-0.35cm}
In practice, the nuisance functions can be estimated parametrically or using flexible methods. For instance, $\mu$ may be estimated by regressing $Y$ on $(X,Z,W)$, $\pi$ by regressing $X$ on $(Z,W)$, and $f_Z$ using either a parametric model or non/semiparametric kernel density methods \citep{benkeser2016highly}; when $Z$ is discrete, $f_Z$ reduces to a regression of $Z$ on $W$. The marginal distribution $p_W$ can be simply replaced by its empirical distribution. Throughout, we allow these regressions to be fit using machine learning methods, with cross-fitting employed as needed for valid asymptotics. We denote the resulting estimates by $\hat{Q}=\{\hat{\mu},\hat{\pi},\hat{ f}_Z,\hat{p}_W\}$. 

Let $\kappa^{\text{pi}}_{x_0,1}(\hat{Q};z) = \frac{1}{n} \sum_{i = 1}^n \hat{\mu}(x_0, z, W_i) \, \hat{\pi}(x_0\, | \, z, W_i)$ and $\kappa^{\text{pi}}_{x_0,2}(\hat{Q};z) = \frac{1}{n}\sum_{i = 1}^n \hat{\pi}(x_0\, | \, z, W_i)$ denote plug-in estimators of $\kappa_{x_0,1}(Q;z)$ and $\kappa_{x_0,2}(Q;z)$, respectively. Let  $\kappa^{\text{pi}}_{x_0,j}(\hat{Q}; \tilde p_z) =  \int  \kappa^{\text{pi}}_{x_0,j}(\hat{Q};z) \, \tilde{p}(z) \, dz $, $j = 1, 2$, where the integral with respect to $\tilde p(Z)$ is evaluated using Monte Carlo or other numerical integration methods. The plug-in estimator of \eqref{eq:ID_tilde_pZ}, denoted $\psi^{\text{pi}}_{x_0}(\hat{Q};\tilde{p}_z)$, is obtained via: 
\vspace{-0.65cm}
\begin{align} 
\psi^{\text{pi}}_{x_0}(\hat{Q};\tilde{p}_z) &=\frac{  \kappa^{\text{pi}}_{x_0,1}(\hat{Q};\tilde{p}_z) }{ \kappa^{\text{pi}}_{x_0,2}(\hat{Q};\tilde{p}_z) }.
\label{eq:est_plug_contZ} 
\end{align}

\vspace{-0.35cm}
As a special case, one may place all mass at a fixed $z^* \in \mathcal Z$ (i.e., $\tilde p(z^*)=1$) and construct the plug-in estimator $\psi^{\text{pi}}_{x_0}(\hat Q; z^*)$ based on \eqref{eq:ID_z*} as $\psi^{\text{pi}}_{x_0}(\hat Q; z^*) = \kappa^{\text{pi}}_{x_0,1}(\hat Q; z^*) \, / \, \kappa^{\text{pi}}_{x_0,2}(\hat Q; z^*)$. 

While the plug-in estimator is straightforward, it is limited in practice. A von Mises expansion shows that $\psi^{\text{pi}}_{x_0}(\hat{Q};\tilde{p}_z) - \psi_{x_0}(Q; \tilde{p}_z) = -P{\Phi_{x_0}(\hat Q; \tilde{p}_z)} + R_2(\hat Q,Q; \tilde{p}_z)$, where $\Phi_{x_0}(\hat Q; \tilde{p}_z)$ denotes a canonical gradient (a.k.a., influence function for differentiable functionals in the sense of \citet{van1991differentiable}) for $\psi_{x_0}(Q; \tilde{p}_z)$ evaluated at $\hat{Q}$, and $R_2(\hat Q,Q; \tilde{p}_z)$ is a second-order remainder. The leading error term, $-P{\Phi_{x_0}(\hat Q; \tilde{p}_z)}$, corresponds to the first-order bias and depends on the quality of the nuisance estimates in $\hat Q$. Achieving asymptotic linearity therefore requires all nuisance estimators to converge to the truth at a sufficiently fast rate of $o_P(n^{-1/2})$, which may be infeasible if the nuisance parameters are highly complex. To address this limitation, we develop estimators that control the first-order bias, attain asymptotic linearity under weaker conditions, and offer robustness properties not shared by the plug-in approach.

Throughout our proposed estimation framework, we require $\tilde p_z$ to be supported on values of $z$ for which the nuisance functions are well defined and estimable from the observed data. Specifically, we assume $f_Z(z\, | \, w) > 0$ for all $z$ in the support of $\tilde{p}_z$ and all $w$ with $p_{W}(w) > 0$. This condition ensures that $\mu(x_0,z,W_i)$ and $\pi(x_0\, | \, z,W_i)$ are identified from the observed data and avoids extrapolation to unsupported $(z,w)$ regions. A more detailed discussion, together with simulations that illustrate the consequences of violations of this condition, is provided in Appendix~\ref{appsubsec:pz-overlap-condition}.

\subsection{Influence-function-based estimators}
\label{subsec:estimators}

We first present an influence function for $\psi_{x_0}(Q;\tilde{p}_z)$, which is uniquely defined under a fully nonparametric model but may not be unique within a semiparametric model. 

\begin{lemma}\label{lem:np_EIF}
For an observation $O_i=(Y_i,X_i,Z_i,W_i)$, a canonical gradient for the functional in \eqref{eq:ID_tilde_pZ}, denoted as $\Phi_{x_0}(Q;\tilde{p}_z)(O_i)$, is given by 

\vspace{-0.5cm}
{\small 
\begin{equation}\label{eq:IF_tilde_pz}
    \begin{aligned}
    \Phi_{x_0}(Q;\tilde{p}_z)(O_i)
    &=\underbrace{\frac{\I(X_i=x_0)}{\kappa_{x_0, 2}(Q;\tilde{p}_z)}  \frac{\tilde{p}(Z_i)}{f_Z(Z_i \mid W_i)}  \Big\{  Y_i -  \mu(x_0, Z_i, W_i) \Big\}}_{\Phi_{Y,x_0}(Q; \, \tilde{p}_z)(O_i)}
    \\[0.35cm] 
    &\hspace{-2.2cm} + \underbrace{\frac{1}{\kappa_{x_0, 2}(Q;\tilde{p}_z)} \frac{\tilde{p}(Z_i)}{f_Z(Z_i \mid W_i)} \Big\{ \mu(x_0, Z_i, W_i) - \psi_{x_0}(Q;\tilde p_z)   \Big\} \Big\{ \I(X_i=x_0) -  \pi(x_0 \mid Z_i, W_i) \Big\}}_{\Phi_{X,x_0}(Q; \, \tilde{p}_z)(O_i)}
     \\[0.35cm]
    &\hspace{-2.2cm} + \frac{1}{\kappa_{x_0, 2}(Q;\tilde{p}_z)} \underbrace{\int  \pi(x_0 \mid z, W_i) \Big\{  \mu(x_0, z, W_i) -  \psi_{x_0}(Q; \tilde p_z) \Big\} \, \tilde{p}(z) dz}_{\Phi_{W,x_0}(Q; \, \tilde{p}_z)(O_i)} \ . 
\end{aligned}
\end{equation}
}
\end{lemma}

See a proof in Appendix~\ref{app:proofs:lem:np_EIF}. 

Under discrete $Z$, $\tilde{p}(Z)$ can be replaced by $\I(Z = z^*)$ in $\Phi_{Y,x_0}$ and $\Phi_{X,x_0}$, while $\Phi_{W,x_0}$ simplifies to $\Phi_{W,x_0}(Q; z^*)(O_i) = \displaystyle \frac{\pi(x_0 \mid z^*, W_i)}{\kappa_{x_0,2}(Q; z^*)} \{ \mu(x_0, z^*, W_i) - \psi_{x_0}(Q; z^*) \}$. 

The influence function $\Phi_{x_0}(Q;\tilde{p}_z)$ admits a decomposition into components corresponding to variationally independent nuisance functionals, with each component lying in a tangent space spanned by the corresponding score function from the observed data distribution factorization $p(o)=p(y\, | \, x, z, w)\, p(x\, | \, z, w) \, p(z\, | \, w)\, p(w)$. This yields contributions associated with $\mu, \pi$, and $p_W$, denoted $\Phi_{Y,x_0}(Q;\tilde{p}_z)$, $\Phi_{X,x_0}(Q;\tilde{p}_z)$, and $\Phi_{W,x_0}(Q;\tilde{p}_z)$, respectively. Importantly, the projection onto the tangent space for $p(z\, | \, w)$ is identically zero, so knowledge of $f_Z$ provides no efficiency gain relative to estimating it from the observed data. 

There are several strategies for addressing the first-order bias term $-P{\Phi_{x_0}(\hat Q;\tilde{p}_z)}$, with $\Phi_{x_0}(Q; \tilde{p}_z)$ defined in \eqref{eq:IF_tilde_pz}. We next describe three estimators that implement these strategies. 

\subsubsection{Estimating equation}  
\label{subsec:est_eq}

The first approach solves the estimating equation $P_n\Phi_{x_0}(\hat Q;\tilde{p}_z)=0$, thereby removing the empirical first-order bias term by construction; here $P_n f \coloneqq \frac{1}{n}\sum_{i = 1}^n f(O_i)$. The resulting estimator of $\psi_{x_0}(Q;\tilde{p}_z)$ in \eqref{eq:ID_tilde_pZ}, denoted $\psi^{\text{ee}}_{x_0}(\hat Q;\tilde{p}_z)$, is given by  

\vspace{-1.5cm}
\begin{align}\label{eq:est_esteq_tilde_pz}
    &\psi^\text{ee}_{x_0}(\hat{Q}; \tilde{p}_z) 
    = {\kappa^{\mathrm{aipw}}_{x_0,1}(\hat Q;\tilde p_z)} \, / \, {\kappa^{\mathrm{aipw}}_{x_0,2}(\hat Q; \tilde p_z)} \ ,
\end{align}%

\vspace{-.75cm}
where the numerator and denominator coincide with the augmented inverse probability weighting (AIPW) estimators for $\kappa_{x_0,1}(Q;\tilde p_z)$ and $\kappa_{x_0,2}(Q;\tilde p_z)$, respectively. That is, 

\vspace{-1.cm}
{\small 
\begin{align*}
    \kappa^{\mathrm{aipw}}_{x_0,1}(\hat Q;\tilde p_z) 
    &= \frac{1}{n}\sum_{i=1}^n \frac{\tilde p(Z_i)}{\hat f_Z(Z_i\mid W_i)}
    \big\{\I(X_i=x_0)Y_i - \hat\mu(x_0,Z_i,W_i)\hat\pi(x_0\mid Z_i,W_i)\big\} 
    + \kappa^{\text{pi}}_{x_0,1}(\hat{Q}; \tilde p_z)  \ , \\[0.5cm]
    \kappa^{\mathrm{aipw}}_{x_0,2}(\hat Q; \tilde p_z) 
    &= \frac{1}{n}\sum_{i=1}^n \frac{\tilde p(Z_i)}{\hat f_Z(Z_i\mid W_i)}
    \big\{\I(X_i=x_0) - \hat\pi(x_0\mid Z_i,W_i)\big\} 
    + \kappa^{\text{pi}}_{x_0,2}(\hat{Q}; \tilde p_z) \ .
\end{align*}
}

\vspace{-0.25cm}
When $Z$ is discrete, we can set $\tilde p(z^*) = 1$ and replace $\tilde p(Z_i)$ with $\I(Z_i = z^*)$. Furthermore, $\kappa^{\text{pi}}_{x_0,2}(\hat{Q}; \tilde p_z) $ and $\kappa^{\text{pi}}_{x_0,2}(\hat{Q}; \tilde p_z) $ would be replaced by $\kappa^{\text{pi}}_{x_0,2}(\hat{Q}; z^*) $ and $\kappa^{\text{pi}}_{x_0,2}(\hat{Q}; z^*)$. 

Detailed derivations are provided in Appendix~\ref{subsubsec:esteq}. 

\subsubsection{One-step corrected plug-in estimator}
\label{subsec:one-step}

An alternative approach augments the plug-in estimator in \eqref{eq:est_plug_contZ} with an empirical estimate of its first-order bias, yielding the one-step estimator $\psi^+_{x_0}(\hat{Q}; \tilde{p}_z)=\psi_{x_0}^{\text{pi}}(\hat{Q};\tilde{p}_z) + P_n\Phi_{x_0}(\hat{Q};\tilde{p}_z)$. Unlike many causal models (e.g., back-door, front-door, and primal-fixable models), where the one-step and estimating equation estimators coincide, they differ under the Napkin model. The explicit form of the one-step estimator for $\psi_{x_0}(Q; \tilde{p}_z)$ in \eqref{eq:ID_tilde_pZ} is given by  

\vspace{-1.25cm}
{\small
\begin{align}
    \psi^+_{x_0}(\hat{Q}; \tilde{p}_z)
    &= \frac{1}{n \, \kappa^{\text{aipw}}_{x_0, 2}(\hat{Q}; \tilde p_z)} \sum_{i=1}^{n}\Bigg\{ \I(X_i=x_0) \frac{\tilde{p}(Z_i)}{\hat{f}_Z(Z_i \mid W_i)} \Big\{  Y_i -  \hat{\mu}(x_0, Z_i, W_i) \Big\} \notag
     \\[0.35cm] 
    &\hspace{1cm} + \frac{\tilde{p}(Z_i)}{\hat{f}_Z(Z_i \mid W_i)} \Big\{ \hat{\mu}(x_0, Z_i, W_i) - \psi_{x_0}^{\text{pi}}(\hat{Q};\tilde p_z)   \Big\} \Big\{ \I(X_i=x_0) -  \hat{\pi}(x_0 \mid Z_i, W_i) \Big\} \label{eq:est_one-step_tilde_pz}
     \\[0.35cm]
    &\hspace{1cm} + \int \hat{\pi}(x_0 \mid z, W_i) \Big\{ \hat{\mu}(x_0, z, W_i)  - \psi_{x_0}^{\text{pi}}(\hat{Q};\tilde p_z) \Big\} \ \tilde{p}(z)\ dz\Bigg\} + \psi_{x_0}^{\text{pi}}(\hat{Q};\tilde{p}_z) \ .  \notag 
\end{align}
}

\vspace{-0.25cm}
When $Z$ is discrete, we can set $\tilde p(z^*) = 1$ and replace $\tilde p(Z_i)$ with $\I(Z_i = z^*)$. Furthermore, $\kappa_{x_0,2}^{\text{aipw}}(\hat{Q};\tilde{p}_z)$ and $\psi_{x_0}^{\text{pi}}(\hat{Q};\tilde{p}_z)$ are replaced by $\kappa_{x_0,2}^{\text{aipw}}(\hat{Q};z^*)$ and $\psi_{x_0}^{\text{pi}}(\hat{Q};z^*)$, respectively. 

\subsubsection{Targeted minimum loss-based estimator}
\label{subsec:tmle}

Our third estimator is a targeted minimum loss-based estimator (TMLE), which updates the initial nuisance estimates $\hat{Q}$ through a targeting step to obtain $\hat{Q}^*$ satisfying $P_n\Phi_{x_0}(\hat{Q}^*; \tilde{p}_z)=\smallO(n^{-1/2})$, thereby rendering the empirical mean of the influence function asymptotically negligible \citep{van2011targeted,van2016one}. 

Let $Q_j$ denote a nuisance parameter in $Q$, and $Q_{-j}$ the collection of all other nuisances; with the corresponding estimates denoted by $\hat{Q}_j$ and $\hat{Q}_{-j}$. Let $\mathcal{M}_{Q_j}$ denote the model space of $Q_j$. To target $\hat{Q}_j$, we define a loss function $L(\tilde{Q}_j;\hat{Q}_{-j})$, for $\tilde{Q}_j \in \mathcal{M}_{Q_j}$, and a parametric submodel $\hat{Q}_j(\varepsilon_j;\hat{Q}_{-j})$, indexed by a univariate real-valued parameter $\varepsilon_j$. We include $\hat{Q}_{-j}$ in the argument to point out that the loss function and submodel may rely on other nuisance estimates apart from $\hat{Q}_j$. A valid parametric submodel and loss function pair needs to satisfy the following three conditions: 
(C1) The submodel passes through the initial estimate at $\varepsilon_j = 0$: $\hat{Q}_j(0;\hat{Q}_{-j}) = \hat{Q}_j$; 
(C2) The true nuisance minimizes the expected loss: $Q_{j} = \arg \min_{\tilde{Q}_{j} \in \mathcal{M}_{Q_{j}}} \int L(\tilde{Q}_{j} ; Q_{-j})(o) \, p(o) \, d o$; and 
(C3) The derivative of the loss function at $\varepsilon_j = 0$ is proportional to the corresponding component of the influence function, denoted by $\Phi_{j}$: $\frac{\partial}{\partial \varepsilon_{j}} L(\hat{Q}_{j}(\varepsilon_{j} ; \hat{Q}_{-j}) ; \hat{Q}_{-j}) \big|_{\varepsilon_{j}=0} \propto \Phi_{j}(\hat{Q})$. 

The empirical first-order bias, $P_n \Phi_{x_0}(\hat{Q};\tilde{p}_z)$, decomposes as $P_n \Phi_{Y,x_0}(\hat{Q};\tilde{p}_z) + P_n \Phi_{X,x_0}(\hat{Q};\tilde{p}_z)+P_n \Phi_{W,x_0}(\hat{Q};\tilde{p}_z)$; see \eqref{eq:IF_tilde_pz} for the definitions of $\Phi_{Y,x_0}, \Phi_{X,x_0}, \Phi_{W,x_0}$. To ensure $P_n \Phi_{x_0}(\hat{Q};\tilde{p}_z)$ is negligible, it suffices to ensure each component is $\smallO(n^{-1/2})$. Given that $\hat{p}_{W}$ is empirically estimated, 
$P_n \Phi_{W,x_0}(\hat{Q};\tilde{p}_z) = \smallO(n^{-1/2})$ once the remaining nuisances are successfully targeted. Updating $\hat{f}_Z$ is also unnecessary, as the influence function projects to zero onto its tangent space. Consequently, the targeting reduces to updating $\hat{\mu}$ and $\hat{\pi}$. With $\hat{Q}^*=\{\hat{\mu}^*, \hat{\pi}^*, \hat{f}_Z, \hat{p}_{W}\}$, the TMLE, denoted by $\psi_{x_0}(\hat{Q}^*,\tilde{p}_z)$, is a plug-in estimator evaluated at $\hat{Q}^*$:
\vspace{-0.25cm}
\begin{align}
    \psi_{x_0}(\hat{Q}^*;\tilde{p}_z) = \frac{ \sum_{i=1}^n \left\{ \int \hat{\mu}^*(x_0, z, W_i) \ \hat{\pi}^*(x_0 \mid z, W_i) \ \tilde{p}(z) \ dz \right\} }{ \sum_{i = 1}^n  \left\{ \int \hat{\pi}^*(x_0 \mid z, W_i) \ \tilde{p}(z) \ dz \right\} } \ .
    \label{eq:est_tmle_tilde_pz}
\end{align}

\vspace{-0.35cm}
We next outline the targeting of $\hat{\mu}$ and $\hat{\pi}$ with their associated loss functions and submodels:
{\small\begin{align*} L_Y(\tilde{\mu};\hat{\pi}, \hat{f}_Z)=\hat{H}_Y(X,Z,W;\tilde{p}_z)\left\{Y_{i}-\tilde{\mu}\left(x_{0}, Z, W\right)\right\}^2, \quad \hat{\mu}(\varepsilon_Y) = \hat{\mu}(x_{0}, Z, W) + \varepsilon_Y,
\end{align*}}
with $\hat{H}_Y(X,Z,W;\tilde{p}_z)= \{\mathbb{I}(X=x_{0}) \, \tilde{p}(Z)\}/ \hat{f}_{Z}\left(Z \mid W\right)$, and 

\vspace{-1.35cm}
{\small\begin{align*}
    L_X(\tilde{\pi};\hat{f}_Z) &= - \{\tilde{p}(Z)/\hat{f}_Z(Z \mid W)\} \, \mathrm{log} \ \tilde{\pi}(X\mid Z,W) \ , \\
    \hat{\pi}(\varepsilon_X; \hat{\mu}, \hat{\pi}, \hat{f}_Z) &= \mathrm{expit}\big\{\mathrm{logit} \ \hat{\pi}(x_0\mid Z,W) + \varepsilon_X \hat{H}_X(Z,W;\tilde{p}_z)\big\} \ ,
\end{align*}
}%
with $\hat{H}_X(Z,W;\tilde{p}_z) = \hat{\mu}(x_0, Z, W) - \psi^{\text{pi}}_{x_0}(\hat{Q}; \tilde p_z) $. The constant term $1/\kappa_{x_0,2}(Q; \tilde{p}_z)$ is omitted when defining the loss function and submodel for $\mu$ and $\pi$, thereby removing the dependence of $L_Y$ on the estimate of $\pi$ and avoiding iterative updates between the nuisance estimates of $\mu$ and $\pi$, while still ensuring that the derivative of the loss function with respect to $\varepsilon_Y$ is proportional to the corresponding influence-function component. 

See Appendix~\ref{app:tmle_validity} for a proof of validity of these submodel–loss function pairs under (C1)-(C3). These pairs only apply when the outcome $Y$ is continuous; extensions for binary $Y$ are discussed in Appendix~\ref{app:tmle_binary}. 

We start by targeting $\hat{\mu}$ to obtain $\hat{\mu}^*$ in one step, and then target $\hat{\pi}$ to obtain $\hat{\pi}^*$, while keeping $\hat{f}_Z$ and $\hat{p}_W$ unchanged. Because the submodel for $\hat{\pi}$ depends on $\hat{\pi}$ itself through $\psi^{\mathrm{pi}}_{x_0}(\hat{Q};\tilde{p}_z)$, within-nuisance targeting is required. As $\hat{\pi}$ is updated, the submodel changes accordingly, necessitating repeated re-targeting until convergence according to a pre-specified stopping criterion $C_{n,\mathrm{stop}} = o(n^{-1/2})$; e.g., $C_{n,\mathrm{stop}} = n^{-1/2}/\log(n)$ for sample size $n$.

\textit{(T1): One-step risk minimization for $\mu$.} 
$\hat{\mu}$ is updated by finding $\hat{\varepsilon}_Y$ that minimizes the loss function $L_Y$. The optimization problem can be solved via a weighted regression: $Y \sim \mathrm{offset}(\hat{\mu}(x_{0}, Z, W))$, with weight $\hat{H}_Y(Z,W;\tilde{p}_z)=\{\mathbb{I}(X=x_{0}) \,  \tilde{p}(Z)\}/ \hat{f}_{Z}\left(Z\, | \, W\right)$. The intercept coefficient gives $\hat{\varepsilon}_Y$, satisfying: $ \hat{\varepsilon}_Y=\underset{\varepsilon_Y \in \mathbb{R}}{\arg \min } \ P_n \Phi_{Y,x_0}(\hat{Q};\tilde{p}_z)$. 

The updates are then made: $\hat{\mu}^*(x_{0}, Z, W)=\hat{\mu}(x_{0}, Z, W) + \hat{\varepsilon}_Y$ and $\hat{Q}=\{\hat{\mu}^*, \hat{\pi}, \hat{f}_Z, \hat{p}_{W}\}$.

Let $\hat{Q}^{(t)} = \{\hat{\mu}^{*}, \hat{\pi}^{(t)}, \hat{f}_Z, \hat{p}_{W}\}$ denote the nuisance estimates after completing the $t$-th within-nuisance targeting iteration. We initialize the process by setting $\hat{Q}^{(0)} = \hat{Q}$.

\textit{(T2): Iterative risk minimization for $\pi$.} 
Given $\hat{\pi}^{(t)}$ as a starting point, the within-nuisance targeting is iteratively performed to update $\hat{\pi}^{(t)}$ to $\hat{\pi}^{(t+1)}$.

We aim to find $\hat{\varepsilon}^{(t+1)}_X$ that minimizes the loss function $L_X$. The optimization problem can be solved via a weighted logistic regression without intercept: $\I(X=x_0) \sim \mathrm{offset}(\hat{\pi}^{(t)}(x_0\, | \, Z,W)) + \hat{H}^{(t)}_X(Z,W)$, with weight ${\tilde{p}(Z)}/{\hat{f}_Z(Z\, | \, W)}$, where $\hat{H}^{(t)}_X(Z,W)$ denotes the auxiliary variable evaluated using $\hat{Q}^{(t)}$. The auxiliary variable's coefficient is $\hat{\varepsilon}^{(t+1)}_X$ which solves $\underset{\varepsilon_X \in \mathbb{R}}{\arg \min } \ P_n \Phi_{X,x_0}(\hat{Q}^{(t)};\tilde{p}_z)$. Following update is then made:
\vspace{-0.35cm}
\begin{align*}
    &\hat{\pi}^{(t+1)}(\hat{\varepsilon}^{(t+1)}_X; \hat{\mu}^{(t)} ,\hat{f}_Z) = \mathrm{expit}\big\{\mathrm{logit} \hat{\pi}^{(t)}(x_0\mid z^*,W) + \hat{\varepsilon}^{(t+1)}_X \ \hat{H}^{(t)}_{X}(Z,W)\big\} \ .
\end{align*}

\vspace{-0.65cm}
This step is repeated until the pre-specified stopping criterion $C_{n, \mathrm{stop}}$ is satisfied at some iteration $t_X^*$, where we have: $P_n \Phi_{X,x_0}(\hat{Q}^{(t_X^*)};\tilde{p}(Z))\leq C_{n, \mathrm{stop}}$. This completes Step (T2), with the final update given by $\hat{\pi}^{(t_X^*)}=\hat{\pi}^*$. We then define $\hat{Q}^* = \{\hat{\mu}^*, \hat{\pi}^*, \hat{f}_Z, \hat{p}_{W}\}$.

A full presentation of the TMLE procedure is provided in Algorithm~\ref{appalg:contiZ} in Appendix~\ref{app:tmle_algo}.

The TMLE procedure for discrete $Z$ proceeds analogously, yielding the plug-in TMLE, denoted by $\psi_{x_0}(\hat{Q}^*; z^*)$:

\vspace{-1.5cm}
\begin{align}
\psi_{x_0}(\hat{Q}^{*}; z^*) 
= {\displaystyle \frac{1}{n}\sum_{i=1}^n \hat{\mu}^{*}(x_0, z^*, W_i) \ \hat{\pi}^{*}(x_0 \mid z^*, W_i) }\Big/{\displaystyle \frac{1}{n}\sum_{i = 1}^n \hat{\pi}^{*}(x_0 \mid z^*, W_i)} \ ,
\label{eq:est_tmle_z^*}
\end{align}

\begin{remark}
The iterative update for $\pi$ can be avoided by using the IPW estimator for $\psi_{x_0}(Q; \tilde{p}_z)$, rather than the plug-in estimator, when defining $\hat{H}_X(Z,W)$. This is because the plug-in estimator for $\psi_{x_0}(Q; \tilde{p}_z)$ depends directly on $\hat{\pi}$, whereas the IPW estimator depends only on the nuisance estimate $\hat{f}_Z$; the IPW estimator is given by 
$\psi_{x_0}^{\text{ipw}}(\hat{Q}; \tilde{p}_z) = \E[\{ {\tilde p(Z)}/{f_Z(Z \mid W)} \} \I(X=x_0) \, Y] \, / \, \E[ \{ {\tilde p(Z)}/{f_Z(Z \mid W)} \} \I(X=x_0)]$. Consequently, updating $\pi$ does not require re-estimating the auxiliary variable $\hat{H}_X(Z,W)$, and iterations can be avoided. Nevertheless, we adopt the plug-in estimator for $\psi_{x_0}(Q;\tilde{p}_z)$ when constructing TMLEs to maintain consistency with our use of the plug-in estimator as the non-IF-based estimator throughout the paper.
\end{remark}

\subsection{Asymptotic linearity} 
\label{subsec:asym}

We now examine the asymptotic behaviors of our proposed estimators in Section~\ref{subsec:estimators}. Given an influence-function based estimator $\psi_{x_0}(\hat{Q})$ of the parameter $\psi_{x_0}(Q)$, we can write the von Mises expansion as: 

\vspace{-0.75cm}
{\small\begin{equation}\label{eq:von_mise}
\begin{aligned}
    \psi_{x_0}(\hat{Q})-\psi_{x_0}(Q)
    = P_n \Phi_{x_0}(Q)
    +\left(P_n-P\right)\{\Phi_{x_0}(\hat{Q})-\Phi_{x_0}(Q)\}
    +R_2(\hat{Q}, Q) \ .
\end{aligned}
\end{equation}
}%

\vspace{-0.25cm}
The first term in the expansion is the sample average of $\Phi_{x_0}(Q)$ which is mean zero, and thus enjoys   $o_P(n^{-1/2})$ asymptotic behavior according to the central limit theorem. The second term is an empirical process term, and enjoys $o_P(n^{-1/2})$ asymptotic behavior if $P\{\Phi_{x_0}(\hat{Q})-\Phi_{x_0}(Q)\}^2=o_P(1)$ and $\Phi_{x_0}(\hat{Q})-\Phi_{x_0}(Q)$ falls in a P-Donsker class with probability tending to 1, where P-Donsker condition limits the complexity of the nuisance models. This condition can be avoided if sample splitting or cross-fitting is adopted for nuisance estimation \citep{zheng2011cross, chernozhukov2018double}. Consequently, the asymptotic linearity of $ \psi_{x_0}(\hat{Q})$  is achieved if the third term enjoys $o_P(n^{-1/2})$ asymptotic behavior, which would then entail rate conditions on nuisance estimates in $\hat{Q}$. Below, we derive the form of the $R_2$ term for our estimators, and determine rate conditions on the nuisance estimates to ensure $R_2 = \smallO(n^{-1/2})$. 

We examine the asymptotic linearity of the three proposed estimators for $\psi_{x_0}(Q; \tilde{p}_z)$ in \eqref{eq:ID_tilde_pZ}: the estimating-equation estimator $\psi^\text{ee}_{x_0}(\hat{Q}; \tilde{p}_z)$ defined in Section~\ref{subsec:est_eq} and \eqref{eq:est_esteq_tilde_pz}; the one-step estimator $\psi^+_{x_0}(\hat{Q}; \tilde{p}_z)$ defined in Section~\ref{subsec:one-step} and \eqref{eq:est_one-step_tilde_pz}; and the TMLE $\psi_{x_0}(\hat{Q}^*; \tilde{p}_z)$ defined in Section~\ref{subsec:tmle} and \eqref{eq:est_tmle_tilde_pz}. We use the one-step estimator as the representative case to present the results, which apply analogously to the other two estimators. The IF-based estimators for discrete $Z$ obtained using a fixed $z^*$ correspond to the choice $\tilde{p}(z^*)=1$. They therefore constitute as a special case, and all results developed below continue to hold.

Let $\|f\| = (Pf^2)^{1/2}$ denote the $L^2(P)$-norm of the function $f$. Under the regularity conditions detailed in  Appendix~\ref{app:proofs_inf_continuous}, $R_2(\hat{Q}, Q)$ can be bounded above by a product of $L^2(P)$-norms, for some finite positive constant $C$: (see a proof in   Appendix~\ref{app:proofs_inf_continuous}.)
\vspace{-0.25cm}
\begin{align*}
    R_2(\hat{Q}, Q) &\leq C\big\{\|\hat{f}_Z-f_Z\| \|\hat{\mu}-\mu\|+\|\hat{f}_Z-f_Z\| \|\hat{\pi}-\pi\|\big\}.
\end{align*}

\vspace{-0.5cm}
We have the following theorem establishing asymptotic linearity of the one-step estimator $\psi^+_{x_0}(\hat{Q}; \tilde{p}_z)$ (and equivalently the other two estimators). 

\begin{theorem}\label{thm:asymp_psi_onestep_contZ}
Assume the \emph{$L^2(P)$} convergence rates of nuisance estimates in $\hat{Q}$ are as follows: 
$|| \hat{\pi} - \pi || =\smallO(n^{-\frac{1}{k}})$, 
$|| \hat{f}_{Z} - f_{Z} || =\smallO(n^{-\frac{1}{b}})$, 
$|| \hat{\mu} - \mu|| = \smallO(n^{-\frac{1}{q}})$. Under the regularity conditions discussed in Appendix~\ref{app:proofs_inf_continuous}, if $\frac{1}{b} + \frac{1}{q} \geq \frac{1}{2}$ and $\frac{1}{k} + \frac{1}{b} \geq \frac{1}{2}$, then the one-step estimator $\psi^{+}_{x_0}(\hat{Q}; \tilde{p}_z)$ is asymptotically linear, that is: $\psi^{+}_{x_0}(\hat{Q}; \tilde{p}_z) - \psi_{x_0}(Q; \tilde{p}_z) = P_n \Phi_{x_0}(Q;\tilde{p}_z) + o_P(n^{-1/2})$ where $\Phi_{x_0}(Q;\tilde{p}_z)$ is given in Lemma~\ref{lem:np_EIF}. 
\end{theorem}
See a proof in Appendix~\ref{app:proofs:thm:asymp_psi_onestep_contZ}. 

The above results suggest that the relevant nuisance parameters can be estimated at rates slower than $\smallO(n^{-1/2})$, thereby broadening the applicability of flexible machine learning and statistical models for nuisance estimation. An immediate corollary of Theorem~\ref{thm:asymp_psi_onestep_contZ} is that our estimators are doubly robust, as formalized below. 

\begin{corollary}\label{cor:robust_psi_onestep_contZ}
$\psi^{+}_{x_0}(\hat{Q}; \tilde{p}_z)$ is consistent for $\psi_{x_0}(Q; \tilde{p}_z)$ if either (i) $||\hat{\pi} - \pi|| = o_P(1) \ \text{and} \ || \hat{\mu} - \mu || = o_P(1)$, or (ii) $|| \hat{f}_Z - f_Z || = o_P(1)$, or both (i) and (ii) hold.
\end{corollary}
See a proof in Appendix~\ref{app:proofs:robust_psi_onestep_contZ}. 

The double robustness property of $\psi^{+}_{x_0}(\hat{Q}; \tilde{p}_z)$ follows from the double robustness of the influence-function-based estimators for $\kappa_{x_0,1}(Q;\tilde{p}_z)$ and $\kappa_{x_0,2}(Q;\tilde{p}_z)$. This is most easily seen in the estimator $\psi_{x_0}^\text{ee}(\hat{Q}; \tilde{p}_z)$, defined as the ratio of two AIPW estimators, $\kappa^{\mathrm{aipw}}_{x_0,1}(\hat{Q};\tilde{p}_z)$ and $\kappa^{\mathrm{aipw}}_{x_0,2}(\hat{Q};\tilde{p}_z)$. Existing theory for AIPW estimators in back-door models \citep{robins1994estimation, scharfstein1999adjusting} implies that $\kappa^{\mathrm{aipw}}_{x_0,1}(\hat{Q};\tilde{p}_z)$ is a consistent estimator of $\kappa_{x_0,1}(Q;\tilde{p}_z)$ if either $\hat{f}_Z$ is consistent, or both $\hat{\mu}$ and $\hat{\pi}$ are consistent. Similarly, $\kappa^{\mathrm{aipw}}_{x_0,2}(\hat{Q};\tilde{p}_z)$ is consistent for $\kappa_{x_0,2}(Q;\tilde{p}_z)$ if either $\hat{f}_Z$ or $\hat{\pi}$ is consistent. Consequently, the estimator $\psi^\text{ee}_{x_0}(\hat{Q}; \tilde{p}_z)$ inherits double robustness as long as conditions are met to ensure the consistency of both $\kappa^{\mathrm{aipw}}_{x_0,1}(\hat{Q};\tilde{p}_z)$ and $\kappa^{\mathrm{aipw}}_{x_0,2}(\hat{Q};\tilde{p}_z)$. The same intuition applies to the other two influence function-based estimators. 

\section{Inference in a semiparametric Napkin model}
\label{sec:efficiency}

Section~\ref{sec:est} developed estimators for $\E(Y^{x_0})$ based on the efficient influence function under the nonparametric observed-data model. Under a semiparametric model, however, influence functions are generally not unique. If $\Phi$ is any influence function and $\mathcal T$ denotes the (model) tangent space, then the efficient influence function is given by the projection of $\Phi$ onto $\mathcal T$. Equivalently, the collection of all influence functions is obtained by adding arbitrary elements of the orthocomplement tangent space $\mathcal T^\perp$ to any valid influence function. Therefore, characterizing $\mathcal T^\perp$ immediately characterizes the full class of influence functions and enables identification of the efficient influence function.

In the Napkin model, a \textit{Verma constraint} (assumption~(iii)) induces additional structure on the observed-data distribution, enlarging $\mathcal T^\perp$ relative to the nonparametric model and creating opportunities for efficiency gains. In this section, we consider a semiparametric model induced by the Verma constraint, characterize the resulting class of influence functions by deriving the orthocomplement of the model tangent space, and identify the corresponding efficient influence function.

\begin{remark}
\label{rem:heuristic}
In Lemma~\ref{lem:id}, the Verma constraint or assumption~(iii) is leveraged to identify $\E (Y^{x_0})$. The Verma constraint in fact over-identifies $\E (Y^{x_0})$, in the sense that we can construct multiple differentiable identification functionals. Based on this reasoning, the Napkin model is a locally semiparametric model and must have a nontrivial tangent space.
\end{remark}

\subsection{A Verma-constrained observed-data model}
\label{sec:verma}

The statistical model associated with a latent DAG is characterized by \textit{equality restrictions} on the observed-data distribution. Such restrictions can take two forms: (i) ordinary conditional independencies implied by d-separation \citep{pearl2009causality}, and (ii) generalized independencies, also known as Verma constraints, which arise in post-intervention distributions obtained by fixing certain variables \citep{verma90equiv, spirtes01causation}.

In a DAG without latent variables, a missing edge implies an ordinary independence constraint on the observed-data distribution through the Markov property, either marginally or conditionally. In latent DAGs, however, a missing edge may or may not induce a restriction on the observed-data distribution, and when it does, the resulting restriction need not be expressible as an ordinary conditional independence \citep{verma90equiv, richardson2023nested}.

Some latent DAGs are \emph{nonparametrically saturated}, meaning that they impose no restrictions whatsoever on the observed-data distribution. In such cases, semiparametric efficiency theory reduces to the unrestricted nonparametric model. A sound and complete graphical characterization of saturation in latent DAGs is provided by \citet{bhattacharya2022semiparametric}.

The Napkin graph is \textbf{not} nonparametrically saturated, as also indicated in Remark~\ref{rem:heuristic}. The missing edge between $Z$ and $Y$ in Figure~\ref{fig:graphs}(a) induces a restriction on the observed-data distribution. However, this restriction is not an ordinary conditional independence. Rather, it is a Verma constraint that becomes visible after intervening on $Z$, as represented by the conditional DAG in Figure~\ref{fig:graphs}(c). Since fixing $Z$ removes all incoming edges into $Z$, d-separation in the resulting graph implies the post-intervention independence $Z \perp Y \mid X$. 

This graphical restriction translates into a probabilistic form through the corresponding post-intervention Markov kernel \citep{richardson2023nested}, denoted $q(Y,X,W\, | \, Z)$:
\vspace{-0.5cm}
\begin{align*} 
q(Y, X, W \mid Z) \coloneqq p(Y \mid X, Z, W)\ p(X \mid Z, W)\ p(W) \ . 
\end{align*}

\vspace{-0.7cm}
The post-intervention independence $Z \perp Y \, | \, X$ implies that the conditional kernel $q(Y \, | \, X,Z)$ is invariant to $Z$. This is the Verma constraint associated with the Napkin graph. 

The conditional kernel $q(Y \, | \, X,Z)$ is derived as: 
\vspace{-0.25cm}
\begin{equation}\label{eq:kernel_verma}
    \begin{aligned}
    q(Y\mid X,Z)  
    =\frac{\int q(Y,X,w\mid Z) \, dw }{\int q(y,X,w\mid Z)\, dy\, dw}
    =\frac{\int p(Y\mid X,Z,w)\, p(X\mid Z,w)\ p(w) \, dw}{\int p(X\mid Z,w)\, p(w) \, dw} \ ,
    \end{aligned}
\end{equation}

\vspace{-0.35cm}\noindent
where $y$ and $w$ are values in the respective domains of $Y$ and $W$. Therefore, the Verma constraint is equivalent to requiring that the right-hand side of \eqref{eq:kernel_verma} be invariant to $Z$. 

The identification functional \eqref{eq:ID_tilde_pZ} is closely connected to this restriction. Indeed, \eqref{eq:kernel_verma} implies
\vspace{-0.75cm}
\begin{align*}
    \E_q(Y\mid X=x_0,Z=z) = \displaystyle \frac{\int y \, p(y\mid x_0,z,w)\, p(x_0\mid z,w)\ p(w) \, dy \, dw}{\int p(x_0\mid z,w)\, p(w) \, dw} \ ,
\end{align*}

\vspace{-0.35cm} \noindent 
which coincides with \eqref{eq:ID_tilde_pZ}. Consequently, the identification result relies only on a mean-scale implication of the full Verma constraint. While the full Verma constraint requires invariance of the entire conditional kernel $q(Y \, | \, X,Z)$, the identification functional depends on this kernel only through its conditional mean. 

Motivated by this observation, we consider the semiparametric observed-data model consisting of distributions for which $\E_q(Y\mid X=x_0,Z=z)$ is invariant to $z$. This mean-scale Verma constraint reduces the size of the observed-data model relative to the unrestricted nonparametric model and thereby creates opportunities for efficiency gains. The remainder of this section characterizes the orthocomplement of the tangent space under this semiparametric statistical model and uses it to derive the class of influence functions and the corresponding semiparametric efficient estimators.

\subsection{The set of influence functions under the semiparametric model}
\label{subsec:weighted_est}

We now characterize the influence functions for $\E(Y^{x_0})$ under the mean-scale Verma model introduced above. Let $\mathcal T$ denote the tangent space of the mean-scale Verma model and let $\mathcal T^\perp$ denote its orthocomplement in $L_0^2(P)$. Since every influence function under the restricted model can be expressed as $\Phi+\kappa$, where $\Phi$ is any valid influence function and $\kappa \in\mathcal T^\perp$, it suffices to characterize $\mathcal T^\perp$. 

In the present setting, this orthocomplement can be derived by expressing the mean-scale Verma restriction as a collection of moment restrictions. Let
$
\psi_{x_0}(P;z) =
\frac{\kappa_{x_0,1}(P;z)}{\kappa_{x_0,2}(P;z)}
$
be the functional in \eqref{eq:ID_z*} as a function of $z$. The mean-scale Verma constraint is equivalent to requiring that $\psi_{x_0}(P;z)$ be constant in $Z$. Equivalently, for every square-integrable function $m$ of $Z$,
\vspace{-0.5cm}
\begin{align}
\E\Big[
\psi_{x_0}(P;Z)\big\{m(Z)-\E[m(Z)]\big\}
\Big]
=0 \ . 
\end{align}

\vspace{-0.5cm}
Differentiating these moment restrictions along regular parametric submodels through $P$ yields the elements orthogonal to the tangent space of the semiparametric model. This argument leads to the following characterization.

\begin{theorem}\label{thm:orthocomp}
Consider the semiparametric observed-data model defined by the mean-scale Verma restriction that $\E_q(Y\mid X=x_0,Z=z)$ is invariant to $z$. The orthocomplement of the model tangent space is 
\vspace{-0.6cm}
\begin{align}
\label{eq:orthocomp}
\mathcal T^\perp
=
\left\{
\int \Phi_{x_0}(Q;z) \, c(z) \, dz 
:
\int c(z)\,dz=0 
\right\},
\end{align}

\vspace{-0.75cm} \noindent 
where $\Phi_{x_0}(Q;z)$ denotes the nonparametric influence function corresponding to the identification functional $\psi_{x_0}(P;z)$ for a fixed value $z$ of $Z$, as introduced after Lemma~\ref{lem:np_EIF} in Section~\ref{subsec:estimators}. Consequently, the set of all influence functions for $\E(Y^{x_0})$ under this semiparametric model, denoted by $\mathcal I(\psi_{x_0})$ is
\vspace{-0.35cm}
\begin{align}
\label{eq:all-IF}
\mathcal I(\psi_{x_0}) = 
\left\{
\int \Phi_{x_0}(Q;z) \, \tilde p(z) \,dz
:
\int \tilde p(z)\,dz=1
\right\}.
\end{align}
\end{theorem}

\vspace{-0.35cm}
See a proof in Appendix~\ref{app:proofs_orthocom}. 

Theorem~\ref{thm:orthocomp} shows that every influence function under the semiparametric model is a weighted average of the fixed-z influence functions. Different weighting schemes therefore give rise to different regular asymptotically linear estimators of the same parameter, potentially with different asymptotic variances. The efficient influence function is the minimum-variance element of this class. 

\subsection{Semiparametric efficient influence functions}

We now identify the minimum-variance element of the influence function class in \eqref{eq:all-IF}, corresponding to the efficient influence function of $\E (Y^{x_0})$. The efficient influence function can help yield estimators that attain the semiparametric efficiency bound. We next divide our characterization of the efficient influence function into two parts, depending on whether $Z$ is discrete or continuous.

\subsubsection{Efficient estimation under discrete $Z$}

We first consider the case where $Z$ is discrete with support $\{1,\ldots,K\}.$
\begin{lemma}\label{lemma:EIF-discreteZ}
Every influence function in \eqref{eq:all-IF} takes the form of a weighted average of $\Phi_{x_0}(Q;z^*=k)$, for $k = 1, \ldots, K$. Let $\alpha_k = \tilde p(z^*=k)$. Thus, the class in \eqref{eq:all-IF} reduces to:
\vspace{-0.25cm}
\begin{align*}
    \mathcal I(\psi_{x_0}) = \Big\{ \ \sum_{k=1}^K \alpha_k \Phi_{x_0}(Q;k) \ : \ \sum_{k=1}^K \alpha_k = 1, \ \alpha_k \in \mathbb R  \ \forall k \  \Big\} \ .
\end{align*}
\vspace{-0.25cm} \noindent
The efficient influence function minimizes the variance of the weighted influence function over  the vector $\alpha = (\alpha_1, \ldots, \alpha_K)$. This corresponds to solving: 
\begin{align*}
    \alpha^{\mathrm{opt}}=\argmin_{\alpha:\,\mathbf 1^\top \alpha = 1} \ \operatorname{Var} \, \Big\{\sum_{k=1}^K\alpha_k\Phi_{x_0}(Q;k)\Big\} \ , 
\end{align*}
where $\mathbf 1=(1,\ldots,1)^\top$. 
Let $\Sigma$ denote the covariance matrix of the basis influence functions with the ij-th element defined as: 
$\Sigma_{ij}=\operatorname{Cov}\{\Phi_{x_0}(Q;i) \,\Phi_{x_0}(Q;j)\}.$ If $\Sigma$ is nonsingular, the efficient influence function is obtained by choosing
$\alpha^{\mathrm{opt}}= \displaystyle\frac{\Sigma^{-1}\mathbf 1}{\mathbf 1^\top \Sigma^{-1}\mathbf 1} \ . $
\end{lemma}
See a proof in Appendix~\ref{app:proofs_eff_gain}.

An estimate of $\alpha^\text{opt}$ is obtained by replacing expectations with empirical averages and substituting $\hat Q$ for $Q$. Thus, we obtain $\hat{\Sigma}_{ij}=\widehat{\operatorname{Cov}}\big\{\Phi_{x_0}(\hat{Q};i) \,\Phi_{x_0}(\hat{Q};j)\big\}$ and $\hat\alpha^\text{opt} = \frac{\hat{\Sigma}^{-1}\mathbf 1}{\mathbf 1^\top \hat{\Sigma}^{-1}\mathbf 1}$. 

The resulting optimal one-step estimator, denoted by $\psi_{x_0}^{+,\mathrm{opt, dis}}(\hat Q)$, is thus given by 
$\psi_{x_0}^{+,\mathrm{opt}, \mathrm{dis}}(\hat Q) = \sum_{k=1}^K \hat{\alpha}_k \Phi_{x_0}(\hat{Q};k)$. The optimal estimating-equation estimator and TMLE are defined analogously and denoted by $\psi_{x_0}^{\mathrm{ee},\mathrm{opt, dis}}(\hat Q)$ and $\psi_{x_0}^{\mathrm{opt, dis}}(\hat Q^*)$, respectively.

The simplification under binary $Z$ is provided in the following corollary. 
\begin{corollary}\label{lemma:EIF-binaryZ}
Suppose $Z\in\{0,1\}$ and write $\alpha=\tilde p(z^*=1).$
Then every influence function in \eqref{eq:all-IF} takes the form $\{ \alpha \,\Phi_{x_0}(Q;z^*=1)+(1-\alpha)\,\Phi_{x_0}(Q;z^*=0) : \alpha \in \mathbb R \}.$ The efficient influence function is obtained using  $\alpha^\text{opt}$, defined as  
\vspace{-0.5cm}
\begin{align}
\alpha^\text{opt}&=\argmin_{\alpha\in\mathbb R}\operatorname{Var}\left\{\alpha \,\Phi_{x_0}(Q;z^*=1)+(1-\alpha)\,\Phi_{x_0}(Q;z^*=0)\right\}\notag
\\
&=\frac{\E\!\left[\Phi_{x_0}(Q;z^*=0)\Big(\Phi_{x_0}(Q;z^*=0)-\Phi_{x_0}(Q;z^*=1)\Big)\right]}{\E\!\left[\Big(\Phi_{x_0}(Q;z^*=0)-\Phi_{x_0}(Q;z^*=1)\Big)^2\right]} \ .
\label{eq:opt_alpha}
\end{align}
\end{corollary}

\subsubsection{Approximating efficient estimation under continuous $Z$}

When $Z$ is continuous, the class of influence functions in \eqref{eq:all-IF} is indexed by the infinite-dimensional weighting function $\tilde p(z)$. Consequently, identifying the efficient influence function requires solving an infinite-dimensional variance minimization problem. Even when a globally optimal weighting function exists, estimating it may require stronger regularity conditions than those needed for estimation of the target parameter itself, potentially limiting its practical utility \citep{young2024rose}. Moreover, any candidate weighting function must satisfy the overlap conditions discussed in Section~\ref{sec:est} to ensure reliable estimation of the nuisance functions.

Rather than pursuing exact efficiency, we pursue a computationally tractable approximation to the efficient influence function, following a common strategy in semiparametric estimation \citep{liu2021efficient,young2024rose,dong2025marginal}. Specifically, we restrict our attention to a finite-dimensional class of weighting functions generated by $K$ prespecified weighting functions $\tilde p_1(z),\ldots,\tilde p_K(z)$, and essentially adopt the sieve GMM approach \citep{chen2007large}. This reduces the original infinite-dimensional optimization problem to a finite-dimensional optimization over the coefficient vector $\alpha=(\alpha_1,\ldots,\alpha_K)^\top,$ with $\alpha_k \in \mathbb R$ and $\sum_{k=1}^K \alpha_k=1.$ The resulting optimization problem is algebraically identical to that of Lemma~\ref{lemma:EIF-discreteZ}, with the basis influence functions playing the role of the fixed-$z$ influence functions.

\begin{lemma}\label{lemma:opt-continuousZ}
Let $G_k(Q)=\int\Phi_{x_0}(Q;z)\,\tilde p_k(z)\,dz, $ 
for $k=1,\ldots,K$, denote the basis influence functions. Consider weighting functions of the form $\tilde p_\alpha(z)=\sum_{k=1}^K\alpha_k\tilde p_k(z)$ where $\alpha_k \in \mathbb R$ and $\sum_{k=1}^K \alpha_k = 1.$ Within this class, the minimum-variance influence function solves
\vspace{-0.35cm}
\begin{align*}
    \alpha^{\mathrm{opt}}=\argmin_{\alpha:\,\mathbf 1^\top\alpha=1} \ \operatorname{Var} \, \Big\{ \ \sum_{k=1}^K\alpha_kG_k(Q) \ \Big\} \ .
\end{align*}
\par\vspace{-0.45cm} \noindent
Let $\Sigma_{ij}=\operatorname{Cov}\{G_i(Q),G_j(Q)\}.$ If $\Sigma$ is nonsingular, the solution is
$\alpha^{\mathrm{opt}}= \displaystyle\frac{\Sigma^{-1}\mathbf 1}{\mathbf 1^\top \Sigma^{-1}\mathbf 1}.$
\end{lemma}

See a proof in Appendix~\ref{app:proofs_eff_gain_conti}.

An estimate $\hat\alpha^{\mathrm{opt}}$ is obtained by replacing $Q$ with $\hat Q$ and empirically estimating $\Sigma$. The resulting weighting function is $\tilde p_{\hat\alpha^{\mathrm{opt}}}(z)=\sum_{k=1}^K\hat\alpha_k^{\mathrm{opt}}\,\tilde p_k(z)$. The resulting locally optimal one-step estimator is $\psi_{x_0}^{+,\mathrm{opt, cont}}(\hat Q)=\int\Phi_{x_0}(\hat Q;z)\,\tilde p_{\hat\alpha^{\mathrm{opt}}}(z)\,dz.$ The corresponding locally optimal estimating-equation estimator and TMLE are defined analogously and denoted by $\psi_{x_0}^{\mathrm{ee},\mathrm{opt, cont}}(\hat Q)$ and $\psi_{x_0}^{\mathrm{opt, cont}}(\hat Q^*)$, respectively.

\section{Simulations}
\label{sec:sims}
We conducted five sets of simulation studies: (1–2) On \textit{theoretical properties}, examining the asymptotic behavior and robustness of the estimators under the proposed conditions; (3) On \textit{weak overlap}, evaluating estimators' performances when the positivity assumption is nearly violated; (4) On \textit{model misspecification}, comparing performances when nuisance parameters are estimated using misspecified parametric models versus flexible machine learning methods; and (5) On \textit{cross-fitting}, investigating the impact of applying versus omitting cross-fitting when flexible machine learning methods are used for nuisance estimation. Simulation~1 considers a well-specified setting satisfying the conditions underlying the theoretical results, whereas Simulations~2--5 examine performance under challenging scenarios that may induce bias or instability. Across all simulations, we focused on the ATE and denoted the corresponding ATE estimators by omitting the subscript ${x_0}$.

The simulation code is available on GitHub at \href{https://github.com/annaguo-bios/Napkin-paper}{\texttt{annaguo-bios/Napkin-paper}} and relies on the \href{https://github.com/annaguo-bios/napkincausal}{\texttt{napkincausal}} package in \textsf{R}, which we developed to implement the proposed estimators for broader use beyond the specific settings considered here.

\textbf{Simulation 1: Asymptotic linearity.}  
We evaluated the asymptotic bias and variance of the proposed estimators under the conditions specified in  Theorem~\ref{thm:asymp_psi_onestep_contZ}. The corresponding data-generating processes (DGPs) are described in Appendix~\ref{appsubsec:dgp_sim1}. For binary $Z$, we assessed the three proposed estimators at $z^* = 1$ and $z^* = 0$, along with the estimators constructed using the optimal weights described in Section~\ref{subsec:weighted_est}. For continuous $Z$, we examined the three estimators under two specifications of $\tilde{p}_z$, namely Beta and Uniform distributions, each defined on the valid support of $Z$, as well as an estimator constructed through an optimal linear combination of these two basis specifications, as described in Section~\ref{subsec:weighted_est}.

All estimators exhibited the expected convergence behavior, with notable efficiency gains achieved through the use of the Verma constraint. Specifically, with binary $Z$, estimators that incorporate the optimal weight attained variances that were approximately three times smaller than those using $z^*=1$ and about 1.5 times smaller than those using $z^*=0$. A similar pattern was observed under continuous $Z$, where the optimally weighted estimator had variance approximately two times smaller than the estimator using a Uniform specification for $\tilde{p}_z$, and about 1.7 times smaller than the estimator using a Beta specification. For brevity, the detailed implementation procedures and results are provided in Appendix~\ref{appsubsec:sim_results}, where Figure~\ref{appfig:sim1-binaryZ} presents the results under binary $Z$, and Figure~\ref{appfig:sim1-contiZ} presents the results under continuous $Z$.

\textbf{Simulation 2: Double robustness.} 
We further evaluated the robustness of our estimators, as established in Corollary~\ref{cor:robust_psi_onestep_contZ}. For continuous $Z$, $\tilde{p}_z$ was set to a Uniform distribution with a valid support on $Z$, using the same DGP as in Simulation~1.

We assessed performance of these estimators at sample sizes $500, 1000$, and $2000$, with $1000$ simulation replicates for each sample size. Nuisance parameters were estimated under three specifications: two implied by the corollaries, which are expected to yield consistent estimators, and one additional specification in which all nuisance models were misspecified. Details on model specifications for each scenario are provided in Appendix~\ref{appsubsec:dgp_sim2}. For all specifications, the estimators were evaluated based on five metrics: the average bias, standard deviation (SD), mean squared error (MSE), 95\% confidence interval (CI) coverage, and average CI width. The same estimators, sample sizes, and evaluation metrics are used throughout the remaining simulations. 

Across all scenarios, the estimators demonstrated the expected robustness property. Specifically, under the two specifications implied by the corollaries, the average bias approached zero as the sample size increased. In contrast, when all models were misspecified, a persistent nonzero bias remained even at large sample sizes. Results for the continuous $Z$ are shown in Table~\ref{table:sim2_continuous}, and those for binary $Z$ appear in Appendix Table~\ref{table:sim2_binary}.

\providecommand{\huxb}[2]{\arrayrulecolor[RGB]{#1}\global\arrayrulewidth=#2pt}
\providecommand{\huxvb}[2]{\color[RGB]{#1}\vrule width #2pt}
\providecommand{\huxtpad}[1]{\rule{0pt}{\dimexpr #1\relax}}
\providecommand{\huxbpad}[1]{\rule[-\dimexpr #1\relax]{0pt}{\dimexpr #1\relax}}

\begin{table}[t]
\centering
\captionsetup{justification=raggedright,singlelinecheck=off}
\caption{Simulation results validating the robustness property of the proposed estimators when Z is continuous, with pre-specified \(\tilde{p}(Z)=\mathrm{Uniform}(0.1,0.25)\).}
 \setlength{\tabcolsep}{3pt}
\resizebox{1\textwidth}{!}{
\renewcommand{\arraystretch}{0.5}
 \setlength{\extrarowheight}{0pt}%
 \setlength{\lineskip}{0pt}\setlength{\lineskiplimit}{0pt}%
 \setlength{\tabcolsep}{0pt}
 \setlength{\arrayrulewidth}{0.5pt}
\begin{tabular}{l l l l l l l l l l l}

\thickline

\multicolumn{1}{!{\huxvb{0, 0, 0}{0}}l!{\huxvb{0, 0, 0}{0}}}{\huxtpad{-1pt + 1em}\raggedright \hspace{-1pt} \textbf{{\fontsize{6pt}{7.2pt}\selectfont }} \hspace{-1pt}\huxbpad{-1pt}} &
\multicolumn{1}{c!{\huxvb{0, 0, 0}{0}}}{\huxtpad{-1pt + 1em}\centering \hspace{-1pt} \textbf{{\fontsize{6pt}{7.2pt}\selectfont }} \hspace{-1pt}\huxbpad{-1pt}} &
\multicolumn{3}{c!{\huxvb{0, 0, 0}{0}}}{\huxtpad{-1pt + 1em}\centering \hspace{-1pt} \textbf{{\fontsize{6pt}{7.2pt}\selectfont TMLEs}} \hspace{-1pt}\huxbpad{-1pt}} &
\multicolumn{3}{c!{\huxvb{0, 0, 0}{0}}}{\huxtpad{-1pt + 1em}\centering \hspace{-1pt} \textbf{{\fontsize{6pt}{7.2pt}\selectfont One-step estimators}} \hspace{-1pt}\huxbpad{-1pt}} &
\multicolumn{3}{c!{\huxvb{0, 0, 0}{0}}}{\huxtpad{-1pt + 1em}\centering \hspace{-1pt} \textbf{{\fontsize{6pt}{7.2pt}\selectfont Estimating equations}} \hspace{-1pt}\huxbpad{-1pt}} \tabularnewline
\end{tabular}}\label{table:sim2_continuous}
\end{table}

\textbf{Simulation 3: Weak overlap.} 
We compared the performance of the proposed estimators under weak overlap (or near-positivity violation), a setting in which TMLEs have previously been shown to have better performance than others \citep{porter2011relative}. Our analysis focused on the case with binary $Z$, as weak overlap in the Napkin model arises through $f_Z$, extreme values of which is more commonly observed when $Z$ is binary. Specifically, we set $f_Z(1\, | \, W)=\expit(-\frac{5}{6}+\frac{5}{3}W)$, which ranges from about 0.007 to 0.993 for $W \sim \text{Unif}(-2.5, 3.5)$, creating a weak-overlap scenario. The detailed DGP is provided in Appendix~\ref{appsubsec:dgp_sim3}. 

Results are summarized in Table~\ref{table:sim3_binary}. Across all sample sizes and fixed levels of $z^*$, the TMLEs consistently outperform the other two estimators, exhibiting smaller bias, standard deviation, MSE, and CI coverage closer to the nominal 95\% level.

  \providecommand{\huxb}[2]{\arrayrulecolor[RGB]{#1}\global\arrayrulewidth=#2pt}
  \providecommand{\huxvb}[2]{\color[RGB]{#1}\vrule width #2pt}
  \providecommand{\huxtpad}[1]{\rule{0pt}{#1}}
  \providecommand{\huxbpad}[1]{\rule[-#1]{0pt}{#1}}

\begin{table}[t]
\centering
\captionsetup{justification=raggedright,singlelinecheck=off}
\caption{Impact of weak overlapping, 
when Z is binary. 
Estimating equation results match the one step estimator to three decimal places and are omitted.}
\setlength{\tabcolsep}{0pt}
\resizebox{1\textwidth}{!}{
\renewcommand{\arraystretch}{1}
 \setlength{\extrarowheight}{0pt}%
 \setlength{\lineskip}{0pt}\setlength{\lineskiplimit}{0pt}%
 \setlength{\tabcolsep}{0pt}
 \setlength{\arrayrulewidth}{0.5pt}
\begin{tabular}{l l l l l l l l}

\thickline

\multicolumn{1}{!{\huxvb{0, 0, 0}{0}}c!{\huxvb{0, 0, 0}{0}}}{\huxtpad{1pt + 1em}\centering \hspace{1pt} \textbf{{\fontsize{8pt}{9.6pt}\selectfont }} \hspace{1pt}\huxbpad{1pt}} &
\multicolumn{1}{c!{\huxvb{0, 0, 0}{0}}}{\huxtpad{1pt + 1em}\centering \hspace{1pt} {\fontsize{8pt}{9.6pt}\selectfont } \hspace{1pt}\huxbpad{1pt}} &
\multicolumn{1}{c!{\huxvb{0, 0, 0}{0}}}{\huxtpad{1pt + 1em}\centering \hspace{1pt} {\fontsize{8pt}{9.6pt}\selectfont \(\psi(\hat{Q}^*;z^*=0)\)} \hspace{1pt}\huxbpad{1pt}} &
\multicolumn{1}{c!{\huxvb{0, 0, 0}{0}}}{\huxtpad{1pt + 1em}\centering \hspace{1pt} {\fontsize{8pt}{9.6pt}\selectfont \(\psi^+(\hat{Q};z^*=0)\)} \hspace{1pt}\huxbpad{1pt}} &
\multicolumn{1}{c!{\huxvb{0, 0, 0}{0}}}{\huxtpad{1pt + 1em}\centering \hspace{1pt} {\fontsize{8pt}{9.6pt}\selectfont \(\psi(\hat{Q}^*;z^*=1)\)} \hspace{1pt}\huxbpad{1pt}} &
\multicolumn{1}{c!{\huxvb{0, 0, 0}{0}}}{\huxtpad{1pt + 1em}\centering \hspace{1pt} {\fontsize{8pt}{9.6pt}\selectfont \(\psi^+(\hat{Q};z^*=1)\)} \hspace{1pt}\huxbpad{1pt}} &
\multicolumn{1}{c!{\huxvb{0, 0, 0}{0}}}{\huxtpad{1pt + 1em}\centering \hspace{1pt} {\fontsize{8pt}{9.6pt}\selectfont \(\psi^{\text{opt}}(\hat{Q}^*;z^*)\)} \hspace{1pt}\huxbpad{1pt}} &
\multicolumn{1}{c!{\huxvb{0, 0, 0}{0}}}{\huxtpad{1pt + 1em}\centering \hspace{1pt} {\fontsize{8pt}{9.6pt}\selectfont \(\psi^{+,\text{opt}}(\hat{Q};z^*)\)} \hspace{1pt}\huxbpad{1pt}} \tabularnewline
\end{tabular}}\label{table:sim3_binary}

\end{table}

\textbf{Simulation 4: Misspecified models vs. flexible estimation.} 
This simulation is motivated by the practical challenge that the true specifications of nuisance models are typically unknown, making model misspecification a common concern. We evaluated the performance of the proposed estimators under model misspecification and assessed whether incorporating more flexible machine learning methods can mitigate this issue, leveraging the fact that our estimators allow for such flexibility in nuisance estimation.

We compared three approaches for fitting nuisance models: linear models, Super Learner, and Super Learner with 10-fold cross-fitting. The linear models are misspecified by omitting key main effects, using incorrect link functions for logistic regression, or specifying an incorrect parametric family for the conditional density $f_Z$. Details on DGPs, model specifications, and library of algorithms used in Super Learner are provided in Appendix~\ref{appsubsec:dgp_sim4}.

The results under continuous $Z$ are shown in Table~\ref{table:sim4_continuous}, and those for binary $Z$ appear in Appendix Table~\ref{table:sim4_binary}. The results indicate that relying solely on main-term linear models led to persistent bias across all sample sizes. In contrast, employing the Super Learner for nuisance estimation reduced both bias and MSE as the sample size increased. Incorporating cross-fitting alongside the Super Learner produced comparable improvements.

  \providecommand{\huxb}[2]{\arrayrulecolor[RGB]{#1}\global\arrayrulewidth=#2pt}
  \providecommand{\huxvb}[2]{\color[RGB]{#1}\vrule width #2pt}
  \providecommand{\huxtpad}[1]{\rule{0pt}{#1}}
  \providecommand{\huxbpad}[1]{\rule[-#1]{0pt}{#1}}

\begin{table}[t]
\centering
\captionsetup{justification=raggedright,singlelinecheck=off}
\caption{Impact of model misspecifications,  
when Z is continuous, with pre-specified \(\tilde{p}(Z)=\mathrm{Uniform}(0.1,0.25)\). Linear refers to generalized linear regressions, SL refers to Super Learner, 
  and CF denotes Super Learner with cross fitting using 10 folds. }
 \setlength{\tabcolsep}{3pt}
\resizebox{1\textwidth}{!}{
\renewcommand{\arraystretch}{0.5}
 \setlength{\extrarowheight}{0pt}%
 \setlength{\lineskip}{0pt}\setlength{\lineskiplimit}{0pt}%
 \setlength{\tabcolsep}{0pt}
 \setlength{\arrayrulewidth}{0.5pt}
\begin{tabular}{l l l l l l l l l l l}

\thickline

\multicolumn{1}{!{\huxvb{0, 0, 0}{0}}l!{\huxvb{0, 0, 0}{0}}}{\huxtpad{-1pt + 1em}\raggedright \hspace{-1pt} \textbf{{\fontsize{6pt}{7.2pt}\selectfont }} \hspace{-1pt}\huxbpad{-1pt}} &
\multicolumn{1}{c!{\huxvb{0, 0, 0}{0}}}{\huxtpad{-1pt + 1em}\centering \hspace{-1pt} \textbf{{\fontsize{6pt}{7.2pt}\selectfont }} \hspace{-1pt}\huxbpad{-1pt}} &
\multicolumn{3}{c!{\huxvb{0, 0, 0}{0}}}{\huxtpad{-1pt + 1em}\centering \hspace{-1pt} \textbf{{\fontsize{6pt}{7.2pt}\selectfont TMLEs}} \hspace{-1pt}\huxbpad{-1pt}} &
\multicolumn{3}{c!{\huxvb{0, 0, 0}{0}}}{\huxtpad{-1pt + 1em}\centering \hspace{-1pt} \textbf{{\fontsize{6pt}{7.2pt}\selectfont One-step estimators}} \hspace{-1pt}\huxbpad{-1pt}} &
\multicolumn{3}{c!{\huxvb{0, 0, 0}{0}}}{\huxtpad{-1pt + 1em}\centering \hspace{-1pt} \textbf{{\fontsize{6pt}{7.2pt}\selectfont Estimating equations}} \hspace{-1pt}\huxbpad{-1pt}} \tabularnewline
\end{tabular}}\label{table:sim4_continuous}

\end{table}

\textbf{Simulation 5: Impact of cross-fitting.} 
We further examined the impact of cross-fitting by focusing on random forests, which prior work suggests perform less effectively when cross-fitting is not applied \citep{chernozhukov2018double, biau2012analysis}.

We analyzed DGPs of varying complexity. For binary $Z$, the DGP matched that in Simulation 4, exhibiting moderate complexity. For continuous $Z$, the DGP incorporated ten confounders with complex interactions and higher-order terms to create a more challenging scenario. Detailed specifications are provided in Appendix~\ref{appsubsec:dgp_sim5}.

Results for continuous $Z$ are reported in Table~\ref{table:sim5_continuous}, with results for binary $Z$ given in Appendix Table~\ref{table:sim5_binary}. Cross-fitting reduced average bias and MSE and improved 95\% CI coverage, with larger gains under the continuous-$Z$ DGP. In both settings, however, it also increased the SD and widened the CIs, reflecting the usual bias--variance trade-off. Overall, these findings highlight the value of cross-fitting for bias reduction when machine learning methods are used for nuisance estimation in complex, high-dimensional settings. This simulation study was designed to evaluate the relative bias reduction achieved by cross-fitting under random forests, rather than nominal asymptotic normality. In practice, however, it is generally recommended to use an ensemble of learners that capture different aspects of the data-generating mechanism.

 \providecommand{\huxb}[2]{\arrayrulecolor[RGB]{#1}\global\arrayrulewidth=#2pt}
  \providecommand{\huxvb}[2]{\color[RGB]{#1}\vrule width #2pt}
  \providecommand{\huxtpad}[1]{\rule{0pt}{#1}}
  \providecommand{\huxbpad}[1]{\rule[-#1]{0pt}{#1}}

\begin{table}[t]
\centering
\captionsetup{justification=raggedright,singlelinecheck=off}
\caption{Impact of cross fitting using random forests under the first DGP, see Appendix~\ref{appsubsec:dgp_sim1}. RF denotes random forest with 500 trees and minimum node sizes of 5 for continuous variables and 1 for binary variables. CF denotes 5 fold cross fitting. $\tilde{p}(Z)$ is set to $\hat{p}(Z)$.}
\resizebox{1\textwidth}{!}{
\renewcommand{\arraystretch}{1.6}
 \setlength{\extrarowheight}{0pt}%
 \setlength{\lineskip}{0pt}\setlength{\lineskiplimit}{0pt}%
 \setlength{\tabcolsep}{0pt}
 \setlength{\arrayrulewidth}{0.5pt}
\begin{tabular}{l l l l l l l l l l l l l l l l l}

\thickline

\multicolumn{1}{!{\huxvb{0, 0, 0}{0}}c!{\huxvb{0, 0, 0}{0}}}{\huxtpad{0pt + 1em}\centering \hspace{0pt} \textbf{{\fontsize{8pt}{9.6pt}\selectfont }} \hspace{0pt}\huxbpad{0pt}} &
\multicolumn{1}{c!{\huxvb{0, 0, 0}{0}}}{\huxtpad{0pt + 1em}\centering \hspace{0pt} \textbf{{\fontsize{8pt}{9.6pt}\selectfont }} \hspace{0pt}\huxbpad{0pt}} &
\multicolumn{5}{c!{\huxvb{0, 0, 0}{0}}}{\huxtpad{0pt + 1em}\centering \hspace{0pt} \textbf{{\fontsize{8pt}{9.6pt}\selectfont n=500}} \hspace{0pt}\huxbpad{0pt}} &
\multicolumn{5}{c!{\huxvb{0, 0, 0}{0}}}{\huxtpad{0pt + 1em}\centering \hspace{0pt} \textbf{{\fontsize{8pt}{9.6pt}\selectfont n=1000}} \hspace{0pt}\huxbpad{0pt}} &
\multicolumn{5}{c!{\huxvb{0, 0, 0}{0}}}{\huxtpad{0pt + 1em}\centering \hspace{0pt} \textbf{{\fontsize{8pt}{9.6pt}\selectfont n=2000}} \hspace{0pt}\huxbpad{0pt}} \tabularnewline[-0.5pt]

\cline{3-17}

\multicolumn{1}{!{\huxvb{0, 0, 0}{0}}c!{\huxvb{0, 0, 0}{0}}}{\huxtpad{0pt + 1em}\centering \hspace{0pt} {\fontsize{8pt}{9.6pt}\selectfont } \hspace{0pt}\huxbpad{0pt}} &
\multicolumn{1}{c!{\huxvb{0, 0, 0}{0}}}{\huxtpad{0pt + 1em}\centering \hspace{0pt} {\fontsize{8pt}{9.6pt}\selectfont } \hspace{0pt}\huxbpad{0pt}} &
\multicolumn{1}{c!{\huxvb{0, 0, 0}{0}}}{\huxtpad{0pt + 1em}\centering \hspace{0pt} {\fontsize{8pt}{9.6pt}\selectfont Bias} \hspace{0pt}\huxbpad{0pt}} &
\multicolumn{1}{c!{\huxvb{0, 0, 0}{0}}}{\huxtpad{0pt + 1em}\centering \hspace{0pt} {\fontsize{8pt}{9.6pt}\selectfont SD} \hspace{0pt}\huxbpad{0pt}} &
\multicolumn{1}{c!{\huxvb{0, 0, 0}{0}}}{\huxtpad{0pt + 1em}\centering \hspace{0pt} {\fontsize{8pt}{9.6pt}\selectfont MSE} \hspace{0pt}\huxbpad{0pt}} &
\multicolumn{1}{c!{\huxvb{0, 0, 0}{0}}}{\huxtpad{0pt + 1em}\centering \hspace{0pt} {\fontsize{8pt}{9.6pt}\selectfont Coverage} \hspace{0pt}\huxbpad{0pt}} &
\multicolumn{1}{c!{\huxvb{0, 0, 0}{0}}}{\huxtpad{0pt + 1em}\centering \hspace{0pt} {\fontsize{8pt}{9.6pt}\selectfont CI width} \hspace{0pt}\huxbpad{0pt}} &
\multicolumn{1}{c!{\huxvb{0, 0, 0}{0}}}{\huxtpad{0pt + 1em}\centering \hspace{0pt} {\fontsize{8pt}{9.6pt}\selectfont Bias} \hspace{0pt}\huxbpad{0pt}} &
\multicolumn{1}{c!{\huxvb{0, 0, 0}{0}}}{\huxtpad{0pt + 1em}\centering \hspace{0pt} {\fontsize{8pt}{9.6pt}\selectfont SD} \hspace{0pt}\huxbpad{0pt}} &
\multicolumn{1}{c!{\huxvb{0, 0, 0}{0}}}{\huxtpad{0pt + 1em}\centering \hspace{0pt} {\fontsize{8pt}{9.6pt}\selectfont MSE} \hspace{0pt}\huxbpad{0pt}} &
\multicolumn{1}{c!{\huxvb{0, 0, 0}{0}}}{\huxtpad{0pt + 1em}\centering \hspace{0pt} {\fontsize{8pt}{9.6pt}\selectfont Coverage} \hspace{0pt}\huxbpad{0pt}} &
\multicolumn{1}{c!{\huxvb{0, 0, 0}{0}}}{\huxtpad{0pt + 1em}\centering \hspace{0pt} {\fontsize{8pt}{9.6pt}\selectfont CI width} \hspace{0pt}\huxbpad{0pt}} &
\multicolumn{1}{c!{\huxvb{0, 0, 0}{0}}}{\huxtpad{0pt + 1em}\centering \hspace{0pt} {\fontsize{8pt}{9.6pt}\selectfont Bias} \hspace{0pt}\huxbpad{0pt}} &
\multicolumn{1}{c!{\huxvb{0, 0, 0}{0}}}{\huxtpad{0pt + 1em}\centering \hspace{0pt} {\fontsize{8pt}{9.6pt}\selectfont SD} \hspace{0pt}\huxbpad{0pt}} &
\multicolumn{1}{c!{\huxvb{0, 0, 0}{0}}}{\huxtpad{0pt + 1em}\centering \hspace{0pt} {\fontsize{8pt}{9.6pt}\selectfont MSE} \hspace{0pt}\huxbpad{0pt}} &
\multicolumn{1}{c!{\huxvb{0, 0, 0}{0}}}{\huxtpad{0pt + 1em}\centering \hspace{0pt} {\fontsize{8pt}{9.6pt}\selectfont Coverage} \hspace{0pt}\huxbpad{0pt}} &
\multicolumn{1}{c!{\huxvb{0, 0, 0}{0}}}{\huxtpad{0pt + 1em}\centering \hspace{0pt} {\fontsize{8pt}{9.6pt}\selectfont CI width} \hspace{0pt}\huxbpad{0pt}} \tabularnewline[-0.5pt]

\cline{3-17}

\multicolumn{1}{!{\huxvb{0, 0, 0}{0}}c!{\huxvb{0, 0, 0}{0}}}{} &
\multicolumn{1}{c!{\huxvb{0, 0, 0}{0}}}{\huxtpad{0pt + 1em}\centering \hspace{0pt} {\fontsize{8pt}{9.6pt}\selectfont RF} \hspace{0pt}\huxbpad{0pt}} &
\multicolumn{1}{c!{\huxvb{0, 0, 0}{0}}}{\huxtpad{0pt + 1em}\centering \hspace{0pt} {\fontsize{8pt}{9.6pt}\selectfont -0.461} \hspace{0pt}\huxbpad{0pt}} &
\multicolumn{1}{c!{\huxvb{0, 0, 0}{0}}}{\huxtpad{0pt + 1em}\centering \hspace{0pt} {\fontsize{8pt}{9.6pt}\selectfont 0.09} \hspace{0pt}\huxbpad{0pt}} &
\multicolumn{1}{c!{\huxvb{0, 0, 0}{0}}}{\huxtpad{0pt + 1em}\centering \hspace{0pt} {\fontsize{8pt}{9.6pt}\selectfont 0.221} \hspace{0pt}\huxbpad{0pt}} &
\multicolumn{1}{c!{\huxvb{0, 0, 0}{0}}}{\huxtpad{0pt + 1em}\centering \hspace{0pt} {\fontsize{8pt}{9.6pt}\selectfont 0.3\%} \hspace{0pt}\huxbpad{0pt}} &
\multicolumn{1}{c!{\huxvb{0, 0, 0}{0.4}}!{\huxvb{0, 0, 0}{0.4}}}{\huxtpad{0pt + 1em}\centering \hspace{0pt} {\fontsize{8pt}{9.6pt}\selectfont 0.08} \hspace{0pt}\huxbpad{0pt}} &
\multicolumn{1}{c!{\huxvb{0, 0, 0}{0}}}{\huxtpad{0pt + 1em}\centering \hspace{0pt} {\fontsize{8pt}{9.6pt}\selectfont -0.366} \hspace{0pt}\huxbpad{0pt}} &
\multicolumn{1}{c!{\huxvb{0, 0, 0}{0}}}{\huxtpad{0pt + 1em}\centering \hspace{0pt} {\fontsize{8pt}{9.6pt}\selectfont 0.076} \hspace{0pt}\huxbpad{0pt}} &
\multicolumn{1}{c!{\huxvb{0, 0, 0}{0}}}{\huxtpad{0pt + 1em}\centering \hspace{0pt} {\fontsize{8pt}{9.6pt}\selectfont 0.14} \hspace{0pt}\huxbpad{0pt}} &
\multicolumn{1}{c!{\huxvb{0, 0, 0}{0}}}{\huxtpad{0pt + 1em}\centering \hspace{0pt} {\fontsize{8pt}{9.6pt}\selectfont 0.3\%} \hspace{0pt}\huxbpad{0pt}} &
\multicolumn{1}{c!{\huxvb{0, 0, 0}{0.4}}!{\huxvb{0, 0, 0}{0.4}}}{\huxtpad{0pt + 1em}\centering \hspace{0pt} {\fontsize{8pt}{9.6pt}\selectfont 0.06} \hspace{0pt}\huxbpad{0pt}} &
\multicolumn{1}{c!{\huxvb{0, 0, 0}{0}}}{\huxtpad{0pt + 1em}\centering \hspace{0pt} {\fontsize{8pt}{9.6pt}\selectfont -0.284} \hspace{0pt}\huxbpad{0pt}} &
\multicolumn{1}{c!{\huxvb{0, 0, 0}{0}}}{\huxtpad{0pt + 1em}\centering \hspace{0pt} {\fontsize{8pt}{9.6pt}\selectfont 0.059} \hspace{0pt}\huxbpad{0pt}} &
\multicolumn{1}{c!{\huxvb{0, 0, 0}{0}}}{\huxtpad{0pt + 1em}\centering \hspace{0pt} {\fontsize{8pt}{9.6pt}\selectfont 0.084} \hspace{0pt}\huxbpad{0pt}} &
\multicolumn{1}{c!{\huxvb{0, 0, 0}{0}}}{\huxtpad{0pt + 1em}\centering \hspace{0pt} {\fontsize{8pt}{9.6pt}\selectfont 0.3\%} \hspace{0pt}\huxbpad{0pt}} &
\multicolumn{1}{c!{\huxvb{0, 0, 0}{0}}}{\huxtpad{0pt + 1em}\centering \hspace{0pt} {\fontsize{8pt}{9.6pt}\selectfont 0.047} \hspace{0pt}\huxbpad{0pt}} \tabularnewline[-0.5pt]

\hhline{>{\huxb{0, 0, 0}{0.4}}||>{\huxb{0, 0, 0}{0.4}}||}

\multicolumn{1}{!{\huxvb{0, 0, 0}{0}}c!{\huxvb{0, 0, 0}{0}}}{\multirow[c]{-2}{*}[0ex]{\huxtpad{0pt + 1em}\centering \hspace{0pt} \rotatebox{90}{{\fontsize{8pt}{9.6pt}\selectfont TMLEs}} \hspace{0pt}\huxbpad{0pt}}} &
\multicolumn{1}{c!{\huxvb{0, 0, 0}{0}}}{\huxtpad{0pt + 1em}\centering \hspace{0pt} {\fontsize{8pt}{9.6pt}\selectfont CF} \hspace{0pt}\huxbpad{0pt}} &
\multicolumn{1}{c!{\huxvb{0, 0, 0}{0}}}{\huxtpad{0pt + 1em}\centering \hspace{0pt} {\fontsize{8pt}{9.6pt}\selectfont -0.052} \hspace{0pt}\huxbpad{0pt}} &
\multicolumn{1}{c!{\huxvb{0, 0, 0}{0}}}{\huxtpad{0pt + 1em}\centering \hspace{0pt} {\fontsize{8pt}{9.6pt}\selectfont 0.221} \hspace{0pt}\huxbpad{0pt}} &
\multicolumn{1}{c!{\huxvb{0, 0, 0}{0}}}{\huxtpad{0pt + 1em}\centering \hspace{0pt} {\fontsize{8pt}{9.6pt}\selectfont 0.051} \hspace{0pt}\huxbpad{0pt}} &
\multicolumn{1}{c!{\huxvb{0, 0, 0}{0}}}{\huxtpad{0pt + 1em}\centering \hspace{0pt} {\fontsize{8pt}{9.6pt}\selectfont 78\%} \hspace{0pt}\huxbpad{0pt}} &
\multicolumn{1}{c!{\huxvb{0, 0, 0}{0.4}}!{\huxvb{0, 0, 0}{0.4}}}{\huxtpad{0pt + 1em}\centering \hspace{0pt} {\fontsize{8pt}{9.6pt}\selectfont 0.547} \hspace{0pt}\huxbpad{0pt}} &
\multicolumn{1}{c!{\huxvb{0, 0, 0}{0}}}{\huxtpad{0pt + 1em}\centering \hspace{0pt} {\fontsize{8pt}{9.6pt}\selectfont -0.07} \hspace{0pt}\huxbpad{0pt}} &
\multicolumn{1}{c!{\huxvb{0, 0, 0}{0}}}{\huxtpad{0pt + 1em}\centering \hspace{0pt} {\fontsize{8pt}{9.6pt}\selectfont 0.194} \hspace{0pt}\huxbpad{0pt}} &
\multicolumn{1}{c!{\huxvb{0, 0, 0}{0}}}{\huxtpad{0pt + 1em}\centering \hspace{0pt} {\fontsize{8pt}{9.6pt}\selectfont 0.042} \hspace{0pt}\huxbpad{0pt}} &
\multicolumn{1}{c!{\huxvb{0, 0, 0}{0}}}{\huxtpad{0pt + 1em}\centering \hspace{0pt} {\fontsize{8pt}{9.6pt}\selectfont 74.4\%} \hspace{0pt}\huxbpad{0pt}} &
\multicolumn{1}{c!{\huxvb{0, 0, 0}{0.4}}!{\huxvb{0, 0, 0}{0.4}}}{\huxtpad{0pt + 1em}\centering \hspace{0pt} {\fontsize{8pt}{9.6pt}\selectfont 0.482} \hspace{0pt}\huxbpad{0pt}} &
\multicolumn{1}{c!{\huxvb{0, 0, 0}{0}}}{\huxtpad{0pt + 1em}\centering \hspace{0pt} {\fontsize{8pt}{9.6pt}\selectfont -0.05} \hspace{0pt}\huxbpad{0pt}} &
\multicolumn{1}{c!{\huxvb{0, 0, 0}{0}}}{\huxtpad{0pt + 1em}\centering \hspace{0pt} {\fontsize{8pt}{9.6pt}\selectfont 0.18} \hspace{0pt}\huxbpad{0pt}} &
\multicolumn{1}{c!{\huxvb{0, 0, 0}{0}}}{\huxtpad{0pt + 1em}\centering \hspace{0pt} {\fontsize{8pt}{9.6pt}\selectfont 0.035} \hspace{0pt}\huxbpad{0pt}} &
\multicolumn{1}{c!{\huxvb{0, 0, 0}{0}}}{\huxtpad{0pt + 1em}\centering \hspace{0pt} {\fontsize{8pt}{9.6pt}\selectfont 75.1\%} \hspace{0pt}\huxbpad{0pt}} &
\multicolumn{1}{c!{\huxvb{0, 0, 0}{0}}}{\huxtpad{0pt + 1em}\centering \hspace{0pt} {\fontsize{8pt}{9.6pt}\selectfont 0.432} \hspace{0pt}\huxbpad{0pt}} \tabularnewline[-0.5pt]

\cdashline{1-17}

\multicolumn{1}{!{\huxvb{0, 0, 0}{0}}c!{\huxvb{0, 0, 0}{0}}}{} &
\multicolumn{1}{c!{\huxvb{0, 0, 0}{0}}}{\huxtpad{0pt + 1em}\centering \hspace{0pt} {\fontsize{8pt}{9.6pt}\selectfont RF} \hspace{0pt}\huxbpad{0pt}} &
\multicolumn{1}{c!{\huxvb{0, 0, 0}{0}}}{\huxtpad{0pt + 1em}\centering \hspace{0pt} {\fontsize{8pt}{9.6pt}\selectfont -0.659} \hspace{0pt}\huxbpad{0pt}} &
\multicolumn{1}{c!{\huxvb{0, 0, 0}{0}}}{\huxtpad{0pt + 1em}\centering \hspace{0pt} {\fontsize{8pt}{9.6pt}\selectfont 0.09} \hspace{0pt}\huxbpad{0pt}} &
\multicolumn{1}{c!{\huxvb{0, 0, 0}{0}}}{\huxtpad{0pt + 1em}\centering \hspace{0pt} {\fontsize{8pt}{9.6pt}\selectfont 0.442} \hspace{0pt}\huxbpad{0pt}} &
\multicolumn{1}{c!{\huxvb{0, 0, 0}{0}}}{\huxtpad{0pt + 1em}\centering \hspace{0pt} {\fontsize{8pt}{9.6pt}\selectfont 0.7\%} \hspace{0pt}\huxbpad{0pt}} &
\multicolumn{1}{c!{\huxvb{0, 0, 0}{0.4}}!{\huxvb{0, 0, 0}{0.4}}}{\huxtpad{0pt + 1em}\centering \hspace{0pt} {\fontsize{8pt}{9.6pt}\selectfont 0.101} \hspace{0pt}\huxbpad{0pt}} &
\multicolumn{1}{c!{\huxvb{0, 0, 0}{0}}}{\huxtpad{0pt + 1em}\centering \hspace{0pt} {\fontsize{8pt}{9.6pt}\selectfont -0.514} \hspace{0pt}\huxbpad{0pt}} &
\multicolumn{1}{c!{\huxvb{0, 0, 0}{0}}}{\huxtpad{0pt + 1em}\centering \hspace{0pt} {\fontsize{8pt}{9.6pt}\selectfont 0.061} \hspace{0pt}\huxbpad{0pt}} &
\multicolumn{1}{c!{\huxvb{0, 0, 0}{0}}}{\huxtpad{0pt + 1em}\centering \hspace{0pt} {\fontsize{8pt}{9.6pt}\selectfont 0.268} \hspace{0pt}\huxbpad{0pt}} &
\multicolumn{1}{c!{\huxvb{0, 0, 0}{0}}}{\huxtpad{0pt + 1em}\centering \hspace{0pt} {\fontsize{8pt}{9.6pt}\selectfont 0.2\%} \hspace{0pt}\huxbpad{0pt}} &
\multicolumn{1}{c!{\huxvb{0, 0, 0}{0.4}}!{\huxvb{0, 0, 0}{0.4}}}{\huxtpad{0pt + 1em}\centering \hspace{0pt} {\fontsize{8pt}{9.6pt}\selectfont 0.068} \hspace{0pt}\huxbpad{0pt}} &
\multicolumn{1}{c!{\huxvb{0, 0, 0}{0}}}{\huxtpad{0pt + 1em}\centering \hspace{0pt} {\fontsize{8pt}{9.6pt}\selectfont -0.396} \hspace{0pt}\huxbpad{0pt}} &
\multicolumn{1}{c!{\huxvb{0, 0, 0}{0}}}{\huxtpad{0pt + 1em}\centering \hspace{0pt} {\fontsize{8pt}{9.6pt}\selectfont 0.07} \hspace{0pt}\huxbpad{0pt}} &
\multicolumn{1}{c!{\huxvb{0, 0, 0}{0}}}{\huxtpad{0pt + 1em}\centering \hspace{0pt} {\fontsize{8pt}{9.6pt}\selectfont 0.162} \hspace{0pt}\huxbpad{0pt}} &
\multicolumn{1}{c!{\huxvb{0, 0, 0}{0}}}{\huxtpad{0pt + 1em}\centering \hspace{0pt} {\fontsize{8pt}{9.6pt}\selectfont 0.5\%} \hspace{0pt}\huxbpad{0pt}} &
\multicolumn{1}{c!{\huxvb{0, 0, 0}{0}}}{\huxtpad{0pt + 1em}\centering \hspace{0pt} {\fontsize{8pt}{9.6pt}\selectfont 0.06} \hspace{0pt}\huxbpad{0pt}} \tabularnewline[-0.5pt]

\hhline{>{\huxb{0, 0, 0}{0.4}}||>{\huxb{0, 0, 0}{0.4}}||}

\multicolumn{1}{!{\huxvb{0, 0, 0}{0}}c!{\huxvb{0, 0, 0}{0}}}{\multirow[c]{-2}{*}[0ex]{\huxtpad{0pt + 1em}\centering \hspace{0pt} \rotatebox{90}{{\fontsize{8pt}{9.6pt}\selectfont One-steps}} \hspace{0pt}\huxbpad{0pt}}} &
\multicolumn{1}{c!{\huxvb{0, 0, 0}{0}}}{\huxtpad{0pt + 1em}\centering \hspace{0pt} {\fontsize{8pt}{9.6pt}\selectfont CF} \hspace{0pt}\huxbpad{0pt}} &
\multicolumn{1}{c!{\huxvb{0, 0, 0}{0}}}{\huxtpad{0pt + 1em}\centering \hspace{0pt} {\fontsize{8pt}{9.6pt}\selectfont 0.019} \hspace{0pt}\huxbpad{0pt}} &
\multicolumn{1}{c!{\huxvb{0, 0, 0}{0}}}{\huxtpad{0pt + 1em}\centering \hspace{0pt} {\fontsize{8pt}{9.6pt}\selectfont 4.159} \hspace{0pt}\huxbpad{0pt}} &
\multicolumn{1}{c!{\huxvb{0, 0, 0}{0}}}{\huxtpad{0pt + 1em}\centering \hspace{0pt} {\fontsize{8pt}{9.6pt}\selectfont 17.279} \hspace{0pt}\huxbpad{0pt}} &
\multicolumn{1}{c!{\huxvb{0, 0, 0}{0}}}{\huxtpad{0pt + 1em}\centering \hspace{0pt} {\fontsize{8pt}{9.6pt}\selectfont 59.1\%} \hspace{0pt}\huxbpad{0pt}} &
\multicolumn{1}{c!{\huxvb{0, 0, 0}{0.4}}!{\huxvb{0, 0, 0}{0.4}}}{\huxtpad{0pt + 1em}\centering \hspace{0pt} {\fontsize{8pt}{9.6pt}\selectfont 1.617} \hspace{0pt}\huxbpad{0pt}} &
\multicolumn{1}{c!{\huxvb{0, 0, 0}{0}}}{\huxtpad{0pt + 1em}\centering \hspace{0pt} {\fontsize{8pt}{9.6pt}\selectfont -0.136} \hspace{0pt}\huxbpad{0pt}} &
\multicolumn{1}{c!{\huxvb{0, 0, 0}{0}}}{\huxtpad{0pt + 1em}\centering \hspace{0pt} {\fontsize{8pt}{9.6pt}\selectfont 0.618} \hspace{0pt}\huxbpad{0pt}} &
\multicolumn{1}{c!{\huxvb{0, 0, 0}{0}}}{\huxtpad{0pt + 1em}\centering \hspace{0pt} {\fontsize{8pt}{9.6pt}\selectfont 0.399} \hspace{0pt}\huxbpad{0pt}} &
\multicolumn{1}{c!{\huxvb{0, 0, 0}{0}}}{\huxtpad{0pt + 1em}\centering \hspace{0pt} {\fontsize{8pt}{9.6pt}\selectfont 56.9\%} \hspace{0pt}\huxbpad{0pt}} &
\multicolumn{1}{c!{\huxvb{0, 0, 0}{0.4}}!{\huxvb{0, 0, 0}{0.4}}}{\huxtpad{0pt + 1em}\centering \hspace{0pt} {\fontsize{8pt}{9.6pt}\selectfont 0.815} \hspace{0pt}\huxbpad{0pt}} &
\multicolumn{1}{c!{\huxvb{0, 0, 0}{0}}}{\huxtpad{0pt + 1em}\centering \hspace{0pt} {\fontsize{8pt}{9.6pt}\selectfont -0.092} \hspace{0pt}\huxbpad{0pt}} &
\multicolumn{1}{c!{\huxvb{0, 0, 0}{0}}}{\huxtpad{0pt + 1em}\centering \hspace{0pt} {\fontsize{8pt}{9.6pt}\selectfont 0.748} \hspace{0pt}\huxbpad{0pt}} &
\multicolumn{1}{c!{\huxvb{0, 0, 0}{0}}}{\huxtpad{0pt + 1em}\centering \hspace{0pt} {\fontsize{8pt}{9.6pt}\selectfont 0.568} \hspace{0pt}\huxbpad{0pt}} &
\multicolumn{1}{c!{\huxvb{0, 0, 0}{0}}}{\huxtpad{0pt + 1em}\centering \hspace{0pt} {\fontsize{8pt}{9.6pt}\selectfont 57.7\%} \hspace{0pt}\huxbpad{0pt}} &
\multicolumn{1}{c!{\huxvb{0, 0, 0}{0}}}{\huxtpad{0pt + 1em}\centering \hspace{0pt} {\fontsize{8pt}{9.6pt}\selectfont 0.808} \hspace{0pt}\huxbpad{0pt}} \tabularnewline[-0.5pt]

\thickline

\end{tabular}}\label{table:sim5_continuous}

\end{table}

\section{Data application}
\label{sec:real-data}
We applied our proposed estimation framework to the Life Course 1971–2002 study from the Finnish Social Science Data Archive \citep{fsd}.\footnote{Available upon application at \url{https://services.fsd.tuni.fi/catalogue/FSD2007}.} This longitudinal cohort follows 634 Finnish children born in 1964–1968 in Jyvaskyla, Finland, and collects detailed information on cognitive ability, family background, and educational outcomes to understand their impact on a person's life course. Verbal intelligence was measured in childhood using the Illinois Test of Psycholinguistic Abilities (ITPA), and information on major life events and socioeconomic outcomes was collected in 1984, 1991, and 2002. 

This dataset has previously been used by \citet{helske2021estimation} to study the causal effect of educational attainment on income under a Napkin graph model using a parametric plug-in estimator. Here, we revisit the same scientific question using our flexible and robust estimation framework. Specifically, using our proposed methods, we estimate the causal effect of educational attainment on income under the DAG in Figure~\ref{fig:graph_realdata}, as introduced and justified by \cite{helske2021estimation}. This graph extends the Napkin graph with measured confounders; identification, estimation, and inference details are provided in Appendix~\ref{app:extension_confounders}. Here, $X$ denotes educational attainment (secondary or less, lower tertiary, and higher tertiary), $Y$ is annual income in 2000 (euros), and covariates include parental socioeconomic status ($W$; low, middle, high), primary school GPA ($Z$), ITPA score ($S$), and sex ($G$; male, female). Relative to the extended Napkin graph, we allow unmeasured confounding between $W$ and $S$ rather than a direct effect of $S$ on $W$. The identification functional remains unchanged, as shown in Appendix~\ref{app:proofs_ID}.

\begin{figure}[t] 
	\begin{center}
    \scalebox{0.75}{
    \begin{tikzpicture}[>=stealth, node distance=2.cm]
        \tikzstyle{format} = [thick, circle, minimum size=1.0mm, inner sep=2pt]
        \tikzstyle{square} = [draw, thick, minimum size=4.5mm, inner sep=2pt]
        
	\begin{scope}[xshift=0cm, yshift=0cm]
		\path[->, thick]
		node[] (w) {$\underset{\substack{\text{Parental SES}}}W$}
  
        node[right of=w, xshift=2cm] (z) {$\underset{\substack{\text{Primary school GPA} }}Z$}
        
        node[right of=z, xshift=3cm] (x) {$\underset{\text{Educational attainment}}{X}$}

        node[right of=x, xshift=2cm] (y) {$\underset{\text{Income}}{Y}$}
        
        node[above of=x, xshift=-1.25cm, yshift=0.25cm] (s) {$\underset{\text{ITPA score}}{S}$}

        node[above of=z,yshift=-0.9cm] (u1) {$U_1$}

        node[above of=z,yshift=0.5cm](u2) {$U_2$}

        node[above of=z,yshift=-0.2cm](u3) {$U_3$}

        node[above of=x, xshift=1.25cm, yshift=0.25cm] (g) {$\underset{\text{Sex}}{G}$}
		 
        (w) edge[blue] (z)
		(z) edge[blue] (x) 
		(x) edge[blue] (y) 
		(s) edge[blue] (z)
        (s) edge[blue] (x)
        (s) edge[blue] (y)
        (g) edge[blue] (z)
        (g) edge[blue] (x)
        (g) edge[blue] (y)
        (u1) edge[red] (s)
        (u1) edge[red] (w)
        (u2) edge[red] (w)
        (u2) edge[red, bend right=12] (x)
        (u3) edge[red] (w)
        (u3) edge[red, bend right=3] (y)
        ;
	\end{scope}
	\end{tikzpicture}}
	\caption{An illustration of the causal relationships among variables in the real data application.} 
	\label{fig:graph_realdata}
	\end{center}
\end{figure}

We conducted both a pooled analysis of the full cohort and subgroup analyses stratified by sex and whether the ITPA score was above or below the sample mean. All three proposed estimators for continuous $Z$ were implemented, with nuisance functions estimated using semiparametric kernel methods and machine learning,  detailed in Appendix~\ref{app:realdata}. 

Table~\ref{table:realdata_fsd} summarizes the results. Overall, higher educational attainment is associated with higher future income, with larger gains observed when comparing higher tertiary to lower tertiary education than when comparing lower tertiary to secondary or lower education. These findings are broadly consistent with those reported by \citet{helske2021estimation}, who estimated mean potential incomes (with 95\% CI) of \texteuro$19500 \ (19484, \ 19515)$, \texteuro$21600 \ (21576, \ 21623)$, and \texteuro$26600 \ (26558, \ 26641)$ under the three educational interventions using a parametric plug-in approach. Similar patterns are observed across most subgroups, although among males with below-average ITPA scores, higher educational attainment is associated with a small and statistically non-significant decrease in income.
\begin{table}[t]

\captionsetup{justification=raggedright,singlelinecheck=off}
\caption{ATEs of educational attainment on income comparing lower tertiary vs secondary or less, and higher vs lower tertiary: pooled and subgroup analyses by sex and ITPA scores. Estimates are shown with 95\% confidence intervals in parentheses. In the pooled analysis, the estimating equation estimator coincides with the one-step estimator and is therefore omitted.}

\resizebox{\textwidth}{!}{
\renewcommand{\arraystretch}{0.5}
 \setlength{\extrarowheight}{0pt}%
 \setlength{\lineskip}{0pt}\setlength{\lineskiplimit}{0pt}%
 \setlength{\tabcolsep}{0pt}%
 \setlength{\arrayrulewidth}{0.5pt}
\begin{tabular}{l l l l l l}

\hhline{>{\huxb{0, 0, 0}{1}}->{\huxb{0, 0, 0}{1}}->{\huxb{0, 0, 0}{1}}->{\huxb{0, 0, 0}{1}}->{\huxb{0, 0, 0}{1}}->{\huxb{0, 0, 0}{1}}-}

\multicolumn{1}{!{\huxvb{0, 0, 0}{0}}l!{\huxvb{0, 0, 0}{0}}}{\huxtpad{1pt + 1em}\raggedright \hspace{1pt} \textbf{{\fontsize{8pt}{9.6pt}\selectfont Sex}} \hspace{1pt}\huxbpad{1pt}} &
\multicolumn{2}{c!{\huxvb{0, 0, 0}{0}}}{\huxtpad{1pt + 1em}\centering \hspace{1pt} {\fontsize{8pt}{9.6pt}\selectfont Female} \hspace{1pt}\huxbpad{1pt}} &
\multicolumn{2}{c!{\huxvb{0, 0, 0}{0}}}{\huxtpad{1pt + 1em}\centering \hspace{1pt} {\fontsize{8pt}{9.6pt}\selectfont Male} \hspace{1pt}\huxbpad{1pt}} &
\multicolumn{1}{c!{\huxvb{0, 0, 0}{0}}}{\huxtpad{1pt + 1em}\centering \hspace{1pt} {\fontsize{8pt}{9.6pt}\selectfont Pooled} \hspace{1pt}\huxbpad{1pt}} \tabularnewline[-0.5pt]

\hhline{}

\multicolumn{1}{!{\huxvb{0, 0, 0}{0}}l!{\huxvb{0, 0, 0}{0}}}{\huxtpad{1pt + 1em}\raggedright \hspace{1pt} \textbf{{\fontsize{8pt}{9.6pt}\selectfont ITPA Score (n)}} \hspace{1pt}\huxbpad{1pt}} &
\multicolumn{1}{c!{\huxvb{0, 0, 0}{0}}}{\huxtpad{1pt + 1em}\centering \hspace{1pt} {\fontsize{8pt}{9.6pt}\selectfont Below avg (137)} \hspace{1pt}\huxbpad{1pt}} &
\multicolumn{1}{c!{\huxvb{0, 0, 0}{0}}}{\huxtpad{1pt + 1em}\centering \hspace{1pt} {\fontsize{8pt}{9.6pt}\selectfont Above avg (133)} \hspace{1pt}\huxbpad{1pt}} &
\multicolumn{1}{c!{\huxvb{0, 0, 0}{0}}}{\huxtpad{1pt + 1em}\centering \hspace{1pt} {\fontsize{8pt}{9.6pt}\selectfont Below avg (126)} \hspace{1pt}\huxbpad{1pt}} &
\multicolumn{1}{c!{\huxvb{0, 0, 0}{0}}}{\huxtpad{1pt + 1em}\centering \hspace{1pt} {\fontsize{8pt}{9.6pt}\selectfont Above avg (113)} \hspace{1pt}\huxbpad{1pt}} &
\multicolumn{1}{c!{\huxvb{0, 0, 0}{0}}}{\huxtpad{1pt + 1em}\centering \hspace{1pt} {\fontsize{8pt}{9.6pt}\selectfont Pooled (137)} \hspace{1pt}\huxbpad{1pt}} \tabularnewline[-0.5pt]

\hhline{>{\huxb{255, 255, 255}{0.4}}->{\huxb{0, 0, 0}{0.4}}->{\huxb{0, 0, 0}{0.4}}->{\huxb{0, 0, 0}{0.4}}->{\huxb{0, 0, 0}{0.4}}->{\huxb{0, 0, 0}{0.4}}-}

\multicolumn{1}{!{\huxvb{0, 0, 0}{0}}l!{\huxvb{0, 0, 0}{0}}}{\huxtpad{1pt + 1em}\raggedright \hspace{1pt} \textbf{{\fontsize{8pt}{9.6pt}\selectfont }} \hspace{1pt}\huxbpad{1pt}} &
\multicolumn{5}{c!{\huxvb{0, 0, 0}{0}}}{\huxtpad{1pt + 1em}\centering \hspace{1pt} \textbf{{\fontsize{8pt}{9.6pt}\selectfont Lower tertiary vs secondary or less}} \hspace{1pt}\huxbpad{1pt}} \tabularnewline[-0.5pt]

\hhline{}

\multicolumn{1}{!{\huxvb{0, 0, 0}{0}}l!{\huxvb{0, 0, 0}{0}}}{\huxtpad{1pt + 1em}\raggedright \hspace{1pt} {\fontsize{8pt}{9.6pt}\selectfont TMLE} \hspace{1pt}\huxbpad{1pt}} &
\multicolumn{1}{c!{\huxvb{0, 0, 0}{0}}}{\huxtpad{1pt + 1em}\centering \hspace{1pt} {\fontsize{8pt}{9.6pt}\selectfont 2751 (533,4970)} \hspace{1pt}\huxbpad{1pt}} &
\multicolumn{1}{c!{\huxvb{0, 0, 0}{0}}}{\huxtpad{1pt + 1em}\centering \hspace{1pt} {\fontsize{8pt}{9.6pt}\selectfont 3750 (-1008,8507)} \hspace{1pt}\huxbpad{1pt}} &
\multicolumn{1}{c!{\huxvb{0, 0, 0}{0}}}{\huxtpad{1pt + 1em}\centering \hspace{1pt} {\fontsize{8pt}{9.6pt}\selectfont -1380 (-9105,6345)} \hspace{1pt}\huxbpad{1pt}} &
\multicolumn{1}{c!{\huxvb{0, 0, 0}{0}}}{\huxtpad{1pt + 1em}\centering \hspace{1pt} {\fontsize{8pt}{9.6pt}\selectfont 3585 (-2334,9504)} \hspace{1pt}\huxbpad{1pt}} &
\multicolumn{1}{c!{\huxvb{0, 0, 0}{0}}}{\huxtpad{1pt + 1em}\centering \hspace{1pt} {\fontsize{8pt}{9.6pt}\selectfont 928 (-2632,4487)} \hspace{1pt}\huxbpad{1pt}} \tabularnewline[-0.5pt]

\hhline{}

\multicolumn{1}{!{\huxvb{0, 0, 0}{0}}l!{\huxvb{0, 0, 0}{0}}}{\huxtpad{1pt + 1em}\raggedright \hspace{1pt} {\fontsize{8pt}{9.6pt}\selectfont Onestep} \hspace{1pt}\huxbpad{1pt}} &
\multicolumn{1}{c!{\huxvb{0, 0, 0}{0}}}{\huxtpad{1pt + 1em}\centering \hspace{1pt} {\fontsize{8pt}{9.6pt}\selectfont 2768 (565,4971)} \hspace{1pt}\huxbpad{1pt}} &
\multicolumn{1}{c!{\huxvb{0, 0, 0}{0}}}{\huxtpad{1pt + 1em}\centering \hspace{1pt} {\fontsize{8pt}{9.6pt}\selectfont 3794 (-1067,8656)} \hspace{1pt}\huxbpad{1pt}} &
\multicolumn{1}{c!{\huxvb{0, 0, 0}{0}}}{\huxtpad{1pt + 1em}\centering \hspace{1pt} {\fontsize{8pt}{9.6pt}\selectfont -1316 (-8563,5931)} \hspace{1pt}\huxbpad{1pt}} &
\multicolumn{1}{c!{\huxvb{0, 0, 0}{0}}}{\huxtpad{1pt + 1em}\centering \hspace{1pt} {\fontsize{8pt}{9.6pt}\selectfont 3552 (-2192,9295)} \hspace{1pt}\huxbpad{1pt}} &
\multicolumn{1}{c!{\huxvb{0, 0, 0}{0}}}{\huxtpad{1pt + 1em}\centering \hspace{1pt} {\fontsize{8pt}{9.6pt}\selectfont 837 (-2734,4409)} \hspace{1pt}\huxbpad{1pt}} \tabularnewline[-0.5pt]

\hhline{}

\multicolumn{1}{!{\huxvb{0, 0, 0}{0}}l!{\huxvb{0, 0, 0}{0}}}{\huxtpad{1pt + 1em}\raggedright \hspace{1pt} {\fontsize{8pt}{9.6pt}\selectfont Est Eq} \hspace{1pt}\huxbpad{1pt}} &
\multicolumn{1}{c!{\huxvb{0, 0, 0}{0}}}{\huxtpad{1pt + 1em}\centering \hspace{1pt} {\fontsize{8pt}{9.6pt}\selectfont 2750 (544,4956)} \hspace{1pt}\huxbpad{1pt}} &
\multicolumn{1}{c!{\huxvb{0, 0, 0}{0}}}{\huxtpad{1pt + 1em}\centering \hspace{1pt} {\fontsize{8pt}{9.6pt}\selectfont 3704 (-1178,8585)} \hspace{1pt}\huxbpad{1pt}} &
\multicolumn{1}{c!{\huxvb{0, 0, 0}{0}}}{\huxtpad{1pt + 1em}\centering \hspace{1pt} {\fontsize{8pt}{9.6pt}\selectfont -1293 (-8495,5909)} \hspace{1pt}\huxbpad{1pt}} &
\multicolumn{1}{c!{\huxvb{0, 0, 0}{0}}}{\huxtpad{1pt + 1em}\centering \hspace{1pt} {\fontsize{8pt}{9.6pt}\selectfont 3733 (-1932,9398)} \hspace{1pt}\huxbpad{1pt}} &
\multicolumn{1}{c!{\huxvb{0, 0, 0}{0}}}{\huxtpad{1pt + 1em}\centering \hspace{1pt} {\fontsize{8pt}{9.6pt}\selectfont } \hspace{1pt}\huxbpad{1pt}} \tabularnewline[-0.5pt]

\hhline{}

\multicolumn{1}{!{\huxvb{0, 0, 0}{0}}l!{\huxvb{0, 0, 0}{0}}}{\huxtpad{1pt + 1em}\raggedright \hspace{1pt} \textbf{{\fontsize{8pt}{9.6pt}\selectfont }} \hspace{1pt}\huxbpad{1pt}} &
\multicolumn{5}{c!{\huxvb{0, 0, 0}{0}}}{\huxtpad{1pt + 1em}\centering \hspace{1pt} \textbf{{\fontsize{8pt}{9.6pt}\selectfont Higher vs lower tertiary}} \hspace{1pt}\huxbpad{1pt}} \tabularnewline[-0.5pt]

\hhline{}

\multicolumn{1}{!{\huxvb{0, 0, 0}{0}}l!{\huxvb{0, 0, 0}{0}}}{\huxtpad{1pt + 1em}\raggedright \hspace{1pt} {\fontsize{8pt}{9.6pt}\selectfont TMLE} \hspace{1pt}\huxbpad{1pt}} &
\multicolumn{1}{c!{\huxvb{0, 0, 0}{0}}}{\huxtpad{1pt + 1em}\centering \hspace{1pt} {\fontsize{8pt}{9.6pt}\selectfont 9864 (1887,17841)} \hspace{1pt}\huxbpad{1pt}} &
\multicolumn{1}{c!{\huxvb{0, 0, 0}{0}}}{\huxtpad{1pt + 1em}\centering \hspace{1pt} {\fontsize{8pt}{9.6pt}\selectfont 8257 (1620,14893)} \hspace{1pt}\huxbpad{1pt}} &
\multicolumn{1}{c!{\huxvb{0, 0, 0}{0}}}{\huxtpad{1pt + 1em}\centering \hspace{1pt} {\fontsize{8pt}{9.6pt}\selectfont -1806 (-12669,9057)} \hspace{1pt}\huxbpad{1pt}} &
\multicolumn{1}{c!{\huxvb{0, 0, 0}{0}}}{\huxtpad{1pt + 1em}\centering \hspace{1pt} {\fontsize{8pt}{9.6pt}\selectfont 7364 (1504,13223)} \hspace{1pt}\huxbpad{1pt}} &
\multicolumn{1}{c!{\huxvb{0, 0, 0}{0}}}{\huxtpad{1pt + 1em}\centering \hspace{1pt} {\fontsize{8pt}{9.6pt}\selectfont 4823 (-1119,10764)} \hspace{1pt}\huxbpad{1pt}} \tabularnewline[-0.5pt]

\hhline{}

\multicolumn{1}{!{\huxvb{0, 0, 0}{0}}l!{\huxvb{0, 0, 0}{0}}}{\huxtpad{1pt + 1em}\raggedright \hspace{1pt} {\fontsize{8pt}{9.6pt}\selectfont Onestep} \hspace{1pt}\huxbpad{1pt}} &
\multicolumn{1}{c!{\huxvb{0, 0, 0}{0}}}{\huxtpad{1pt + 1em}\centering \hspace{1pt} {\fontsize{8pt}{9.6pt}\selectfont 9174 (-488,18835)} \hspace{1pt}\huxbpad{1pt}} &
\multicolumn{1}{c!{\huxvb{0, 0, 0}{0}}}{\huxtpad{1pt + 1em}\centering \hspace{1pt} {\fontsize{8pt}{9.6pt}\selectfont 8007 (1235,14780)} \hspace{1pt}\huxbpad{1pt}} &
\multicolumn{1}{c!{\huxvb{0, 0, 0}{0}}}{\huxtpad{1pt + 1em}\centering \hspace{1pt} {\fontsize{8pt}{9.6pt}\selectfont -1902 (-12127,8324)} \hspace{1pt}\huxbpad{1pt}} &
\multicolumn{1}{c!{\huxvb{0, 0, 0}{0}}}{\huxtpad{1pt + 1em}\centering \hspace{1pt} {\fontsize{8pt}{9.6pt}\selectfont 7101 (1406,12796)} \hspace{1pt}\huxbpad{1pt}} &
\multicolumn{1}{c!{\huxvb{0, 0, 0}{0}}}{\huxtpad{1pt + 1em}\centering \hspace{1pt} {\fontsize{8pt}{9.6pt}\selectfont 4757 (-1236,10750)} \hspace{1pt}\huxbpad{1pt}} \tabularnewline[-0.5pt]

\hhline{}

\multicolumn{1}{!{\huxvb{0, 0, 0}{0}}l!{\huxvb{0, 0, 0}{0}}}{\huxtpad{1pt + 1em}\raggedright \hspace{1pt} {\fontsize{8pt}{9.6pt}\selectfont Est Eq} \hspace{1pt}\huxbpad{1pt}} &
\multicolumn{1}{c!{\huxvb{0, 0, 0}{0}}}{\huxtpad{1pt + 1em}\centering \hspace{1pt} {\fontsize{8pt}{9.6pt}\selectfont 9387 (-582,19355)} \hspace{1pt}\huxbpad{1pt}} &
\multicolumn{1}{c!{\huxvb{0, 0, 0}{0}}}{\huxtpad{1pt + 1em}\centering \hspace{1pt} {\fontsize{8pt}{9.6pt}\selectfont 8348 (1615,15082)} \hspace{1pt}\huxbpad{1pt}} &
\multicolumn{1}{c!{\huxvb{0, 0, 0}{0}}}{\huxtpad{1pt + 1em}\centering \hspace{1pt} {\fontsize{8pt}{9.6pt}\selectfont -2002 (-12391,8387)} \hspace{1pt}\huxbpad{1pt}} &
\multicolumn{1}{c!{\huxvb{0, 0, 0}{0}}}{\huxtpad{1pt + 1em}\centering \hspace{1pt} {\fontsize{8pt}{9.6pt}\selectfont 6899 (1123,12674)} \hspace{1pt}\huxbpad{1pt}} &
\multicolumn{1}{c!{\huxvb{0, 0, 0}{0}}}{\huxtpad{1pt + 1em}\centering \hspace{1pt} {\fontsize{8pt}{9.6pt}\selectfont } \hspace{1pt}\huxbpad{1pt}} \tabularnewline[-0.5pt]

\hhline{>{\huxb{0, 0, 0}{1}}->{\huxb{0, 0, 0}{1}}->{\huxb{0, 0, 0}{1}}->{\huxb{0, 0, 0}{1}}->{\huxb{0, 0, 0}{1}}->{\huxb{0, 0, 0}{1}}-}

\end{tabular}}\label{table:realdata_fsd}

\end{table}

\section{Discussion}
\label{sec:conc}

In this work, we proposed a flexible estimation framework for the average treatment effect under the Napkin graph. We developed three influence function-based estimators, namely the one-step estimator, the estimating equation estimator, and TMLE, and established their robustness properties along with the nuisance rate conditions required for achieving asymptotic linearity. These theoretical results were complemented by simulation studies, addressing the gap in the literature, where estimation under the Napkin graph has largely been limited to parametric approaches. All proposed estimators are implemented in the \href{https://github.com/annaguo-bios/napkincausal}{\texttt{napkincausal}} package in \textsc{R}, enabling practitioners to readily apply our methods in practice.

Another important contribution of this work is the development of semiparametric efficiency theory under a Verma constraint encoded in the Napkin model. Our developments inform the construction of IF-based estimators that achieve semiparametric efficiency when the trapdoor variable $Z$ is binary and approximate efficiency when $Z$ is continuous. As demonstrated in the simulation studies, these estimators can yield considerable efficiency gains under simple data-generating processes. Beyond the specific setting of the Napkin graph, our proposals also contribute to the broader semiparametric efficiency literature, since Verma constraints arise in a wide class of causal latent DAG models, and semiparametric efficiency theory under such constraints remains largely underdeveloped.

This work also opens several avenues for future research. First, while this work develops semiparametric efficiency theory under the Verma constraint in the specific setting of the Napkin graph, an important direction for future work is to extend the theoretical strategies developed here toward a general framework for deriving efficient influence functions in the presence of Verma constraints. Such developments would broaden semiparametric efficiency theory beyond settings characterized by ordinary equality constraints and inform the construction of efficient estimators in more general causal models. Second, our framework assumes that the Napkin graph correctly represents the underlying data-generating process. Developing statistical tests for the Verma constraint and other implications of the graph would provide an important complement to expert-driven causal model specification. While related testability results have recently been established for front-door models \citep{bhattacharya2022testability, guo2023flexible}, extending such ideas to the Napkin graph and more general latent variable DAGs remains an important direction for future work. Finally, while we focused on the Napkin graph, many latent variable DAGs induce more complex Verma constraints and corresponding identification functionals. Extending the estimation and efficiency theory developed here to broader classes of latent causal models would substantially expand the scope of semiparametric causal inference under hidden variable structures. In addition, when the trapdoor variable in the Napkin graph is continuous, our approach targets local rather than global efficiency. Characterizing globally efficient influence functions in this setting remains an important open problem and may require new methods for infinite-dimensional variance optimization under overlap constraints.


\vspace{0.5cm}
\begingroup
\renewcommand{\baselinestretch}{1}
\selectfont  
\setlength{\bibsep}{10pt}    
\bibliography{references}
\endgroup

\newpage
\appendix

\setcounter{page}{1}

\setcounter{section}{0}
\renewcommand{\thesection}{S\arabic{section}}
\renewcommand{\thesubsection}{S\arabic{section}.\arabic{subsection}}
\renewcommand{\thesubsubsection}{S\arabic{section}.\arabic{subsection}.\arabic{subsubsection}}

\setcounter{figure}{0}
\renewcommand{\thefigure}{S\arabic{figure}}

\setcounter{table}{0}
\renewcommand{\thetable}{S\arabic{table}}

\setcounter{equation}{0}
\renewcommand{\theequation}{S\arabic{equation}}

\begin{center}
\noindent {\bf \LARGE Supplementary Materials}
\end{center}

\vspace{0.5cm}

\begin{spacing}{1.6}

\begin{description}
    \item[GitHub repository:] The GitHub repository \href{https://github.com/annaguo-bios/Napkin-paper}{\texttt{annaguo-bios/Napkin-paper}} provides the code needed to reproduce the numerical results in this paper, including the simulation studies and real-data analyses.
    
    \vspace{0.25cm}
    \item[R package:] The R package \href{https://github.com/annaguo-bios/napkincausal}{\texttt{napkincausal}} provides a general-purpose implementation of the estimators proposed in this work for the causal effects under the Napkin graph, and is intended for broader use beyond the specific settings considered in this paper.
    
    \vspace{0.25cm}
    \item[Supplementary appendix:] This appendix contains additional methodological details, theoretical results, and supporting analyses. It is organized as follows.
\end{description}

\noindent 
Appendix~\ref{app:notation} offers a summary of the notations used throughout the manuscript to aid in understanding and reference.

\noindent
Appendix~\ref{app:proofs} provides proofs for the identification results, nonparametric influence function derivations, and all claims related to inference and robustness.

\noindent
Appendix~\ref{app:extension_confounders} extends the main results to include measured confounders that affect all observed variables in the Napkin graph. It details how the identification, estimation, inference, and robustness results generalize to this broader setting.

\noindent
Appendix~\ref{app:tmle} provides additional details on the TMLE procedures and the extensions to binary outcomes.

\noindent
Appendices~\ref{app:sim} and~\ref{app:realdata} present additional details for the simulation studies and the real data application, respectively.

\end{spacing}

\vspace{0.5cm}
\noindent For ease of reference, a table of contents for the supplementary materials is provided below.

\pagebreak

\startcontents[supp]
{
\small 
\renewcommand{\baselinestretch}{0.95}\selectfont
\printcontents[supp]{}{1}{}
}

\pagebreak

\section{Glossary of terms and notations} 
\label{app:notation} 

\begin{table}[H]
\begin{center}
\caption{\centering Glossary of terms and notations}
\label{tab:notations}
\addtolength{\tabcolsep}{8pt}
\resizebox{\textwidth}{!}{
{\small
\begin{tabular}{ ll | ll}
\hline
\textbf{Symbol} & \textbf{Definition}
&
\textbf{Symbol} & \textbf{Definition}
\\
\hline
$X, x_0$
&
Treatment, fixed assignment
&
$\pi(X\mid Z,W)$
&
Propensity score
\\[0.05cm]
$Y, Y^{x_0}$
&
Outcome, potential outcome
&
$\mu(X,Z,W)$
&
Outcome regression
\\[0.05cm]
$W, Z$
&
Observed covariates
&
$f_Z(Z\mid W)$
&
Conditional density of $Z$
\\[0.05cm]
$U=\{U_1,U_2\}$
&
Unmeasured variables
&
$p_W$
&
Marginal density of $W$
\\[0.05cm]
$O,P$
&
Observed data, its distribution
&
$Q$
&
Collection of nuisance functions
\\[0.05cm]
$\mathcal M$
&
Model space
&
$\mathcal Z$
&
Support of $Z$
\\[0.05cm]
$\hat Q^*$
&
Targeted nuisance estimates
&
$C_{n,\mathrm{stop}}$
&
Stopping criterion in TMLE
\\[0.05cm]
$L_Y,\hat\mu(\varepsilon_Y)$
&
Loss function and fluctuation model for $\mu$
&
$L_X,\hat\pi(\varepsilon_X;\cdot)$
&
Loss function and fluctuation model for $\pi$
\\[0.05cm]
$H_Y(X,Z,W;\tilde{p}_z)$
&
Weight function in $L_Y$
&
$H_X(Z,W;\tilde{p}_z)$
&
Clever covariate in the submodel for $\pi$
\\[0.05cm]
$\hat{Q}^{(t)}$
&
Nuisance estimates at TMLE iteration $t$
&
$R_2(\hat Q,Q;\tilde p_z)$
&
Second-order remainder
\\[0.05cm]
$\|f\|$
&
$(Pf^2)^{1/2}$
&
$P_n f$
&
$1/n\sum_{i=1}^n f(O_i)$
\\[0.05cm]
$\mathcal{T}$
& 
Model tangent space
&
$\mathcal{T}^\perp$
& 
Orthocomplement tangent space
\\[0.05cm]
$q(Y,X,W\mid Z)$
&
Post-intervention Markov kernel
&
$\E_q$
&
Expectation with respect to $q(Y,X,W\mid Z)$
\\[0.05cm]
$\Sigma$
&
Covariate matrix of basis IF
&
$\tilde{p}_{\hat{\alpha}^\mathrm{opt}}(z)$
&
Estimate of the optimal weighting function
\\[0.05cm]
$G_k$
& 
Basis IF under continuous $Z$
&
$\alpha,\alpha^{\text{opt}}$
&
Weight and optimal weight
\\[0.05cm]
$m(Z)$
&
Any square-integrable function of $Z$
&
$c(Z)$
&
A mean-zero square-integrable function of $Z$
\\[0.05cm]
$\mathcal{I}(\psi_{x_0})$
&
Class of IF for $\E(Y^{x_0})$
&
&
\\[0.35cm]
\multicolumn{2}{c|}{\textit{Fixed-$z^*$ formulation}}
&
\multicolumn{2}{c}{\textit{Weighted-$\tilde p_z$ formulation}}
\\[0.15cm]
$\psi_{x_0}(P;z^*)$
&
Target parameter at $z^*$
&
$\psi_{x_0}(P;\tilde p_z)$
&
Target parameter indexed by $\tilde p_z$
\\[0.20cm]
$\kappa_{x_0,1}(P;z^*)$
&
Numerator functional
&
$\kappa_{x_0,1}(P;\tilde p_z)$
&
Weighted numerator functional
\\[0.20cm]
$\kappa_{x_0,2}(P;z^*)$
&
Denominator functional
&
$\kappa_{x_0,2}(P;\tilde p_z)$
&
Weighted denominator functional
\\[0.20cm]
$\Phi_{x_0}(Q;z^*)$
&
Influence function at $z^*$
&
$\Phi_{x_0}(Q;\tilde p_z)$
&
Influence function indexed by $\tilde p_z$
\\[0.20cm]
$\kappa^{\text{pi}}_{x_0,1}(\hat Q;z^*)$
&
Plug-in estimator of $\kappa_{x_0,1}$
&
$\psi^{\text{pi}}_{x_0}(\hat Q;\tilde p_z)$
&
Plug-in estimator of $\psi_{x_0}$
\\[0.20cm]
$\kappa^{\text{pi}}_{x_0,2}(\hat Q;z^*)$
&
Plug-in estimator of $\kappa_{x_0,2}$
&
$\psi^{\text{ee}}_{x_0}(\hat Q;\tilde p_z)$
&
Estimating-equation estimator
\\[0.20cm]
$\kappa^{\text{aipw}}_{x_0,1}(\hat Q;z^*)$
&
AIPW estimator of $\kappa_{x_0,1}$
&
$\psi^{+}_{x_0}(\hat Q;\tilde p_z)$
&
One-step estimator
\\[0.20cm]
$\kappa^{\text{aipw}}_{x_0,2}(\hat Q;z^*)$
&
AIPW estimator of $\kappa_{x_0,2}$
&
$\psi_{x_0}(\hat Q^*;\tilde p_z)$
&
TMLE
\\[0.20cm]
$\psi_{x_0}^{\text{ee,opt,dis}}(\hat Q)$
&
Optimal estimating-equation estimator
&
$\psi_{x_0}^{\text{ee,opt,cont}}(\hat Q)$
&
Optimal estimating-equation estimator
\\[0.20cm]
$\psi_{x_0}^{+,\text{opt,dis}}(\hat Q)$
&
Optimal one-step estimator
&
$\psi_{x_0}^{+,\text{opt,cont}}(\hat Q)$
&
Optimal one-step estimator
\\[0.15cm]
$\psi_{x_0}^{\text{opt,dis}}(\hat Q^*)$
&
Optimal TMLE
&
$\psi_{x_0}^{\text{opt,cont}}(\hat Q^*)$
&
Optimal TMLE
\\[0.15cm]
&
&
$\psi_{x_0}^{ipw}(\hat{Q};\tilde{p}_z)$
&
IPW estimator of $\psi_{x_0}$
\\
\hline
\end{tabular}
}
}
\end{center}
\end{table}

\section{Proofs} 
\label{app:proofs} 

\subsection{Lemma~\ref{lem:id}: Identification} 
\label{app:proofs_ID}

The identification functional for $\E(Y^{x_0})$, equivalently denoted by $\E(Y\, | \, \doo(x_0))$ in do-calculus notation, can be derived using either Pearl's do-calculus rules \citep{pearl2009causality} or the assumptions stated in Lemma~\ref{lem:id} expressed in potential outcome notation. Let $z^*$ be a value in $\mathcal{Z}$. 

Following the probability and do-calculus rules, the identification proceeds as follows:
\vspace{-0.5cm}
\begin{align*}
    p(y \mid \doo(x_0)) &= p(y \mid \doo(x_0), \doo(z^*)) \qquad & \text{(3rd rule of do-calculus)} \\
    &= p(y \mid x_0, \doo(z^*)) &\text{(2nd rule of do-calculus)}\\
    &= \frac{p(y, x_0 \mid \doo(z^*))}{p(x_0 \mid \doo(z^*))} &\text{(Bayes rule)}\\
    &= \frac{\int p(y, x_0 \mid z^*, w) \ p(w) \ dw}{\int p(x_0 \mid z^*,w) \ p(w) \ dw}  \ . &\text{(back-door rule)}
\end{align*}

In potential outcome notation, we have: 

\vspace{-1.75cm}
\begin{align*}
    p(y^{x_0}) &= p(y^{z^*, x_0}) \qquad & \text{(no direct effect)} \\
    &= p(y^{z^*, x_0} \mid x_0^{z^*}) &\text{(conditional ignorability)}\\
    &= \frac{p(y^{z^*, x_0}, x_0^{z^*})}{p(x_0^{z^*})} &\text{(Bayes rule)}\\
    &= \frac{\int p(y^{z^*, x_0}, x_0^{z^*} \mid w) \, p(w) \, dw }{\int p(x_0^{z^*} \mid w) \, p(w) \, dw} &\text{(probability rules)}\\
     &= \frac{\int p(y^{z^*, x_0}, x_0^{z^*} \mid z^*, w) \, p(w) \, dw }{\int p(x_0^{z^*} \mid z^*, w) \, p(w) \, dw} &\text{(conditional ignorability)}\\
    &= \frac{\int p(y, x_0 \mid z^*, w) \ p(w) \ dw}{\int p(x_0 \mid z^*,w) \ p(w) \ dw}  \ . &\text{(consistency)}
\end{align*}

The above result can also be derived directly using the identification algorithms of \citet{richardson2023nested} and \citet{bhattacharya2022semiparametric}. 

Given our nuisance notations, our parameter of interest is therefore identified as: 

\vspace{-1.5cm}
\begin{align*}
    \E(Y^{x_0}) &= \frac{\int \mu(x_0, z^*, w) \ \pi(x_0 \mid z^*, w) \ p(w) \ dw }{\int \pi(x_0 \mid z^*,w) \ p(w) \ dw} 
    = \frac{\kappa_{x_0, 1}(Q; z^*)}{\kappa_{x_0, 2}(Q; z^*)}\ . 
\end{align*}

\vspace{-0.5cm}
The invariance of the identification functional to the choice of $z^* \in \cal Z$, a consequence of the Verma constraint implied by the Napkin graph, yields the following algebraically equivalent representation: (referred to as the \textit{nested-IPW} functional by \citet{bhattacharya2022semiparametric})
\begin{align*}
    \E(Y^{x_0}) &= \int  \frac{\int \mu(x_0, z, w) \ \pi(x_0 \mid z, w) \ p(w) \ dw  }{ \int \pi(x_0 \mid z, w) \ p(w) \ dw} \ p^\dagger(z) \ dz 
    = \int \frac{ \kappa_{x_0,1}(Q;z) }{ \kappa_{x_0,2}(Q;z) } \, p^\dagger(z) \, dz \ , 
\end{align*}%
where the indexing density $p^\dagger$ is an auxiliary distribution over $\mathcal Z$ and need not coincide with the true marginal density of $Z$. The density $p^\dagger$ must be chosen such that for every $z$ in its support, the nuisances $\mu$ and $\pi$ are well defined and can be reliably estimated from the observed data. To guarantee this, we require $f_Z(z\, | \, w) > 0$ for all $z$ in the support of $p^\dagger$ and all $w$ with $p_{W}(w) > 0$. 

We can also rewrite the above as: 
\vspace{-0.5cm}
\begin{align*}
    \E(Y^{x_0}) &= \frac{\int \mu(x_0, z, w) \ \pi(x_0 \mid z, w) \ p(w) \ \tilde p(z) \ dw \ dz }{ \int \pi(x_0 \mid z, w) \ p(w) \ \tilde p(z) \ dw \ dz} 
    = \frac{\int \kappa_{x_0,1}(Q;z)\, \tilde p(z) \, dz}{\int \kappa_{x_0,2}(Q;z) \, \tilde p(z) \, dz}
    = \frac{\kappa_{x_0,1}(Q; \tilde p_z)}{\kappa_{x_0,2}(Q; \tilde p_z)} \ , 
\end{align*}
where 
\vspace{-1.5cm}
\begin{align*}
    \tilde p(z) = \displaystyle \frac{1}{ \kappa_{x_0,2}(Q; z) \int \frac{p^\dagger(z)}{\kappa_{x_0,2}(Q; z)} \, dz} \,  p^\dagger(z) \ . 
\end{align*}
The indexing density $\tilde p$ is an auxiliary distribution over $\mathcal Z$  with the same support requirement as outlined for $p^\dagger$. 
To verify the equivalence, observe that
\vspace{-0.5cm}
\begin{align*}
\frac{\int \kappa_{x_0,1}(Q;z)\, \tilde p(z) \, dz}{\int \kappa_{x_0,2}(Q;z) \, \tilde p(z) \, dz} 
& = \frac{\cancel{ \frac{1}{ \int \frac{p^\dagger(z)}{\kappa_{x_0,2}(Q; z)} \, dz}}\ \int \frac{\kappa_{x_0,1}(Q;z)}{\kappa_{x_0,2}(Q;z)}\ p^\dagger(z) \, dz}{\cancel{ \frac{1}{ \int \frac{p^\dagger(z)}{\kappa_{x_0,2}(Q; z)} \, dz}}\ \cancelto{1}{\int p^\dagger(z)\ dz}}
= \int \frac{ \kappa_{x_0,1}(Q;z) }{ \kappa_{x_0,2}(Q;z) } \, p^\dagger(z) \, dz \ . 
\end{align*}

\subsection{Estimation} 
\label{app:proofs_est}

\subsubsection{Lemma~\ref{lem:np_EIF}: Nonparametric influence function} 
\label{app:proofs:lem:np_EIF}

In the following, let $\tilde{P}(Z)$ denote the cumulative distribution function corresponding to $\tilde{p}(Z)$, and $o=(y,x,z,w)$ denote a realization of the random vector $O=(Y,X,Z,W)$. We derive the pathwise derivative of $\psi_{x_0}\left(P;\tilde{p}_z\right)$ as follows: 
{\small\begin{align}
    \left.\frac{\partial}{\partial \varepsilon} \psi_{x_0}\left(P_{\varepsilon};\tilde{p}_z\right)\right|_{\varepsilon=0}
    &=\frac{\partial}{\partial \varepsilon}\int \frac{y \ dP_{\varepsilon}(y\mid x_0, z, w)\ dP_{\varepsilon}\left(x_0 \mid z, w\right)}{\int dP_{\varepsilon}\left(x_0 \mid z, w\right) \ dP_{\varepsilon}(w)\ d\tilde{P}(z) }\ dP_{\varepsilon}(w)\ d\tilde{P}(z)\ \Bigg|_{\varepsilon=0}\notag
    \\
    &\hspace{-2cm}=\int y \ S(y\mid x_0, z, w)\ dP(y\mid x_0, z, w) \frac{dP\left(x_0 \mid z, w\right)}{\int dP\left(x_0 \mid z, w\right) \ dP(w) \ d\tilde{P}(z)}\ dP(w)\ d\tilde{P}(z)\label{eq:contZ_EIF_y}
    \\
    &\hspace{-1.75cm}+\int y \ dP(y\mid x_0, z, w) \ \frac{\partial}{\partial \varepsilon}\Big(\frac{dP_{\varepsilon}\left(x_0 \mid z, w\right)}{\int dP_{\varepsilon}\left(x_0 \mid z, w\right) \ dP_{\varepsilon}(w)\ d\tilde{P}(z)}\Big)\Bigg|_{\varepsilon=0} \ dP(w)\ d\tilde{P}(z)\label{eq:contZ_EIF_ratio}
    \\
    &\hspace{-1.75cm}+\int y \ S(w)\ dP(y\mid x_0, z, w) \frac{dP\left(x_0 \mid z, w\right)}{\int dP\left(x_0 \mid z, w\right) \ dP(w) \ d\tilde{P}(z)}\ dP(w)\ d\tilde{P}(z) \ . \label{eq:contZ_EIF_w}
\end{align}}

Line \eqref{eq:contZ_EIF_y} simplifies to
\begin{align}
    \eqref{eq:contZ_EIF_y}&=\int \frac{\I(x=x_0)}{\kappa_{x_0, 2}(Q; \tilde{p}_z)}\frac{\tilde{p}(z)}{f_Z(z\mid w)}\big(y-\mu(x_0,z,w)\big) \ S(o) \ dP(o) \ . \label{eq:contZ_EIF_y_simple}
\end{align}

Line \eqref{eq:contZ_EIF_ratio} simplifies to
{\small\begin{align}
    \eqref{eq:contZ_EIF_ratio}&=\int \mu(x_0,z,w)\Big\{\frac{S(x_0\mid z,w) dP\left(x_0 \mid z, w\right)}{\kappa_{x_0, 2}(Q; \tilde{p}_z)}\notag
    \\
    &\hspace{3cm}-\frac{dP\left(x_0 \mid z, w\right)\int \big(S(x_0\mid z,w)+S(w)\big)\ dP\left(x_0 \mid z, w\right) \ dP(w)\ d\tilde{P}(z) }{\kappa^2_{x_0, 2}(Q; \tilde{p}_z)}\Big\} dP(w) \ d\tilde{P}(z)\notag
    \\
    &=\int \mu(x_0,z,w)\frac{S(x_0\mid z,w) dP\left(x_0 \mid z, w\right)}{\kappa_{x_0, 2}(Q; \tilde{p}_z)}\ dP(w) \ d\tilde{P}(z) \label{eq:contZ_EIF_ratio_1}
    \\
    &\hspace{0.25cm}-\int \mu(x_0,z,w)\frac{dP\left(x_0 \mid z, w\right)\int S(x_0\mid z,w)dP\left(x_0 \mid z, w\right) \ dP(w)\ d\tilde{P}(z)}{\kappa^2_{x_0, 2}(Q; \tilde{p}_z)}\ dP(w) \ d\tilde{P}(z) \label{eq:contZ_EIF_ratio_2}
    \\
    &\hspace{0.25cm}-\int \mu(x_0,z,w)\frac{dP\left(x_0 \mid z, w\right)\int S(w)dP\left(x_0 \mid z, w\right) \ dP(w) \ d\tilde{P}(z)}{\kappa^2_{x_0, 2}(Q; \tilde{p}_z)}\ dP(w) \ d\tilde{P}(z) \ , \label{eq:contZ_EIF_ratio_3}
\end{align}}
where \eqref{eq:contZ_EIF_ratio_1}, \eqref{eq:contZ_EIF_ratio_2}, and \eqref{eq:contZ_EIF_ratio_3} can be further simplified as follows
\begin{align}
    \eqref{eq:contZ_EIF_ratio_1}&=\int \frac{1}{\kappa_{x_0, 2}(Q; \tilde{p}_z)}\frac{\tilde{p}(z)}{f_Z(z\mid w)}\mu(x_0,z,w)\Big(\I(x=x_0)-\pi(x_0\mid z,w)\Big)\ S(o)\ dP(o) \ . \label{eq:contZ_EIF_ratio_1_simple}
    \\
    \eqref{eq:contZ_EIF_ratio_2}&=-\int \frac{\psi_{x_0}(Q;\tilde{p}_z)}{\kappa_{x_0, 2}(Q; \tilde{p}_z)} \ S(x_0\mid z,w)\ dP\left(x_0 \mid z, w\right) \ dP(w)\ d\tilde{P}(z) \notag
    \\
    &=-\int \frac{\psi_{x_0}(Q;\tilde{p}_z)}{\kappa_{x_0, 2}(Q; \tilde{p}_z)}\frac{\tilde{p}(z)}{f_Z(z\mid w)}\big(\I(x=x_0)-\pi(x_0\mid z,w)\big)\ S(o)\ dP(o) \ . \label{eq:contZ_EIF_ratio_2_simple}
    \\
    \eqref{eq:contZ_EIF_ratio_3}&=-\int \frac{\psi_{x_0}(Q;\tilde{p}_z)}{\kappa_{x_0, 2}(Q; \tilde{p}_z)}\ S(w)\ dP\left(x_0 \mid z, w\right) \ dP(w) \ d\tilde{P}(z) \notag
    \\
    &=-\int  \Big\{ \frac{\psi_{x_0}(Q;\tilde{p}_z)}{\kappa_{x_0, 2}(Q; \tilde{p}_z)} \int \big( \pi(x_0\mid z,w) - \kappa_{x_0,2}(Q;z) \big) \ \tilde{p}(z) \ dz  \Big\} \ S(w) \ dP(w) \ . \label{eq:contZ_EIF_ratio_3_simple}
\end{align}

Line \eqref{eq:contZ_EIF_w} simplifies to
\begin{align}
    \eqref{eq:contZ_EIF_w}&=\int \Big\{ \frac{1}{\kappa_{x_0,2}(Q,\tilde{p}_z)}\int \big( \mu(x_0,z,w)\ \pi(x_0\mid z,w)  - \ \kappa_{x_0,1}(Q,z)\big) \ \tilde{p}(z) \ dz \Big\} \  S(w)\ dP(w) \ . \label{eq:contZ_EIF_w_simple}
\end{align}

Combining \eqref{eq:contZ_EIF_y_simple}, \eqref{eq:contZ_EIF_ratio_1_simple}, \eqref{eq:contZ_EIF_ratio_2_simple}, \eqref{eq:contZ_EIF_ratio_3_simple}, and \eqref{eq:contZ_EIF_w_simple} yields the EIF as follows: 
{\small\begin{align*}
    \Phi_{x_0}(Q;\tilde{p}_z(O_i)
    &=\underbrace{\frac{\I(X_i=x)}{\kappa_{x_0, 2}(Q;\tilde{p}_z)}  \frac{\tilde{p}(Z_i)}{f_Z(Z_i \mid W_i)} \Big\{  Y_i -  \mu(x_0, Z_i, W_i) \Big\}}_{\phi_{Y,x_0}(Q;\tilde{p}_z(O_i)} 
    \\ 
    &\hspace{-1cm} + \underbrace{\frac{1}{\kappa_{x_0, 2}(Q;\tilde{p}_z)} \frac{\tilde{p}(Z_i)}{f_Z(Z_i \mid W_i)} \Big\{ \mu(x_0, Z_i, W_i) - \psi_{x_0}(Q;\tilde{p}_z)   \Big\} \Big\{ \I(X_i=x_0) -  \pi(x_0 \mid Z_i, W_i) \Big\}}_{\phi_{X,x_0}(Q;\tilde{p}_z(O_i)}  
     \\
    &\hspace{-1cm} + \underbrace{\frac{1}{\kappa_{x_0, 2}(Q; \tilde{p}_z)}\int   \pi(x_0 \mid z, W_i)  \Big\{ \mu(x_0, z, W_i) - \psi_{x_0}(Q;\tilde{p}_z)    \Big\}\ \tilde{p}(z) \ dz}_{\phi_{W,x_0}(Q;\tilde{p}_z(O_i)} \ . 
\end{align*}}
Here, the first line corresponds to \eqref{eq:contZ_EIF_y_simple}, 
the second line corresponds to \eqref{eq:contZ_EIF_ratio_1_simple} and \eqref{eq:contZ_EIF_ratio_2_simple}, 
and the third line corresponds to \eqref{eq:contZ_EIF_ratio_3_simple} and  \eqref{eq:contZ_EIF_w_simple}. 

\subsubsection*{\underline{Nonparametric influence function for $\psi_{x_0}(Q; z^*)$} }

The nonparametric influence function (np-IF) for $\psi_{x_0}(Q; z^*)$ arises as a special case of $\Phi_{x_0}(Q;\tilde{p}_z$, as discussed in Section~\ref{sec:est}. It can also be derived directly from the characterization below. 

Let $\Phi_{x_0, 1}(Q; z^*)$ and $\Phi_{x_0, 2}(Q; z^*)$ denote the np-IF for $\kappa_{x_0, 1}(Q; z^*)$ and $\kappa_{x_0, 2}(Q; z^*)$, respectively. Following standard pathwise derivations, we have: 

\vspace{-1.25cm}
{\small 
\begin{align*}
    \Phi_{x_0, 1}(Q; z^*)(O_i) &= \frac{\I(Z_i=z^*)}{f_Z(z^* \mid W_i)} \Big( \I(X_i=x_0) \ Y_i -  \mu(x_0, z^*, W_i) \pi(x_0 \mid z^*, W_i) \Big) \\ 
    &\hspace{4cm} + \mu(x_0, z^*, W_i) \pi(x_0 \mid z^*, W_i) - \kappa_{x_0, 1}(Q; z^*) \ ,  \\
    \Phi_{x_0, 2}(Q; z^*)(O_i) &= \frac{\I(Z_i=z^*)}{f_Z(z^* \mid W_i)} \Big( \I(X_i=x_0) -  \pi(x_0 \mid z^*, W_i) \Big)  + \pi(x_0 \mid z^*, W_i) - \kappa_{x_0, 2}(Q; z^*) \ .  
\end{align*}
}

Using the delta method, the np-IF for $\psi_{x_0}(Q; z^*)$, denoted by $\Phi_{x_0}(Q; z^*)$, is expressed as:
\begin{align*}
    \Phi_{x_0}(Q; z^*)(O_i)
    &= \frac{1}{\kappa_{x_0, 2}(Q; z^*)} \Phi_{x_0, 1}(Q; z^*)(O_i) - \frac{\kappa_{x_0, 1}(Q; z^*)}{\kappa^2_{x_0, 2}(Q; z^*)} \Phi_{x_0, 2}(Q; z^*)(O_i)  \ .
\end{align*} 
Substituting $\kappa_{x_0,1}$, $\kappa_{x_0,2}$, $\Phi_{x_0, 1}$, and $\Phi_{x_0, 2}$ with their explicit forms concludes the result. 

\subsubsection{Construction of the estimating equation estimator}\label{subsubsec:esteq}
The estimating equation estimator aims to solve $P_n \Phi_{x_0}(\hat{Q};\tilde{p}_z)=0$. That is,  
\begin{align*}
    &\frac{1}{n}\sum_{i = 1}^n \Big[\frac{\I(X_i=x_0)\, \tilde{p}(Z_i)}{\hat{f}_Z(Z_i \mid W_i)}  \big\{  Y_i -  \hat{\mu}(x_0, Z_i, W_i) \big\}
    \\ 
    &+  \frac{\tilde{p}(Z_i)}{\hat{f}_Z(Z_i \mid W_i)} \big\{ \hat{\mu}(x_0, Z_i, W_i) - \psi^\text{ee}_{x_0}(\hat{Q}; \tilde{p}(Z))   \big\} \big\{ \I(X_i=x_0) -  \hat{\pi}(x_0 \mid Z_i, W_i) \big\}
     \\
    &+ \int  \hat{\pi}(x_0 \mid z, W_i)  \big\{  \hat{\mu}(x_0, z, W_i) -  \psi^\text{ee}_{x_0}(\hat{Q}; \tilde{p}(Z)) \big\} \ \tilde{p}(z) dz\Big]=0 \ . 
\end{align*}
Collecting the terms that involve $\psi^\text{ee}_{x_0}(\hat{Q}; \tilde{p}_z)$ on the right-hand side yields
\begin{align*}
    &\sum_{i = 1}^n \Big[\frac{\I(X_i=x_0)\,\tilde{p}(Z_i)}{\hat{f}_Z(Z_i \mid W_i)}  \big\{  Y_i -  \hat{\mu}(x_0, Z_i, W_i) \big\}
    \\
    &\hspace{0.5cm}+\frac{\tilde{p}(Z_i)}{\hat{f}_Z(Z_i \mid W_i)} \hat{\mu}(x_0, Z_i, W_i)\big\{ \I(X_i=x_0) -  \hat{\pi}(x_0 \mid Z_i, W_i) \big\}
    \\
    &\hspace{0.5cm}+\int  \hat{\pi}(x_0 \mid z, W_i) \ \hat{\mu}(x_0, z, W_i) \ \tilde{p}(z) dz\Big]
    \\
    &=\psi^\text{ee}_{x_0}(\hat{Q}; \tilde{p}_z) \ \sum_{i = 1}^n\Big\{\frac{\tilde{p}(Z_i)}{\hat{f}_Z(Z_i \mid W_i)}\ \big\{ \I(X_i=x_0) -  \hat{\pi}(x_0 \mid Z_i, W_i) \big\} +\int  \hat{\pi}(x_0 \mid z, W_i)\ \tilde{p}(z) dz\Big\} \ .
\end{align*}

Solving the above equation for $\psi^\text{ee}_{x_0}(\hat{Q}; \tilde{p}_z)$ yields
{\small\begin{align*}
    \psi^\text{ee}_{x_0}(\hat{Q}; \tilde{p}_z) 
    & = \frac{\frac{1}{n}\sum_{i = 1}^n  \frac{\tilde{p}(Z_i)}{\hat{f}_Z(Z_i \mid W_i)}  \big\{  \I(X_i=x_0)\ Y_i -  \hat{\mu}(x_0, Z_i, W_i)\ \hat{\pi}(x_0 \mid Z_i, W_i) \big\}+ \kappa_{x_0,1}^{\text{pi}}(\hat{Q};\tilde{p}_z) }{\frac{1}{n}\sum_{i = 1}^n \frac{\tilde{p}(Z_i)}{\hat{f}_Z(Z_i \mid W_i)}\ \big\{ \I(X_i=x_0) -  \hat{\pi}(x_0 \mid Z_i, W_i) \big\}+ \kappa_{x_0,2}^{\text{pi}}(\hat{Q};\tilde{p}_z)} \ .
\end{align*}}

\subsection{Inference}
\label{app:proofs_inf}

\subsubsection{Second-order remainder term and regularity conditions}
\label{app:proofs_inf_continuous} 

\subsubsection*{\underline{Second-order remainder term}}

The second-order remainder term for the one-step estimator $\psi^+_{x_0}(\hat{Q};\tilde{p}_z)$, denoted by $R_2(\hat{Q},Q;\tilde{p}_z)$, can be derived as follows. 
{\small\begin{align}
    R_2(\hat{Q},Q;\tilde{p}_z)
    &=\psi^{\text{pi}}_{x_0}(\hat{Q};\tilde{p}_z)-\psi_{x_0}(Q)+\int \Phi_{x_0}(\hat{Q};\tilde{p}_z) \ dP(O)\notag
    \\
    &=\psi^{\text{pi}}_{x_0}(\hat{Q};\tilde{p}_z)-\psi_{x_0}(Q) \notag
    \\
    &\hspace{0.25cm}+\frac{1}{\kappa^{\text{aipw}}_{x_0,2}(\hat{Q} ; \tilde{p}_z)}\bigg\{\int \frac{\I(x=x_0)\,\tilde{p}(z)}{\hat{f}_Z(z \mid w)}\big(y-\hat{\mu}(x_0, z, w)\big) \ d P(o)\label{eq:r2_continuous_1}
    \\
    &\hspace{0.25cm}+\int \frac{\tilde{p}(z)}{\hat{f}_Z(z \mid w)}\big(\hat{\mu}\left(x_0, z, w\right)-\psi^{\text{pi}}_{x_0}(\hat{Q} ; \tilde{p}_z)\big)\big(\I\left(x=x_0\right)-\hat{\pi}(x_0 \mid z,w)\big) \ dP(o) \label{eq:r2_continuous_2}
    \\
    &\hspace{0.25cm}+\int \Big\{\int\hat{\pi}\left(x_0 \mid z, w\right) \big( \hat{\mu}(x_0, z, w)-\psi^{\text{pi}}_{x_0}(\hat{Q};\tilde{p}_z)\big) \ \tilde{p}(z) \ d z\Big\} \ d P(w)\bigg\} \ . \label{eq:r2_continuous_3}
\end{align}}

Line \eqref{eq:r2_continuous_1} can be reformulated as
{\small\begin{align}
    \eqref{eq:r2_continuous_1}
    &=\int \frac{\pi\left(x_0\mid z,w\right)}{\kappa^{\text{aipw}}_{x_0,2}(\hat{Q} ; \tilde{p}_z)} \frac{\tilde{p}\left(z\right)\ \big(f_Z(z\mid w)-\hat{f}_Z(z\mid w)\big)}{\hat{f}_Z(z\mid w)\ f_Z(z\mid w)}\big(\mu(x_0,z,w)-\hat{\mu}(x_0,z,w)\big) \ d P(z,w)
    \\
    &\hspace{0.25cm}+\int \frac{\pi\left(x_0\mid z,w\right)}{\kappa^{\text{aipw}}_{x_0,2}(\hat{Q} ; \tilde{p}_z)} \frac{\tilde{p}\left(z\right)}{f_Z(z \mid w)}\ \mu(x_0,z,w)\ d P(z,w) \label{eq:r2_continuous_1_2_1}
    \\
    &\hspace{0.25cm}-\int \frac{\pi\left(x_0\mid z,w\right)}{\kappa^{\text{aipw}}_{x_0,2}(\hat{Q} ; \tilde{p}_z)} \frac{\tilde{p}\left(Z\right)}{f_Z(z \mid w)}\ \hat{\mu}(x_0, z, w) \ d P(z,w) \ . \label{eq:r2_continuous_1_2_2}
\end{align}}

Line \eqref{eq:r2_continuous_2} can be reformulated as
{\small\begin{align}
    &\int \frac{\hat{\mu}(x_0,z,w)-\psi^{\text{pi}}_{x_0}(\hat{Q} ; \tilde{p}_z)}{\kappa^{\text{aipw}}_{x_0,2}(\hat{Q}; \tilde{p}_z)} \frac{\tilde{p}\left(z\right)\ \big(f_Z(z\mid w)-\hat{f}_Z(z\mid w)\big)}{\hat{f}_Z(z\mid w)\ f_Z(z\mid w)}\ (\pi(x_0\mid z,w)-\hat{\pi}(x_0\mid z,w)) \ d P(z,w)
    \\
    &\hspace{0.25cm}-\int \frac{\hat{\mu}(x_0,z,w)-\psi^{\text{pi}}_{x_0}(\hat{Q} ; \tilde{p}_z)}{\kappa^{\text{aipw}}_{x_0,2}(\hat{Q}; \tilde{p}_z)} \frac{\tilde{p}\left(z\right)}{f_Z(z\mid w)}\hat{\pi}(x_0\mid z,w) \ dP(z,w) \label{eq:r2_continuous_2_2_1}
    \\
    &\hspace{0.25cm}+\int \frac{1}{\kappa^{\text{aipw}}_{x_0,2}(\hat{Q}; \tilde{p}_z)} \frac{\tilde{p}\left(z\right)}{f_Z(z\mid w)}\ \hat{\mu}(x_0,z,w)\ \pi(x_0\mid z,w)\ dP(z,w) \label{eq:r2_continuous_2_2_2}
    \\
    &\hspace{0.25cm}-\int \frac{1}{\kappa^{\text{aipw}}_{x_0,2}(\hat{Q}; \tilde{p}_z)} \frac{\tilde{p}\left(z\right)}{f_Z(z\mid w)}\ \ \psi^{\text{pi}}_{x_0}(\hat{Q};\tilde{p}_z)\ \pi(x_0\mid z,w)\ dP(z,w) \ . \label{eq:r2_continuous_2_2_23}
\end{align}}

Note that \eqref{eq:r2_continuous_2_2_1} cancels out \eqref{eq:r2_continuous_3}, and \eqref{eq:r2_continuous_2_2_2} cancels out \eqref{eq:r2_continuous_1_2_2}.

Combining \eqref{eq:r2_continuous_1_2_1} with term $-\psi_{x_0}(Q)$ results in 
\begin{align}
    \eqref{eq:r2_continuous_1_2_1}-\psi_{x_0}(Q)&=\int\Big[\frac{\kappa_{x_0,2}(Q;\tilde{p}_z) - \kappa^{\text{aipw}}_{x_0,2}(\hat{Q};\tilde{p}_z)}{\kappa_{x_0,2}(Q;\tilde{p}_z)\ \kappa^{\text{aipw}}_{x_0,2}(\hat{Q};\tilde{p}_z)}\ \mu(x_0,z,w)\ \pi(x_0\mid z,w) \ \tilde{p}(z)\ dz\Big] \ dP(w)\notag
    \\
    &=\frac{\kappa_{x_0,2}(Q;\tilde{p}_z) - \kappa^{\text{aipw}}_{x_0,2}(\hat{Q};\tilde{p}_z)}{\kappa^{\text{aipw}}_{x_0,2}(\hat{Q};\tilde{p}_z)}\ \psi_{x_0}(Q;\tilde{p}_z) \ . \label{eq:r2_continuous_kappa_1}
\end{align}

Combining \eqref{eq:r2_continuous_2_2_23} with term $\psi^{\text{pi}}_{x_0}(\hat{Q})$ results in 
\begin{align}
    \psi^{\text{pi}}_{x_0}(\hat{Q};\tilde{p}_z)\frac{\kappa^{\text{aipw}}_{x_0,2}(\hat{Q};\tilde{p}_z)-\kappa_{x_0,2}(Q;\tilde{p}_z)}{\kappa^{\text{aipw}}_{x_0,2}(\hat{Q};\tilde{p}_z)} \ .\label{eq:r2_continuous_kappa_2}
\end{align}

Therefore, combining \eqref{eq:r2_continuous_1_2_1} and \eqref{eq:r2_continuous_2_2_23} with $\psi^{\text{pi}}_{x_0}(\hat{Q})-\psi_{x_0}(Q)$ results in 
\begin{align}
    \left(\psi^{\text{pi}}_{x_0}(\hat{Q};\tilde{p}_z)-\psi_{x_0}(\hat{Q};\tilde{p}_z)\right)\frac{\kappa^{\text{aipw}}_{x_0,2}(\hat{Q};\tilde{p}_z)-\kappa_{x_0,2}(Q;\tilde{p}_z)}{\kappa^{\text{aipw}}_{x_0,2}(\hat{Q};\tilde{p}_z)} \ .\label{eq:r2_continuous_kappa}
\end{align}

Combining all terms yields the expression for the second-order remainder term as follows:
{\footnotesize
\begin{equation}
    \begin{aligned}
    R_2(\hat{Q},Q;\tilde{p}_z)&=\int \frac{\pi\left(x_0\mid z,w\right)}{\kappa^{\text{aipw}}_{x_0,2}(\hat{Q} ; \tilde{p}_z)} \frac{\tilde{p}\left(z\right)\ \big(f_Z(z\mid w)-\hat{f}_Z(z\mid w)\big)}{\hat{f}_Z(z\mid w)\ f_Z(z\mid w)}\big(\mu(x_0,z,w)-\hat{\mu}(x_0,z,w)\big) \ d P(O)
    \\
    &\hspace{-1.5cm}+\int \frac{\hat{\mu}(x_0,z,w)-\psi^{\text{pi}}_{x_0}(\hat{Q} ; \tilde{p}_z)}{\kappa^{\text{aipw}}_{x_0,2}(\hat{Q}; \tilde{p}_z)} \frac{\tilde{p}\left(z\right)\ \big(f_Z(z\mid w)-\hat{f}_Z(z\mid w)\big)}{\hat{f}_Z(z\mid w)\ f_Z(z\mid w)}\ (\pi(x_0\mid z,w)-\hat{\pi}(x_0\mid z,w)) \ d P(z,w)
    \\
    &\hspace{-1.5cm}+\frac{\psi^{\text{pi}}_{x_0}(\hat{Q};\tilde{p}_z)-\psi_{x_0}(\hat{Q};\tilde{p}_z)}{\kappa^{\text{aipw}}_{x_0,2}(\hat{Q};\tilde{p}_z)}\int \frac{\tilde{p}(z)\,(f_Z(z \mid w)-\hat{f}_Z(z \mid w))}{\hat{f}_Z(z \mid w)}\ \big(\pi(x_0 \mid z, w) -  \hat{\pi}(x_0 \mid z, w)\big)\ dz\ dP(w)
    \\
    &\hspace{-1.5cm}+o_P(n^{-1/2}) \ .
\end{aligned}\label{appeq:r2_contZ}
\end{equation}
}
The last two lines of \eqref{appeq:r2_contZ} are obtained via a von Mises expansion of $\kappa^{\text{aipw}}_{x_0,2}(\hat{Q};\tilde{p}_z)$ in \eqref{eq:r2_continuous_kappa} under the Donsker conditions.

\subsubsection*{\underline{Regularity conditions}}

We assume the following regularity conditions:
\begin{equation}\label{appeq:regularity_contZ}
    \begin{aligned}
\inf_{z\in\mathcal{Z},\ w \in \mathcal{W}} \hat{f}_Z(z|w)>0 \quad 
\text{and} \quad \inf _{x_0\in\{0,1\}} \kappa^{\text{aipw}}_{x_0,2}(\hat{Q};\tilde{p}_z)>0 \ .
\end{aligned}
\end{equation}

Given the boundedness conditions in \eqref{appeq:regularity_contZ}, applying the Cauchy–Schwarz inequality to each term in \eqref{appeq:r2_contZ} yields the following bound for some sufficiently large constant $C \in \mathbb{R}$:
\begin{align*}
    R_2\left(\hat{Q}, Q;\tilde{p}_z\right) &\leq C\Big[\left\|\hat{f}_Z-f_Z\right\| \times\left\|\hat{\mu}-\mu\right\|+\left\|\hat{f}_Z-f_Z\right\| \times\left\|\hat{\pi}-\pi\right\|\Big] \ .
\end{align*}

\subsubsection{Theorem~\ref{thm:asymp_psi_onestep_contZ}: Asymptotic linearity}
\label{app:proofs:thm:asymp_psi_onestep_contZ}

By the von Mises expansion in \eqref{eq:von_mise},
\[
\psi^{+}_{x_0}(\hat Q;\tilde p_z)-\psi_{x_0}(Q;\tilde p_z)
=
P_n\Phi_{x_0}(Q;\tilde p_z)
+
(P_n-P)\{\Phi_{x_0}(\hat Q;\tilde p_z)-\Phi_{x_0}(Q;\tilde p_z)\}
+
R_2(\hat Q,Q;\tilde p_z) \ .
\]
Under the stated empirical process conditions, or under cross-fitting, the second term is
\(o_P(n^{-1/2})\). It remains to show that the second-order remainder is also
\(o_P(n^{-1/2})\).

From Appendix~\ref{app:proofs_inf_continuous}, under the regularity conditions in
\eqref{appeq:regularity_contZ},
\[
R_2(\hat Q,Q;\tilde p_z)
\leq
C\left\{
\|\hat f_Z-f_Z\|\,\|\hat\mu-\mu\|
+
\|\hat f_Z-f_Z\|\,\|\hat\pi-\pi\|
\right\} \ .
\]
By the assumed rates,
\[
\|\hat f_Z-f_Z\|\,\|\hat\mu-\mu\|
=
o_P\left(n^{-1/b-1/q}\right),
\qquad
\|\hat f_Z-f_Z\|\,\|\hat\pi-\pi\|
=
o_P\left(n^{-1/b-1/k}\right) \ .
\]
If \(1/b+1/q\geq 1/2\) and \(1/b+1/k\geq 1/2\), both products are
\(o_P(n^{-1/2})\). Hence
$
R_2(\hat Q,Q;\tilde p_z)=o_P(n^{-1/2}).
$
Combining these terms gives
\[
\psi^{+}_{x_0}(\hat Q;\tilde p_z)-\psi_{x_0}(Q;\tilde p_z)
=
P_n\Phi_{x_0}(Q;\tilde p_z)
+
o_P(n^{-1/2}) \ ,
\]
which establishes asymptotic linearity. The same argument applies to the estimating-equation estimator and TMLE, since their expansions have the same first-order term and second-order remainder structure.

\subsubsection{Corollary~\ref{cor:robust_psi_onestep_contZ}: Robustness properties}
\label{app:proofs:robust_psi_onestep_contZ}

The robustness property follows from the form of the second-order remainder. Under the boundedness conditions in \eqref{appeq:regularity_contZ},
\[
R_2(\hat Q,Q;\tilde p_z)
\leq
C\left\{
\|\hat f_Z-f_Z\|\,\|\hat\mu-\mu\|
+
\|\hat f_Z-f_Z\|\,\|\hat\pi-\pi\|
\right\} \ .
\]
If \(\|\hat\pi-\pi\|=o_P(1)\) and \(\|\hat\mu-\mu\|=o_P(1)\), then both products are
\(o_P(1)\), provided the remaining nuisance estimates are uniformly bounded as assumed in the regularity conditions. Thus \(R_2(\hat Q,Q;\tilde p_z)=o_P(1)\).

Alternatively, if \(\|\hat f_Z-f_Z\|=o_P(1)\), then both products are again \(o_P(1)\), even if \(\hat\mu\) and \(\hat\pi\) are not consistently estimated, provided they remain bounded. Hence \(R_2(\hat Q,Q;\tilde p_z)=o_P(1)\) under either condition.

Therefore the first-order bias vanishes under either: (i) consistent estimation of both \(\mu\) and \(\pi\), or (ii) consistent estimation of \(f_Z\). It follows that
$
\psi^{+}_{x_0}(\hat Q;\tilde p_z)
\overset{P}{\longrightarrow}
\psi_{x_0}(Q;\tilde p_z),
$
establishing the stated double robustness property. The same reasoning applies to the estimating-equation estimator and TMLE because they are based on the same influence-function correction and have the same second-order bias structure.

\subsection{Semiparametric inference under a Verma constraint}

\subsubsection{Theorem~\ref{thm:orthocomp}: Orthocomplement tangent space}\label{app:proofs_orthocom}

Let $P_t$ be a regular parametric submodel through $P$ at $t=0$, with $S(o)= \left.\frac{\partial }{\partial t} \log p_t(o) \right|_{t=0}$. Write
\begin{align*}
    \psi_t(z) \equiv \psi_{x_0}(P_t;z)=\frac{\kappa_{x_0,1}(P_t;z)}{\kappa_{x_0,2}(P_t;z)} \ .
\end{align*}
Under the mean-scale Verma model, $\psi_t(z)$ is constant in $z$ for every $t$ in a neighborhood of zero. Equivalently, for every square-integrable function $m$ of $Z$,
\begin{align}
    M_t(m)\equiv \E_t\left[\psi_t(Z)\{m(Z)-E_t[m(Z)]\}\right]=0 \ .\label{appeq:moment-constraint}
\end{align}

Let
\begin{align*}
    \bar m_t(Z)=m(Z)-E_t[m(Z)],\qquad\bar m(Z)=\bar m_0(Z) \ .
\end{align*}
Differentiating \eqref{appeq:moment-constraint} at $t=0$ gives
\begin{align*}0=\left.\frac{\partial}{\partial t}M_t(m)\right|_{t=0}&=\E\{\psi(Z)\bar m(Z)S(O)\}+\E\{\dot\psi(Z)\bar m(Z)\}+\E\{\psi(Z)\dot{\bar m}(Z)\} \ ,
\end{align*}
where
\begin{align*}
    \dot\psi(z)=\left.\frac{\partial}{\partial t}\psi_t(z)\right|_{t=0},\qquad \dot{\bar m}(Z)=\left.\frac{\partial}{\partial t}\bar m_t(Z)\right|_{t=0} \ .
\end{align*}
Because $\psi(Z)$ is constant in $Z$ under the restricted model, the first and third terms simplify as follows:
\begin{align*}
    (\text{first term})\ \E\{\psi(Z)\ \bar{m}(Z)\ S(O)\} & = \psi(Z) \{ \E[m(Z)\ S(O)] - \E[m(Z)] \ \cancelto{0}{\E[S(O)]} \} \ ,
    \\
    (\text{third term})\ E\Big\{\psi(Z)\frac{\partial \bar{m}_t(Z)}{\partial t}\Big\vert_{t=0}\Big\} & = - \psi(Z) \ \E\big[ \E\big(m(Z)\ S(O)\big)\big] = -\psi(Z) \E[m(Z)\ S(O)] \ .
\end{align*}
Thus, the first and third terms cancel. Hence
\begin{align*}
0=\left.\frac{\partial}{\partial t}M_t(m)\right|_{t=0}=\E\{\dot\psi(Z)\bar m(Z)\} \ .
\end{align*}

For each fixed $z$, $\Phi_{x_0}(Q;z)$ is the nonparametric influence function for
$\psi_{x_0}(P;z)$. Therefore, along the submodel $P_t$, $\dot\psi(z)=\E\{\Phi_{x_0}(Q;z)(O) \, S(O)\} \ .$

Substituting this representation into the preceding display, 
\begin{align*}
0&=\int \E\{\Phi_{x_0}(Q;z)(O)S(O)\}\,\bar m(z) \, p(z)\,dz \\
&= \int \left\{ \int  \Phi_{x_0}(Q;z)(o) \, S(o) \, p(o) \, d(o)   \right\}  \bar{m}(z) \, p(z)\,dz \\ 
&= \int \left\{ \int  \Phi_{x_0}(Q;z)(o) \, \bar{m}(z) \, p(z)\,dz \right\} S(o) \, p(o) \, d(o) \\ 
&=\E\left[\left\{\int \Phi_{x_0}(Q;z)\bar m(z)p(z)\,dz\right\}S(O)\right] \ .
\end{align*}
Define $c(z)=\bar m(z)\, p(z)$, which satisfies $\int c(z)\,dz=\int \{m(z)-\E[m(Z)]\}p(z)\,dz=0 \ .$ Thus, for every score $S\in\mathcal T$ and every function $c$ of this form,
\begin{align*}
    \E\left[\left\{\int \Phi_{x_0}(Q;z)c(z)\,dz\right\}S(O)\right]=0 \ .
\end{align*}
Therefore,
\begin{align*}
    \left\{\int \Phi_{x_0}(Q;z)c(z)\,dz:\int c(z)\,dz=0\right\}\subseteq\mathcal T^\perp \ .
\end{align*}

Conversely, the mean-scale Verma model is defined by the collection of moment restrictions
\eqref{appeq:moment-constraint} over all square-integrable functions $m$ of $Z$. The
corresponding linearized restrictions are exactly the orthogonality conditions displayed
above. Since the set of functions $\bar m(z)\,p(z)$ ranges over all square-integrable functions
$c$ satisfying $\int c(z)\,dz=0$, these linearized restrictions generate the full normal space
to the restricted model. Hence no additional orthogonal directions are present, and
\begin{align*}
    \mathcal T^\perp=\left\{\int \Phi_{x_0}(Q;z)c(z)\,dz:\int c(z)\,dz=0\right\} \ .
\end{align*}
For each fixed $z$, $\Phi_{x_0}(Q;z)$ is an influence function for the same target
parameter $\E(Y^{x_0})$. Therefore, for any score $S \in \mathcal T$, $\E\{\Phi_{x_0}(Q;z)S\}=\dot\psi(S),$
where $\dot\psi(S)$ denotes the pathwise derivative of the target parameter in the
direction $S$. For any weighting function $\tilde p$ satisfying $\int \tilde p(z)\,dz = 1,$
it follows that
\begin{align*}
\E\left[\left\{\int \Phi_{x_0}(Q;z)\tilde p(z)\,dz\right\}S\right]&=\int \tilde p(z)\E\{\Phi_{x_0}(Q;z)S\}\,dz 
\\
&=\dot\psi(S)\int \tilde p(z)\,dz \\
&=\dot\psi(S) \ .
\end{align*}
Hence $\int \Phi_{x_0}(Q;z)\tilde p(z)\,dz$ is itself an influence function.

Moreover, if $\tilde p_1$ and $\tilde p_2$ both integrate to one, then
\begin{align*}
    \int \Phi_{x_0}(Q;z)\tilde p_1(z)\,dz-\int \Phi_{x_0}(Q;z)\tilde p_2(z)\,dz=\int \Phi_{x_0}(Q;z)c(z)\,dz \ ,
\end{align*}
where $c(z)=\tilde p_1(z)-\tilde p_2(z)$ satisfies $\int c(z)\,dz=0.$ Therefore the difference between any two such influence functions belongs to $\mathcal T^\perp$. 

Combining this observation with the characterization of $\mathcal T^\perp$ above yields
\begin{align*}
    \mathcal I(\psi_{x_0})=\left\{\int \Phi_{x_0}(Q;z)\ \tilde p(z)\,dz:\int \tilde p(z)\,dz=1\right\} \ .
\end{align*}

When $Z$ is binary, the condition $\sum_z c(z)=0$ gives $c(0)=-c(1)$. Therefore, let $\alpha=c(1)$, we have
\begin{align*}
    \mathcal I(\psi_{x_0})=\left\{\alpha\ \Phi_{x_0}(Q;z^*=1) + (1-\alpha)\ \Phi_{x_0}(Q;z^*=0) : \alpha\in\mathbb R\right\} \ .
\end{align*}
A similar result extends to discrete $Z$ with more than two levels.

\textbf{Construction of explicit parametric submodels satisfying the Verma constraint}\\
A rigorous proof requires constructing explicit parametric submodels and showing that the resulting score is orthogonal to every element of $\mathcal T^{\perp}$. We construct such parametric submodels by perturbing the observed data distribution as follows:
\begin{align*}
p_t(o)=p_t(y\mid x,z,w)\,p_t(x\mid z,w)\,p_t(z\mid w)\,p_t(w) \ .
\end{align*}
For the marginal and conditional distributions of $W$, $Z$, and $X$, define
\begin{align*}
p_t(w)&=p(w)\{1+t s_W(w)\},\quad \E[s_W(W)] = 0 \ , \\
p_t(z \mid w)&=p(z \mid w)\{1+t s_Z(z, w)\},\quad \E[s_Z(Z,W)\mid W] = 0 \ , \\
p_t(x \mid z, w)&=p(x \mid z, w)\{1+t s_X(x, z, w)\},\quad \E[s_X(X,Z,W)\mid Z,W] = 0 \ ,
\end{align*}
where $s_W$, $s_Z$, and $s_X$ are the score functions corresponding to these submodels.

We now construct a submodel for $p_t(y\mid x,z,w)$ that induces a distribution $P_t$ under which the target parameter $\psi_t(z)$ does not depend on $z$. We achieve this using an exponential tilting submodel:
\begin{align*}
p_t(y\mid x,z,w)&= p(y\mid x,z,w)\frac{\exp\{h_{t}(y,x,z,w)\}}{\E[\exp\{h_{t}(y,x,z,w)\}\mid x,z,w]}, \text{ where}
\\
h_{t}(y,x,z,w)&=y\,\eta_t(x,z,w), \text{ with }\eta_{t=0}(x,z,w)=0 \ .
\end{align*}
The resulting score function for the outcome model is
\begin{align*}
s_Y(y,x,z,w)&=\frac{\partial}{\partial t}\log p_t(y\mid x,z,w)\Big|_{t=0} \\
&=\frac{\partial h_{t}}{\partial t}\Big|_{t=0}-\E\!\left[\frac{\partial h_{t}}{\partial t}\Big|_{t=0}\mid x,z,w\right] \\
&=\eta'_t(x,z,w)\Big|_{t=0}\left\{y-\E[Y\mid x,z,w]\right\} \ .
\end{align*}
The perturbed outcome regression is
\begin{align*}
\mu_t(x,z,w)=\frac{\E[Y\exp\{Y\eta_t(x,z,w)\}\mid x,z,w]}{\E[\exp\{Y\eta_t(x,z,w)\}\mid x,z,w]} \ .
\end{align*}

Differentiating with respect to $\eta_t$ and evaluating at $t=0$ yields
\begin{align*}
    \frac{\partial \mu_t(x,z,w)}{\partial \eta_t}\Big\vert_{t=0} & =  \frac{\E[Y^2 \,\eta_t(x,z,w)\ \exp(Y\,\eta_t(x,z,w))\mid x,z,w]}{\E[\exp(Y\,\eta_t(x,z,w))\mid x,z,w]}
    \\
    &\hspace{1cm}- \frac{\E[Y \exp(Y\,\eta_t(x,z,w))\mid x,z,w]\ \E[Y\,\eta_t(x,z,w)\, \exp(Y\,\eta_t(x,z,w))\mid x,z,w]}{\E^2[\exp(Y\,\eta_t(x,z,w))\mid x,z,w]}
    \\
    & = \eta_0(x,z,w) \underbrace{\{\E[Y^2\mid x,z,w] - \E^2[Y\mid x,z,w]\}}_{\sigma^2(x,z,w)} \ .
\end{align*}

We approach the proof by constructing $\mu_t(x,z,w)$ so that the constraint that $\psi_t(z)$ is independent of $z$ is satisfied, and then prove the existence of $\eta_t$ under which $\mu_t(x,z,w)$ admits this form. To carry out this construction, let $\psi_t$ follow the scalar path:
\vspace{-0.5cm}
\begin{align*}
\psi_t(z)=\psi(z) + t c.
\end{align*}
\vspace{-0.5cm}
We further introduce the following definitions:
\vspace{-0.5cm}
\begin{align*}
r(x,z,w)&=\mu(x,z,w)-\psi,
\\
C_t(x,z)&=\frac{\int r(x,z,w)\, p_t(x\mid z,w)p_t(w)\,dw}{\int p_t(x\mid z,w)p_t(w)\,dw}.
\end{align*}
Based on these quantities, we construct the outcome regression under the submodel $P_t$ as
\begin{align}
\mu_t(x,z,w)=\psi_t + r(x,z,w) - C_t(x,z) \ ,
\label{eq:sub_mu}
\end{align}
which satisfies $\mu_{t=0}(x,z,w)=\mu(x,z,w)$.

Next, we verify that this construction satisfies the Verma constraint. Define
\begin{align*}
\kappa_{1,t}(z)&=\int \mu_t(x,z,w)\,p_t(x\mid z,w)p_t(w)\,dw \ , \\
\kappa_{2,t}(z)&=\int p_t(x\mid z,w)p_t(w)\,dw \ .
\end{align*}
One can show that the following equality holds, which implies that $\psi_t=\kappa_{1,t}(z)/\kappa_{2,t}(z)$ does not depend on $z$. Therefore, the constraint is satisfied.
\begin{align*}
\kappa_{1,t}(z)&=\int \{\psi_t+r(x,z,w)-C_t(x,z)\}p_t(x\mid z,w)p_t(w)\,dw \\
&=\psi_t\,\kappa_{2,t}(z) \ ,
\end{align*}

The next step is to show that there exists $\eta_t(x,z,w)$ in a neighborhood of $t=0$ such that $\mu_t(x,z,w)$ admits the form in \eqref{eq:sub_mu}. Define
\begin{align*}
F(\eta_t,t,x,z,w)=\frac{\E[Y\exp\{Y\eta_t(x,z,w)\}\mid x,z,w]}{\E[\exp\{Y\eta_t(x,z,w)\}\mid x,z,w]}-\mu_t(x,z,w) \ .
\end{align*}
The function $F$ has the following properties:
\begin{align*}
F(0,0,x,z,w)=0,\quad \frac{\partial F}{\partial \eta_t}\Big|_{\eta_t=0,t=0}=\sigma^2(x,z,w) > 0 \ .
\end{align*}
Consequently, by the implicit function theorem, there exists a unique differentiable function $\eta_t(x,z,w)$ in a neighborhood of $t=0$ such that
\begin{align*}
F(\eta_t(x,z,w),t,x,z,w)=0 \ ,
\end{align*}
which completes the proof.

\subsubsection{Lemma~\ref{lemma:EIF-binaryZ}: Discrete trapdoor variable}\label{app:proofs_eff_gain}

Assume $Z$ has $K$ categories $\{1,\ldots,K\}$. Consider a class of influence functions, defined as linear combinations of basis influence functions $\Phi_{x_0}(Q;k)$, with weight $\alpha \coloneqq (\alpha_k)_{k=1}^K$ subject to $\alpha_k\in\mathbb{R}$ and $\sum_{k=1}^K \alpha_k=1$: $\sum_{k=1}^K \alpha_k \Phi_{x_0}(Q;k).$ Let $\Sigma$ denote the covariance matrix of the basis influence functions, with $\Sigma_{ij}=\operatorname{Cov}\{\Phi_{x_0}(Q;i),\Phi_{x_0}(Q;j)\}.$

The objective is to find the optimal coefficients
\begin{align*}
    \alpha^{\mathrm{opt}} & = \argmin_{\alpha:\,\mathbf 1^\top\alpha=1} \operatorname{Var}
        \left\{\sum_{k=1}^K \ \alpha_k  \ \Phi_{x_0}(Q;k) \right\} 
        \\
    &= \argmin_{\alpha:\,\mathbf 1^\top\alpha=1} \alpha^\top \Sigma \alpha\ .
\end{align*}
Define the Lagrangian: $\mathcal{L}(\alpha,\lambda)
    = \alpha^\top \Sigma \alpha - 2\lambda (\mathbf{1}^\top \alpha - 1)$. Taking derivative with respect to $\alpha$ yields ${\partial \mathcal{L}}/{\partial \alpha}
    = 2\Sigma \alpha - 2\lambda \mathbf{1}.$
Setting this equal to zero. If $\Sigma$ is invertible, we obtain 
\begin{align*}
    \alpha = \lambda \Sigma^{-1} \mathbf{1} \ .
\end{align*}
To solve for $\lambda$, apply the constraint $\mathbf{1}^\top \alpha = 1$:
\begin{align*}
    1 = \mathbf{1}^\top \alpha
    = \lambda \mathbf{1}^\top \Sigma^{-1} \mathbf{1},\text{ thus } \lambda = \frac{1}{\mathbf{1}^\top \Sigma^{-1} \mathbf{1}}.
\end{align*}
The optimal weight and the resulting optimal influence function within this class is
\begin{align*}
    \sum_{k=1}^K \alpha_k^{\mathrm{opt}} \Phi_{x_0}(Q; k),\text{ where } \alpha^{\mathrm{opt}}
    =
    \frac{\Sigma^{-1} \mathbf{1}}{\mathbf{1}^\top \Sigma^{-1} \mathbf{1}} \ .
\end{align*}

\underline{\textbf{Binary $Z$ as a special case}}\\
We show that, when $Z$ is binary, the closed-form expression for $\alpha^{\mathrm{opt}}$ implies the result stated in Lemma~\ref{lemma:EIF-binaryZ}. For notational simplicity, let $\Phi_j= \Phi_{x_0}(Q; z^*=j)$, for $j\in\{0,1\}$. Let $a \coloneqq \E[\Phi_1^2]$, $b \coloneqq \E[\Phi_1\Phi_0]$, and $c \coloneqq \E[\Phi_0^2]$. Assuming $\Sigma$ is invertible, we have 
\begin{align*}
\Sigma
=
\begin{bmatrix}
a & b \\
b & c
\end{bmatrix}, \quad 
\Sigma^{-1}\mathbf{1}
=
\frac{1}{ac-b^2}
\begin{pmatrix}
c-b \\
a-b
\end{pmatrix},\quad\text{and } \mathbf{1}^\top \Sigma^{-1}\mathbf{1}
=
\frac{a+c-2b}{ac-b^2} \ .
\end{align*}

It follows that the first component of the optimal weight vector is $\frac{c-b}{a+c-2b}$. Substituting the definitions of $a$, $b$, and $c$ gives $\alpha^\mathrm{opt}$ in Lemma~\ref{lemma:EIF-binaryZ}.

\subsubsection{Lemma~\ref{lemma:opt-continuousZ}:  Continuous trapdoor variable}
\label{app:proofs_eff_gain_conti}

Let $\tilde{p}_1(z), \ldots, \tilde{p}_K(z)$ denote $K$ prespecified basis weighting functions. Consider a class of weighting functions: $\tilde{p}_{\alpha}(Z)=\sum_{k=1}^K \alpha_k\ \tilde{p}_k(Z)$  where $\alpha_k\in\R$ and $\sum_{k=1}^K\alpha_k=1$. Define the basis influence functions: $G_k(Q)=\int \Phi_{x_0}(Q; z) \, \tilde{p}_k(z) \, dz$ for $k\in\{1,\cdots,K\}$, and the covariance matrix $\Sigma \in \mathbb{R}^{K\times K}$, for which $\Sigma_{ij} = \E[ G_i(O) G_j(O)].$

Derivation of $\alpha^\mathrm{opt}$ follows analogously from the arguments in Appendix~\ref{app:proofs_eff_gain}. The resulting optimal weight and influence function in this class are
\begin{align*}
    \sum_{k=1}^K \alpha_k^{\mathrm{opt}} \int \tilde{p}_k(z) \Phi_{x_0}(Q; z) dz,\text{ where } \alpha^{\mathrm{opt}}
    =
    \frac{\Sigma^{-1} \mathbf{1}}{\mathbf{1}^\top \Sigma^{-1} \mathbf{1}} \ .
\end{align*}

\subsection{On the validity of the indexing density $\tilde{p}_z$}\label{appsubsec:pz-overlap-condition}

We use the DGP for Simulation 1 described in Section~\ref{appsubsec:dgp_sim1} as a working example to illustrate validity requirements for the weighting function $\tilde{p}(Z)$ used in constructing the IF-based estimators in Section~\ref{subsec:estimators}. As discussed in Section~\ref{subsec:one-step}, $\tilde{p}(Z)$ is valid if its support overlaps sufficiently with the observed data, in the sense that $f_Z(z \mid w) > 0$ for all $z$ in the support of $\tilde{p}(Z)$ and all $w$ such that $p_W(w) > 0$. Here, the validity of $\tilde{p}(Z)$ implies that the identification functional \eqref{eq:ID_tilde_pZ} admits a finite semiparametric efficiency bound.

Under the considered DGP, the conditional distribution of $Z$ depends on $W$: specifically, $Z \mid W=0 \sim \mathrm{Unif}(0.1,0.25)$, whereas $Z \mid W=1 \sim \mathrm{Unif}(0.1,0.5)$. As a result, although the marginal support of $Z$ is $(0.1,0.5)$, the common support across levels of $W$ is only $(0.1,0.25)$. Consequently, weighting functions $\tilde{p}(Z)$ supported outside this common region, such as $\mathrm{Unif}(0.1,0.5)$ or the marginal density $p(Z)$, are not valid because observations with $(z,w=0)$ for $z>0.25$ do not occur in the data. In such regions, the nuisance functions $\mu(x,z,w)$ and $\pi(x \mid z,w)$ cannot be identified from the observed data and therefore necessarily rely on extrapolation, rendering them not well-defined under nonparametric and semiparametric models. Consequently, the resulting influence function is not regular, and variances obtained from this non-regular influence function should not be interpreted as valid efficiency bounds.

To illustrate this point, we replicate Simulation 1 under continuous $Z$, but instead choose $\tilde{p}(Z)$ as $\mathrm{Unif}(0.1,0.5)$, and examine the asymptotic behavior of the IF-based estimators, including the one-step estimator, estimating equation estimator, and TMLE. The results are summarized in Figure~\ref{appfig:sim1-contiZ-unif05}. Although the variance obtained from the corresponding influence function is close to $2$, the empirical $n$-scaled variance of the implemented estimators does not converge to this value. Instead, it converges to approximately $4$. Since nuisance estimation in this simulation is performed using correctly specified parametric models, this limiting variance corresponds to the parametric delta-method variance of the extrapolated functional rather than the semiparametric efficiency bound. The practical relevance of the parametric delta-method variance is limited because correct specification of all nuisance models is generally difficult to guarantee in real applications.

\section{On the inclusion of measured confounders}
\label{app:extension_confounders} 

We extend the nuisance functions from the main manuscript to incorporate  measured confounders $C$. Let $\mu(x,z,w,c)\coloneqq\E(Y\mid X=x,Z=z,W=w,C=c)$, $\pi(x\mid z,w,c)\coloneqq P(X=x\mid Z=z,W=w,C=c)$, $f_Z(z\mid w,c)\coloneqq p(Z=z\mid W=w,C=c)$, $f_W(w\mid c)\coloneqq p(W=w\mid C=c)$, and $p_C(c)\coloneqq p(C=c)$. 

\subsection{Identification}
\label{app:extension_confounders_ID} 

When measured confounders $C$ are present, the identification functional can be modified accordingly as follows:
\begin{align*}
    p(y \mid \doo(x_0)) &= \int p(y \mid \doo(x_0),c)\ p(c) \ dc \\
    &= \int p(y \mid \doo(x_0), \doo(z^*),c) \ p(c) \ dc \qquad & \text{(3rd rule of do-calculus)} \\
    &= \int p(y \mid x_0, \doo(z^*), c) \ p(c) \ dc &\text{(2nd rule of do-calculus)}\\
    &= \int \frac{p(y, x_0, c \mid \doo(z^*))}{p(x_0, c \mid \doo(z^*))} \ p(c) \ dc &\text{(Bayes rule)}\\
    &= \int \frac{\int p(y, x_0 \mid z^*, w,c) \ p(w\mid c) \ dw}{\int p(x_0 \mid z^*,w,c) \ p(w\mid c) \ dw} \ p(c) \ dc \ . &\text{(back-door rule)}
\end{align*}

Thus, our parameter of interest is identified as: 
\begin{align}
    \E(Y^{x_0}) &= \int \frac{\int \mu(x_0, z^*, w,c)\ \pi(x_0 \mid z^*,w,c) \ p(w\mid c) \ dw}{\int \pi(x_0 \mid z^*,w,c) \ p(w\mid c) \ dw} \ p(c) \ dc  \ . 
    \label{app:eq:id_z*}
\end{align}

Algorithm 2 in \cite{bhattacharya2022semiparametric} yields the following equivalent functional, known as the \textit{nested-IPW}, more suitable for continuous-valued $Z$: 
\begin{align}
    \E(Y^{x_0}) & = \int \frac{\int \mu(x_0, z, w,c)\ \pi(x_0 \mid z,w,c) \ p(w\mid c) \ dw}{\int \pi(x_0 \mid z,w,c) \ p(w\mid c) \ dw} \ p(c) \ p^\dagger(z)\ dc\ dz  \ . 
\end{align}%

Equivalently, we can write: 
\begin{align}
    \E(Y^{x_0}) & = \int \frac{\int \mu(x_0, z, w,c)\ \pi(x_0 \mid z,w,c) \ p(w\mid c)\ \tilde{p}(z) \ dw\ dz}{\int \pi(x_0 \mid z,w,c) \ p(w\mid c) \ \tilde{p}(z)\ dw\ dz} \ p(c) \ dc \ , 
    \label{app:eq:id_tildeZ}
\end{align}%
where
\vspace{-1.5cm}
\begin{align*}
    \tilde p(z) = \displaystyle \frac{1}{ \kappa_{x_0,2}(c,Q;z) \int \frac{p^\dagger(z)}{\kappa_{x_0,2}(c,Q;z)} \, dz} \,  p^\dagger(z) \ . 
\end{align*}
The indexing density $\tilde p$ is an auxiliary distribution over $\mathcal Z$  with the the same support requirement as outlined for $p^\dagger$. 

To verify the equivalence, observe that 
\vspace{-0.5cm}
\begin{align*}
\frac{\int \kappa_{x_0,1}(c,Q;z)\, \tilde p(z) \, dz}{\int \kappa_{x_0,2}(c,Q;z) \, \tilde p(z) \, dz} 
& = \frac{\cancel{ \frac{1}{ \int \frac{p^\dagger(z)}{\kappa_{x_0,2}(c,Q;z)} \, dz}}\ \int \frac{\kappa_{x_0,1}(c,Q;z)}{\kappa_{x_0,2}(c,Q;z)}\ p^\dagger(z) \, dz}{\cancel{ \frac{1}{ \int \frac{p^\dagger(z)}{\kappa_{x_0,2}(c,Q;z)} \, dz}}\ \cancelto{1}{\int p^\dagger(z)\ dz}}
= \int \frac{ \kappa_{x_0,1}(c,Q;z) }{ \kappa_{x_0,2}(c,Q;z) } \, p^\dagger(z) \, dz \ . 
\end{align*}

\subsection{Estimation} 
\label{app:extension_est}

\subsubsection*{\underline{Plug-in estimation in the presence of confounders $C$}}

We focus primarily on the case of continuous $Z$, discussing the construction of plug-in and influence-function-based estimators, together with their asymptotic properties. These results extend directly to the discrete-$Z$ setting, for which only minor modifications are required and will be noted as needed. 

Define the nuisance functions
\begin{align*}
    \kappa_{x_0,1}(c,Q;z) & = \int \mu(x_0,z, w, c) \ \pi(x_0 \mid z, w, c) \ f_W(w\mid c) \ dw, 
    \\
    \kappa_{x_0,2} (c,Q;z) & = \int \pi(x_0 \mid z, w, c) \ f_W(w\mid c) \ dw.
\end{align*}
The corresponding plug-in estimators, $\kappa_{x_0,1}^{\text{pi}}(c,\hat{Q};z)$ and $\kappa_{x_0,2}^{\text{pi}}(c,\hat{Q};z)$, are given by:
\begin{equation}\label{appeq:c-plugin-disZ}
    \begin{aligned}
    \kappa^{\text{pi}}_{x_0,1}(c,\hat{Q};z) & = \int \hat{\mu}(x_0,z, w, c) \ \hat{\pi}(x_0 \mid z, w, c) \ \hat{f}_W(w\mid c) \ dw, 
    \\
    \kappa^{\text{pi}}_{x_0,2} (c,\hat{Q};z) & = \int \hat{\pi}(x_0 \mid z, w, c) \ \hat{f}_W(w\mid c) \ dw.
\end{aligned}
\end{equation}

The plug-in estimator for $\E(Y^{x_0})$ is then constructed as
\begin{align*}
    \psi_{x_0}^{\text{pi}}(\hat{Q};\tilde{p}_z)&=\frac{1}{n}\sum_{i=1}^{n}\left\{\kappa_{x_0,1}^{\text{pi}}(C_i,\hat{Q};\tilde{p}_z)/\kappa_{x_0,2}^{\text{pi}}(C_i,\hat{Q};\tilde{p}_z)\right\}\ .
\end{align*}
Here, $\kappa_{x_0,1}^{\text{pi}}(c,\hat{Q};\tilde{p}_z)$ and $\kappa_{x_0,2}^{\text{pi}}(c,\hat{Q};\tilde{p}_z)$ are obtained by integrating $\kappa^\mathrm{pi}_{x_0,1}(c,\hat{Q};z)$ and $\kappa^\mathrm{pi}_{x_0,2}(c,\hat{Q};z)$, respectively, with respect to the pre-specified $\tilde{p}(z)$.

When $W$ is non-discrete, construction of the estimators $\kappa_{x_0,1}^{\mathrm{pi}}(c,\hat{Q};\tilde{p}_z)$ and $\kappa_{x_0,2}^{\mathrm{pi}}(c,\hat{Q};\tilde{p}_z)$ requires estimation of the conditional density $f_W(W=w \mid C=c)$ through its appearance in \eqref{appeq:c-plugin-disZ}, which can be computationally demanding. An alternative approach that avoids density estimation is the following regression-based method:
\begin{align*}
\kappa^{\text{pi}}_{x_0,1}(c,\hat{Q};\tilde{p}_z) & = \hat{\E}\Big\{\int \hat{\mu}(x_0,z, W, C) \, \hat{\pi}(x_0 \mid z, W, C) \, \tilde{p}(z)\ dz \, \Big| \, C=c \Big\} \ , \\
\kappa^{\text{pi}}_{x_0,2}(c,\hat{Q};\tilde{p}_z) & = \hat{\E}\Big\{\int\hat{\pi}(x_0 \mid z, W, C) \, \tilde{p}(z)\ dz\, \Big| \, C=c \Big\} \ ,
\end{align*}
where these quantities are estimated by regressing the corresponding pseudo-outcomes on $C$ and evaluating the fitted regressions at $C=c$.

These estimators, $\kappa^{\text{pi}}_{x_0,1}(c,\hat{Q};\tilde{p}_z)$ and $\kappa^{\text{pi}}_{x_0,2}(c,\hat{Q};\tilde{p}_z)$, are also applicable when $Z$ is discrete. In this case, $\tilde{p}_z$ is specified as a probability mass function for $Z$, and integration with respect to $\tilde{p}_z$ reduces to a finite summation over the support of $Z$.

\subsubsection*{\underline{IF-based estimation in the presence of confounders $C$}}

We first derive the nonparametric influence function for $\psi_{x_0}(Q; \tilde{p}_z)$ in \eqref{app:eq:id_tildeZ}. According to the Leibniz rule, we have
{\small\begin{align*}
    \frac{\partial}{\partial \varepsilon}\psi_{x_0}(P_{\varepsilon}; \tilde{p}_z) & = \int \frac{\partial}{\partial\varepsilon}\frac{\kappa_{x_0,1}(c,P_{\varepsilon};\tilde{p}_z)}{\kappa_{x_0,2}(c,P_{\varepsilon};\tilde{p}_z)}\ d P_{\varepsilon}(c)
    \\
    &=\int \Big(\frac{\frac{\partial}{\partial\varepsilon} \kappa_{x_0,1}(c,P_{\varepsilon};\tilde{p}_z)}{ \kappa_{x_0,2}(c,P;\tilde{p}_z)} - \frac{ \kappa_{x_0,1}(c,P;\tilde{p}_z)\ \frac{\partial}{\partial\varepsilon} \kappa_{x_0,2}(c,P_{\varepsilon};\tilde{p}_z)}{ \kappa^2_{x_0,2}(c,P;\tilde{p}_z)}\Big) \ dP(c)
    \\
    &\hspace{0.5cm}+\int \frac{\kappa_{x_0,1}(c,P;\tilde{p}_z)}{\kappa_{x_0,2}(c,P;\tilde{p}_z)}\ \frac{\partial}{\partial\varepsilon} dP_{\varepsilon}(c),
\end{align*}}
where
{\small\begin{align*}
  &\frac{\partial}{\partial\varepsilon} \kappa_{x_0,1}(c,P_{\varepsilon};\tilde{p}_z)  = \int \frac{\I(x=x_0)\ \tilde{p}(z)}{f_Z(z\mid w,c)}\ (y-\mu(x_0,z,w,c)) \ S(y, x,z, w, c) \ dP(y,x,z,w\mid c)
  \\
  &\hspace{1.6cm}+\int \frac{\tilde{p}(z)}{f_Z(z\mid w,c)}\ \mu(x_0,z,w,c)\ (\I(x=x_0)- \pi(x_0\mid z,w,c)) \ S(x,z,w, c) \ dP(x,z,w\mid c)
  \\
  &\hspace{3cm}+\int \big(\int \mu(x_0,z,w,c)\ \pi(x_0\mid z,w,c)\ \tilde{p}(z)\ dz - \kappa_{x_0,1}(c,P;\tilde{p}_z)\big) \ S(w, c)\ dP(w\mid c),
  \\[1ex]
  &\frac{\partial}{\partial\varepsilon} \kappa_{x_0,2}(c,P_{\varepsilon};\tilde{p}_z)  = \int \frac{\tilde{p}(z)}{f_Z(z\mid w,c)}\ (\I(x=x_0)- \pi(x_0\mid z,w,c)) \ S(x,z,w, c) \ dP(x,z,w\mid c)
  \\
  &\hspace{3cm}+\int \big(\int\pi(x_0\mid z,w,c)\ \tilde{p}(z)\ dz - \kappa_{x_0,2}(c,P;\tilde{p}_z)\big) \ S(w, c)\ dP(w\mid c),
  \\[1ex]
  &\int \frac{\kappa_{x_0,1}(c,P;\tilde{p}_z)}{\kappa_{x_0,2}(c,P;\tilde{p}_z)}\ \frac{\partial}{\partial\varepsilon} dP_{\varepsilon}(c) =\int (\frac{\kappa_{x_0,1}(c,P;\tilde{p}_z)}{\kappa_{x_0,2}(c,P;\tilde{p}_z)} - \psi_{x_0}(P;z))\ S(c) \ dP(c).
  \end{align*}}

By substituting the derivatives with their explicit forms and simplifying the resulting expressions, the nonparametric influence function is given as
{\small\begin{align*}
    \Phi_{x_0}(Q; \tilde{p}_z)(O_i)
    &= \underbrace{\frac{\I(X_i=x_0)}{\kappa_{x_0, 2}(C_i,Q; \tilde{p}_z)}  \frac{\tilde{p}(Z_i)}{f_Z(Z_i \mid W_i,C_i)} \{  Y_i -  \mu(x_0, Z_i, W_i,C_i) \}}_{\Phi_{Y,x_0}(Q; \tilde{p}_z)(O_i)} 
    \\ 
    &\hspace{-2.5cm} + \underbrace{\frac{1}{\kappa_{x_0, 2}(C_i,Q; \tilde{p}_z)} \frac{\tilde{p}(Z_i)}{f_Z(Z_i \mid W_i,C_i)} \{ \mu(x_0, Z_i, W_i,C_i) - \frac{\kappa_{x_0, 1}(C_i,Q; \tilde{p}_z)}{\kappa_{x_0, 2}(C_i,Q; \tilde{p}_z)}   \} \{ \I(X_i=x_0) -  \pi(x_0 \mid Z_i, W_i,C_i) \}}_{\Phi_{X,x_0}(Q; \tilde{p}_z)(O_i)}  
     \\
    &\hspace{-1.5cm} + \underbrace{\frac{1}{\kappa_{x_0, 2}(C_i,Q; \tilde{p}_z)}\int\pi(x_0 \mid z, W_i,C_i) \Big\{ \mu(x_0, z, W_i,C_i)  - \frac{\kappa_{x_0, 1}(C_i,Q; \tilde{p}_z)}{\kappa_{x_0, 2}(C_i,Q; \tilde{p}_z)}   \Big\} \ \tilde{p}(z)\ dz}_{\Phi_{W,x_0}(Q; \tilde{p}_z)(O_i)}
    \\
    &\hspace{-1.5cm} + \underbrace{\frac{\kappa_{x_0, 1}(C_i,Q; \tilde{p}_z)}{\kappa_{x_0, 2}(C_i,Q; \tilde{p}_z)} - \psi_{x_0}(Q;\tilde{p}_z)}_{\Phi_{C,x_0}(Q; \tilde{p}_z)(O_i)}\ .
\end{align*}}

\underline{\textbf{Construction of the one-step estimator:}}\\
The one-step estimator coincides with the estimating equation estimator and is given as
{\small
\begin{equation}\label{eq:app:one-step}
\begin{aligned}
    \psi^{+}_{x_0}(\hat{Q};\tilde{p}_z) & = \frac{1}{n}\sum_{i=1}^{n} \Bigg\{\frac{\I(X_i=x_0)}{\kappa^{\text{aipw}}_{x_0, 2}(C_i,\hat{Q}; \tilde{p}_z)}  \frac{\tilde{p}(Z_i)}{\hat{f}_Z(Z_i \mid W_i,C_i)} \{  Y_i -  \hat{\mu}(x_0, Z_i, W_i,C_i) \}
    \\
    &\hspace{-1.5cm}+\frac{1}{\kappa^{\text{aipw}}_{x_0, 2}(C_i,\hat{Q}; \tilde{p}_z)} \frac{\tilde{p}(Z_i)}{\hat{f}_Z(Z_i \mid W_i,C_i)} \{ \hat{\mu}(x_0, Z_i, W_i,C_i) - \frac{\kappa^{\text{pi}}_{x_0, 1}(C_i,\hat{Q}; \tilde{p}_z)}{\kappa^{\text{pi}}_{x_0, 2}(C_i,\hat{Q}; \tilde{p}_z)}   \}  \times \{ \I(X_i=x_0) -  \hat{\pi}(x_0 \mid Z_i, W_i,C_i) \} 
    \\
    &\hspace{1.5cm}+\frac{1}{\kappa^{\text{aipw}}_{x_0, 2}(C_i,\hat{Q}; \tilde{p}_z)}\int\hat{\pi}(x_0 \mid z, W_i,C_i) \Big \{\hat{\mu}(x_0, z, W_i,C_i)  - \frac{\kappa^{\text{pi}}_{x_0, 1}(C_i,\hat{Q}; \tilde{p}_z)}{\kappa^{\text{pi}}_{x_0, 2}(C_i,\hat{Q}; \tilde{p}_z)}\Big\}\ \tilde{p}(z)\ dz
    \\
    &\hspace{1.5cm}+\frac{\kappa^{\text{pi}}_{x_0, 1}(C_i,\hat{Q}; \tilde{p}_z)}{\kappa^{\text{pi}}_{x_0, 2}(C_i,\hat{Q}; \tilde{p}_z)}
    \Bigg\}\ .
\end{aligned}
\end{equation}}%

Here, $\kappa^{\text{aipw}}_{x_0, 2}(C_i,\hat{Q}; \tilde{p}_z)$ is defined as
{\small\begin{align*}
    \kappa^{\text{aipw}}_{x_0, 2}(C_i,\hat{Q}; \tilde{p}_z)=\hat{\E}\bigg\{\frac{\tilde{p}(Z)}{\hat{f}_Z(Z\mid W,C)}\ (\I(X=x_0)- \hat{\pi}(x_0\mid Z,W,C))+\int \hat{\pi}(x_0\mid z,W,C)\tilde{p}(z)\ dz \ \bigg|\ C=C_i \bigg\} \ .
\end{align*}}
It is estimated by regressing the target quantity inside the conditional expectation on $C$ and then predicting at $C=C_i$.

\underline{\textbf{Construction of the TMLE:}}\\
The TMLE procedure is more complex than in the setting without measured confounders considered in the main manuscript. This additional complexity arises from two sources. First, the term $1/\kappa^{\mathrm{aipw}}_{x_0,2}(C,Q;\tilde{p}_z)$ is no longer constant and therefore cannot be omitted when defining the loss functions and submodels for the nuisance parameters. Second, an additional targeting step is required to ensure that $\Phi_{W,x_0}(\hat{Q}^*;\tilde{p}_z)$ is negligible, a property that is no longer guaranteed by successful targeting of $\mu$ and $\pi$ alone. The first issue necessitates iterative targeting between $\mu$ and $\pi$, since $1/\kappa^{\mathrm{aipw}}_{x_0,2}(C,Q;\tilde{p}_z)$ depends on $\pi$. To address the second, we target the ratio ${\kappa_{x_0,1}(C,Q;\tilde{p}_z)}/{\kappa_{x_0,2}(C,Q;\tilde{p}_z)}$. For notational simplicity, we denote this quantity by $\psi_{x_0}(C,Q;\tilde{p}_z)$.

We define the loss functions and submodels for nuisances $\mu$, $\pi$, and $\psi_{x_0}(C,Q;z^*)$ as follows:
\vspace{-0.5cm}
{\small\begin{align*}
   L_Y(\tilde{\mu};\hat{\pi}, \hat{f}_Z)&=\hat{H}_Y(X,Z,W,C;\tilde{p}_z)\left\{Y_{i}-\tilde{\mu}\left(x_{0}, Z, W,C\right)\right\}^2 \ , \\
   \hat{\mu}(\varepsilon_Y) &= \hat{\mu}(x_{0}, Z, W,C) + \varepsilon_Y \ ,
\end{align*}
}%

\vspace{-.65cm}
with $\hat{H}_Y(X,W,C;z^*)=\{\I(X=x_{0}) \,\tilde{p}(Z)\}/\{\kappa^{\text{aipw}}_{x_0,2}(C,\hat{Q};\tilde{p}_z) \, \hat{f}_{Z}\left(Z \mid W,C\right)\}$, 

\vspace{-1.35cm}
{\small\begin{align*}
    L_X(\tilde{\pi};\hat{f}_Z) &= - \{\tilde{p}(Z)/\hat{f}_Z(Z \mid W,C)\} \, \mathrm{log} \ \tilde{\pi}(X\mid Z,W,C) \ , \\
    \hat{\pi}(\varepsilon_X; \hat{\mu}, \hat{\pi}, \hat{f}_Z) &= \mathrm{expit}\big\{\mathrm{logit} \ \hat{\pi}(x_0\mid Z,W,C) + \varepsilon_X \hat{H}_X(W,C;\tilde{p}_z)\big\} \ ,
\end{align*}
}%
with $\hat{H}_X(W,C;\tilde{p}_z) = \{ \hat{\mu}(x_0, Z, W,C) - \psi^{\text{pi}}_{x_0}(C,\hat{Q}; \tilde{p}_z) \} / \kappa^{\text{aipw}}_{x_0,2}(C,\hat{Q};\tilde{p}_z)$, and 
{\small\begin{align*}
   L_W(\tilde{\psi}_{x_0};\hat{\mu},\hat{\pi}, \hat{f}_Z)&=\hat{H}_W(X,Z,W,C;\tilde{p}_z)\left\{I(W_i,C_i;\tilde{p}_z)-\tilde{\psi}_{x_0}(C,\hat{Q};\tilde{p}_z)\right\}^2 \ , \\
   {\psi}_{x_0}(\varepsilon_W,C,\hat{Q};\tilde{p}_z) &= {\psi}_{x_0}(C,\hat{Q};\tilde{p}_z) + \varepsilon_W \ ,
\end{align*}
}%
with $\hat{H}_W(W,C;\tilde{p}_z) = \{\int \hat{\pi}(x_0\mid z,W,C)\, \tilde{p}(z)\, dz\}/\{\kappa^\mathrm{aipw}_{x_0,2}(C_i,\hat{Q};\tilde{p}_z)\}$, and the pseudo-outcome $I(W,C;\tilde{p}_z)=\{\int \hat{\mu}(x_0,z,W,C)\, \hat{\pi}(x_0\mid z,W,C)\, \tilde{p}(z)\, dz\}/\{\int \hat{\pi}(x_0\mid z,W,C)\, \tilde{p}(z)\, dz\}$.

We begin by targeting $\hat{\pi}$ through within-nuisance iterative updates, which we label as step (T1). Once convergence is reached according to a pre-specified stopping criterion $C_{n, \mathrm{stop}}=\smallO(n^{-1/2})$, we target $\hat{\mu}$ in one step, labeled as (T2), then re-target $\hat{\pi}$ with the updated $\hat{\mu}$. This alternating procedure between (T1) and (T2) is iterated until the nuisance estimates render the empirical mean of both $\Phi_{Y,x_0}$ and $\Phi_{X,x_0}$ below $C_{n, \mathrm{stop}}$.

Let $\hat{Q}^{(t,0)} = \{\hat{\mu}^{(t)}, \hat{\pi}^{(t,0)}, \hat{f}_Z, \hat{p}_{W}\}$ denote the nuisance estimates after completing the $t$-th across-nuisance targeting iteration between $\mu$ and $\pi$. We initialize the process by setting $\hat{\mu}^{(0)} = \hat{\mu}$ and $\hat{\pi}^{(0,0)} = \hat{\pi}$, so $\hat{Q}^{(0,0)} = \hat{Q}$. The second index in the superscript of $\hat{\pi}$ and $\hat{Q}$ is reserved to track the within-nuisance targeting updates for the estimate of $\pi$. Specifically, $\hat{\pi}^{(t,t')}$ denotes the estimate of $\pi$ obtained after the $t'$-th within-nuisance targeting step, which initialized at $\hat{\pi}^{(t,0)}$. Correspondingly, $\hat{Q}^{(t,t')}$ denotes the collection of nuisance estimates that incorporates $\hat{\pi}^{(t,t')}$ as the estimate of $\pi$.

The targeting procedures for $\mu$ and $\pi$ are analogous to those described in the main draft and are therefore omitted here. Suppose convergence is achieved at the $t^*$-th across-nuisance targeting iteration, yielding the nuisance estimates $\hat{Q}^{(t^*,0)}$. The updated estimates are then used to update the plug-in estimator of the ratio, denoted by $\psi(C_i,\hat{Q}^{(t^*,0)};\tilde{p}_z)$. We then proceed to target this ratio, labeled (T3). This is done through a weighted regression, $I^{(t^*,0)}(W,C;\tilde{p}_z) \sim \operatorname{offest}(\psi(C_i,\hat{Q}^{(t^*,0)};\tilde{p}_z))$, with weight $H^{(t^*,0)}_W(W,C;\tilde{p}_z)$, where $I^{(t^*,0)}$ and $H^{(t^*,0)}$ denote the pseudo-outcome and weight, respectively, evaluated at $\hat{Q}^{(t^*,0)}$.
The estimate $\psi(C_i,\hat{Q}^{(t^*,0)};\tilde{p}_z)$ is then updated by adding the fitted intercept coefficient from this regression. 

After this update, we return to the iterative targeting between (T1) and (T2). Once convergence is achieved, we proceed to (T3), and this cycle is repeated until the empirical first-order bias is smaller than $C_{n, \mathrm{stop}}$. At convergence, we denote the resulting nuisance estimates by $\hat{Q}^*$, and the TMLE is obtained as the empirical mean of $\kappa^\mathrm{pi}_{x_0,1}(C,\hat{Q}^*;\tilde{p}_z)/\kappa^\mathrm{pi}_{x_0,2}(C,\hat{Q}^*;\tilde{p}_z)$.

\subsection{Inference} 
\label{app:extension_inf}

\subsubsection*{\underline{Second-order remainder term for $\psi^+_{x_0}(\hat{Q} ; \tilde{p}_z)$}}

The second-order remainder term for $\psi^+_{x_0}(\hat{Q}; \tilde{p}_z)$ in \eqref{eq:app:one-step}, is given by

\vspace{-1.5cm}
{\small\begin{align}
    R_2(\hat{Q},Q;\tilde{p}_z)&=\psi^{\text{pi}}_{x_0}(\hat{Q};\tilde{p}_z)-\psi_{x_0}(Q;\tilde{p}_z)+\int \Phi_{x_0}(\hat{Q} ; \tilde{p}_z)(o) \ dP(o)
    \\
    &=\psi^{\text{pi}}_{x_0}(\hat{Q};\tilde{p}_z)-\psi_{x_0}(Q;\tilde{p}_z)\label{eq:r2_zconti_0}
    \\
    &\hspace{0.5cm}+\int \frac{\I(x=x_0)}{\kappa^{\text{aipw}}_{x_0, 2}(c,\hat{Q}; \tilde{p}_z)}  \frac{\tilde{p}(z)}{\hat{f}_Z(z \mid w,c)} \Big\{  y -  \hat{\mu}(x_0, z, w,c) \Big\} \ dP(o) \label{eq:r2_zconti_1}
    \\
    &\hspace{0.5cm}+\int \frac{1}{\kappa^{\text{aipw}}_{x_0, 2}(c,Q; \tilde{p}_z)} \frac{\tilde{p}(z)}{\hat{f}_Z(z \mid w,c)} \Big\{ \hat{\mu}(x_0, z, w,c) -{\psi^{\text{pi}}_{x_0}(c,\hat{Q}; \tilde{p}_z)}   \Big\} \notag
    \\
    &\hspace{5.5cm}\times \Big\{ \I(x=x_0) -  \hat{\pi}(x_0 \mid z, w,c) \Big\}\ dP(o) \label{eq:r2_zconti_2}
    \\
    &\hspace{0.5cm}+\int \frac{\hat{\pi}(x_0 \mid z, w,c)}{\kappa^{\text{aipw}}_{x_0, 2}(c,\hat{Q}; \tilde{p}_z)} \Big\{ \hat{\mu}(x_0, z, w,c)  - {\psi^{\text{pi}}_{x_0}(c,\hat{Q}; \tilde{p}_z)}   \Big\}\ \tilde{p}(z)\ dz\  dP(w,c) \label{eq:r2_zconti_3}
    \\
    &\hspace{0.5cm}+\int {\psi^{\text{pi}}_{x_0}(c,\hat{Q}; \tilde{p}_z)} p(c)\ dc - \psi^{\text{pi}}_{x_0}(\hat{Q};\tilde{p}_z) \ . \label{eq:r2_zconti_4}
\end{align}}
Line \eqref{eq:r2_zconti_1} can be reformulated as
\begin{align}
    \eqref{eq:r2_zconti_1} & = \int \frac{\pi(x_0\mid z,w,c)}{\kappa^{\text{aipw}}_{x_0, 2}(c,\hat{Q}; \tilde{p}_z)}  \frac{\tilde{p}(z)}{\hat{f}_Z(z \mid w,c)\ f_Z(z \mid w,c)} \notag
    \\
    &\hspace{1cm}\times\big\{f_Z(z \mid w,c)-\hat{f}_Z(z \mid w,c)\big\} \big\{  \mu(x_0, z, w,c) -  \hat{\mu}(x_0, z, w,c) \big\} \ dP(o)\notag
    \\
    &\hspace{0.25cm}+\int \frac{\pi\left(x_0\mid z,w,c\right)}{\kappa^{\text{aipw}}_{x_0,2}(c,\hat{Q} ; \tilde{p}_z)} \frac{\tilde{p}\left(z\right)}{f_Z(z \mid w,c)}\ \mu(x_0,z,w,c)\ d P(z,w,c) \label{eq:r2_zconti_1_2_1}
    \\
    &\hspace{0.25cm}-\int \frac{\pi\left(x_0\mid z,w,c\right)}{\kappa^{\text{aipw}}_{x_0,2}(c,\hat{Q} ; \tilde{p}_z)} \frac{\tilde{p}\left(z\right)}{f_Z(z \mid w,c)}\ \hat{\mu}(x_0, z, w,c) \ d P(z,w,c) \ . \label{eq:r2_zconti_1_2_2}
\end{align}

Line \eqref{eq:r2_zconti_2} can be reformulated as
{\small\begin{align}
    \eqref{eq:r2_zconti_2}
    &=\int \frac{\hat{\mu}(x_0,z,w,c)-{\psi^{\text{pi}}_{x_0}(c,\hat{Q}; \tilde{p}_z)}}{\kappa^{\text{aipw}}_{x_0,2}(c,\hat{Q}; \tilde{p}_z)} \frac{\tilde{p}\left(z\right)}{\hat{f}_Z(z\mid w,c)\ f_Z(z\mid w,c)}\notag
    \\
    &\hspace{1cm}\times\big\{f_Z(z\mid w,c)-\hat{f}_Z(z\mid w,c)\big\}(\pi(x_0\mid z,w,c)-\hat{\pi}(x_0\mid z,w,c)) \ d P(z,w,c)\notag
    \\
    &\hspace{0.25cm}-\int \frac{\hat{\mu}(x_0,z,w,c)-{\psi^{\text{pi}}_{x_0}(c,\hat{Q}; \tilde{p}_z)}}{\kappa^{\text{aipw}}_{x_0,2}(c,\hat{Q}; \tilde{p}_z)} \frac{\tilde{p}\left(z\right)}{f_Z(z\mid w,c)}\hat{\pi}(x_0\mid z,w,c) \ dP(z,w,c) \label{eq:r2_zconti_2_2_1}
    \\
    &\hspace{0.25cm}+\int \frac{1}{\kappa^{\text{aipw}}_{x_0,2}(c,\hat{Q}; z)} \frac{\tilde{p}\left(z\right)}{f_Z(z\mid w,c)}\ \hat{\mu}(x_0,z,w,c)\ \pi(x_0\mid z,w,c)\ dP(z,w,c) \label{eq:r2_zconti_2_2_2}
    \\
    &\hspace{0.25cm}-\int \frac{1}{\kappa^{\text{aipw}}_{x_0,2}(c,\hat{Q}; z)} \frac{\tilde{p}\left(z\right)}{f_Z(z\mid w,c)}\ \ {\psi^{\text{pi}}_{x_0}(c,\hat{Q}; \tilde{p}_z)}\ \pi(x_0\mid z,w,c)\ dP(z,w,c)\ . \label{eq:r2_zconti_2_2_23}
\end{align}}
\noindent Note that \eqref{eq:r2_zconti_2_2_1} cancels out \eqref{eq:r2_zconti_3}, and \eqref{eq:r2_zconti_2_2_2} cancels out \eqref{eq:r2_zconti_1_2_2}.

Combining lines \eqref{eq:r2_zconti_0}, \eqref{eq:r2_zconti_4}, \eqref{eq:r2_zconti_1_2_1}, and \eqref{eq:r2_zconti_2_2_23} results in
{\small\begin{align}
    &\int \Big({\psi^{\text{pi}}_{x_0}(c,\hat{Q}; \tilde{p}_z)}-{\psi_{x_0}(c,Q; \tilde{p}_z)}+\frac{\kappa_{x_0, 1}(c,Q; \tilde{p}_z)}{\kappa^{\text{aipw}}_{x_0, 2}(c,\hat{Q}; \tilde{p}_z)}-\frac{\kappa_{x_0, 2}(c,Q; \tilde{p}_z)}{\kappa^{\text{aipw}}_{x_0, 2}(c,\hat{Q}; \tilde{p}_z)}\ {\psi^{\text{pi}}_{x_0}(c,\hat{Q}; \tilde{p}_z)}\Big)\ p(c)\ dc\notag
    \\
    &=\int \frac{\psi^{\text{pi}}_{x_0}(c,\hat{Q}; \tilde{p}_z)-\psi_{x_0}(c,Q; \tilde{p}_z)}{\kappa^{\text{aipw}}_{x_0, 2}(c,\hat{Q}; \tilde{p}_z)}\Big({\kappa^{\text{aipw}}_{x_0, 2}(c,\hat{Q}; \tilde{p}_z)}-{\kappa_{x_0, 2}(c,Q; \tilde{p}_z)}\Big)\ p(c)\ dc \ .\label{appeq:C-r2-last-term}
 \end{align}}

Combining all terms results in the expression for the second-order remainder
term as follows:
{\footnotesize
\begin{equation}
    \begin{aligned}
    R_2(\hat{Q},Q;\tilde{p}_z)&=\int \frac{\pi(x_0\mid z,w,c)}{\kappa^{\text{aipw}}_{x_0, 2}(c,\hat{Q}; \tilde{p}_z)}  \frac{\tilde{p}(z)}{\hat{f}_Z(z \mid w,c)\ f_Z(z \mid w,c)}
    \\
    &\hspace{1cm}\times\big\{f_Z(z \mid w,c)-\hat{f}_Z(z \mid w,c)\big\} \big\{  \mu(x_0, z, w,c) -  \hat{\mu}(x_0, z, w,c) \big\} \ dP(o)
    \\
    &\hspace{-1.5cm}+\int \frac{\hat{\mu}(x_0,z,w,c)-{\psi^{\text{pi}}_{x_0}(c,\hat{Q}; \tilde{p}_z)}}{\kappa^{\text{aipw}}_{x_0,2}(c,\hat{Q}; \tilde{p}_z)} \frac{\tilde{p}\left(z\right)}{\hat{f}_Z(z\mid w,c)\ f_Z(z\mid w,c)}
    \\
    &\hspace{1cm}\times\big\{f_Z(z\mid w,c)-\hat{f}_Z(z\mid w,c)\big\}(\pi(x_0\mid z,w,c)-\hat{\pi}(x_0\mid z,w,c)) \ d P(z,w,c)
    \\
    &\hspace{-1.5cm}+\int \frac{\psi^{\text{pi}}_{x_0}(c,\hat{Q}; \tilde{p}_z)-\psi_{x_0}(c,Q; \tilde{p}_z)}{\kappa^{\text{aipw}}_{x_0, 2}(c,\hat{Q}; \tilde{p}_z)}\Big({\kappa^{\text{aipw}}_{x_0, 2}(c,\hat{Q}; \tilde{p}_z)}-{\kappa_{x_0, 2}(c,Q; \tilde{p}_z)}\Big)\ p(c)\ dc \ .
\end{aligned}\label{appeq:ext_r2_contZ}
\end{equation}
}

\subsubsection*{\underline{Asymptotic linearity}}

Assume the following regularity conditions:
\begin{equation}\label{appeq:regularity_ext_contZ}
    \begin{aligned}
\inf_{z\in\mathcal{Z},\ w \in \mathcal{W},\ c \in \mathcal{C}} \hat{f}_Z(z|w,c)>0, \quad \max _{x_0\in\{0,1\},\ c \in\mathcal{C}} \Big|\frac{\psi^{\text{pi}}_{x_0}(c,\hat{Q}; \tilde{p}_z)-\psi_{x_0}(c,Q; \tilde{p}_z)}{\kappa^{\text{aipw}}_{x_0, 2}(c,\hat{Q}; \tilde{p}_z)}\Big|<C,\quad \exists C\in\R.
\end{aligned}
\end{equation}

Given the boundedness conditions in \eqref{appeq:regularity_ext_contZ}, applying the Cauchy–Schwarz inequality to each term in \eqref{appeq:ext_r2_contZ} yields the following bound for some sufficiently large constant $C \in \mathbb{R}$:
{\small\begin{align*}
    R_2\left(\hat{Q}, Q;\tilde{p}_z\right) &\leq C\Big[\left\|\hat{f}_Z-f_Z\right\| \times\left\|\hat{\mu}-\mu\right\|+\left\|\hat{f}_Z-f_Z\right\| \times\left\|\hat{\pi}-\pi\right\| + \left\|{\kappa^{\text{aipw}}_{x_0, 2}(c,\hat{Q}; \tilde{p}_z)}-{\tilde{\kappa}^{\text{aipw}}_{x_0, 2}(c,\hat{Q}; \tilde{p}_z)}\right\| \Big].
\end{align*}}
Here, we define $\tilde{\kappa}^{\text{aipw}}_{x_0, 2}(c,\hat{Q}; \tilde{p}_z)$ as a modified version of $\kappa^{\text{aipw}}_{x_0, 2}(c,\hat{Q}; \tilde{p}_z)$ in which the conditional expectation is taken with respect to the true conditional distribution, rather than estimated. Thus, $\tilde{\kappa}^{\text{aipw}}_{x_0, 2}(c,\hat{Q}; \tilde{p}_z)$ represents the limiting target of $\kappa^{\text{aipw}}_{x_0, 2}(c,\hat{Q}; \tilde{p}_z)$ when the pseudo-outcome regression used in its construction is correctly specified.
{\small\begin{align*}
    \tilde{\kappa}^{\text{aipw}}_{x_0, 2}(c,\hat{Q}; \tilde{p}_z) & = \E\Big(\frac{\tilde{p}(Z)}{\hat{f}_Z(Z\mid W,C)}\ (\I(X=x_0)- \hat{\pi}(x_0\mid Z,W,C))+\int \hat{\pi}(x_0\mid z,W,C)\tilde{p}(z)\ dz \ \big|\ C=C_i \Big).
\end{align*}}
With this definition, and the regularity condition in \eqref{appeq:regularity_ext_contZ}, \eqref{appeq:C-r2-last-term} can be re-formulated as
\begin{align*}
    &\int \frac{\psi^{\text{pi}}_{x_0}(c,\hat{Q}; \tilde{p}_z)-\psi_{x_0}(c,Q; \tilde{p}_z)}{\kappa^{\text{aipw}}_{x_0, 2}(c,\hat{Q}; \tilde{p}_z)}\Big({\kappa^{\text{aipw}}_{x_0, 2}(c,\hat{Q}; \tilde{p}_z)}-{\tilde{\kappa}^{\text{aipw}}_{x_0, 2}(c,\hat{Q}; \tilde{p}_z)}\Big)\ p(c)\ dc 
    \\
    &\hspace{0.5cm}+\int \frac{\psi^{\text{pi}}_{x_0}(c,\hat{Q}; \tilde{p}_z)-\psi_{x_0}(c,Q; \tilde{p}_z)}{\kappa^{\text{aipw}}_{x_0, 2}(c,\hat{Q}; \tilde{p}_z)}\Big({\tilde{\kappa}^{\text{aipw}}_{x_0, 2}(c,\hat{Q}; \tilde{p}_z)}-{\kappa_{x_0, 2}(c,Q; \tilde{p}_z)}\Big)\ p(c)\ dc 
    \\
    &\leq C \left\|{\kappa^{\text{aipw}}_{x_0, 2}(C,\hat{Q}; \tilde{p}_z)}-{\tilde{\kappa}^{\text{aipw}}_{x_0, 2}(C,\hat{Q}; \tilde{p}_z)}\right\|
    \\
    &\hspace{0.5cm}+C\left\|\hat{f}_Z(Z|W,C)-f_Z(Z|W,C)\right\| \times\left\|\hat{\pi}(x_0\mid Z,W,C)-\pi(x_0\mid Z,W,C)\right\| \ .
\end{align*}

The asymptotic behavior of the estimator $\psi^+_{x_0}(\hat{Q};\tilde{p}_z)$ is characterized in Theorem~\ref{thm:ext_asymp_psi_onestep_contZ}.

\begin{theorem}\label{thm:ext_asymp_psi_onestep_contZ}
Assume the \emph{$L^2(P)$} convergence rates of nuisance estimates in $\hat{Q}$ are as follows: 
$|| \hat{\pi} - \pi || =\smallO(n^{-\frac{1}{k}})$, 
$|| \hat{f}_{Z} - f_{Z} || =\smallO(n^{-\frac{1}{b}})$, 
$|| \hat{\mu} - \mu|| = \smallO(n^{-\frac{1}{q}})$, $\left\|{\kappa^{\text{aipw}}_{x_0, 2}(C,\hat{Q}; \tilde{p}_z)}-{\tilde{\kappa}^{\text{aipw}}_{x_0, 2}(C,\hat{Q}; \tilde{p}_z)}\right\| = \smallO(n^{-\frac{1}{p}})$. Under the regularity conditions discussed in Appendix~\ref{appeq:regularity_ext_contZ}, if $\frac{1}{p}\geq \frac{1}{2}$, $\frac{1}{b} + \frac{1}{q} \geq \frac{1}{2}$, and $\frac{1}{k} + \frac{1}{b} \geq \frac{1}{2}$, then the one-step estimator $\psi^{+}_{x_0}(\hat{Q}; \tilde{p}_z)$ is asymptotically linear, that is: $\psi^{+}_{x_0}(\hat{Q}; \tilde{p}_z) - \psi_{x_0}(Q; \tilde{p}_z) = P_n \Phi_{x_0}(Q;\tilde{p}_z) + o_P(n^{-1/2})$ where $\Phi_{x_0}(Q;\tilde{p}_z)$ is given in Section~\ref{app:extension_est}. 
\end{theorem}

The above results requires ${\kappa^{\text{aipw}}_{x_0, 2}(C,\hat{Q}; \tilde{p}_z)}$ to converge to ${\tilde{\kappa}^{\text{aipw}}_{x_0, 2}(C,\hat{Q}; \tilde{p}_z)}$ at a rate faster than $\smallO(n^{-1/2})$. This condition can be achieved, for example, by using a correctly specified parametric model for the pseudo-outcome regression, while the other nuisance parameters may be estimated at rates slower than $\smallO(n^{-1/2})$. 

\section{Details and extensions of the TMLE procedures}
\label{app:tmle} 

\subsection{Validity of loss function and submodel combinations}\label{app:tmle_validity}

We establish the validity of the loss function and submodel combinations used for targeting the estimates of $\mu$ and $\pi$ with continuous $Z$ outlined in Section~\ref{subsec:tmle} and detailed in Algorithm~\ref{appalg:contiZ}. The proof for settings with discrete $Z$ follows an analogous argument. 

\underline{Loss function and submodel combination for targeting estimate of $\mu$}

\vspace{-1.5cm}
{\small\begin{align*}
   L_Y(\tilde{\mu};\hat{\pi}, \hat{f}_Z) &=\hat{H}_Y(X,Z,W;\tilde{p}_z)\ \left\{Y_{i}-\tilde{\mu}\left(x_{0}, Z, W\right)\right\}^2 \ , \\
   \hat{\mu}(\varepsilon_Y) &= \hat{\mu}(x_{0}, Z, W) + \varepsilon_Y \ ,\\
\end{align*}
}%

\vspace{-1.75cm}
with $\hat{H}_Y(X,Z,W;\tilde{p}_z)=\{\mathbb{I}(X=x_{0}) \, \tilde{p}(Z)\}/\hat{f}_{Z}\left(Z \mid W\right)$. 

\textit{Proof of (C1):} $ \hat{\mu}(\varepsilon_Y=0) = \hat{\mu}(x_{0}, Z, W) $.\\
\textit{Proof of (C2):}

\vspace{-1.8cm}
\begin{align*}
    \E(L_Y(\tilde{\mu};\hat{\pi}, \hat{f}_Z))&=\E\big(\hat{H}_Y(X,Z,W;\tilde{p}_z)\ \{Y-\tilde{\mu}\left(x_{0}, Z, W\right)\}^2\big)
    \\
    &=\E\big(\hat{H}_Y(X,Z,W;\tilde{p}_z)\ \{Y-\mu(x_{0}, Z, W)+\mu(x_{0}, Z, W)-\tilde{\mu}\left(x_{0}, Z, W\right)\}^2\big)
    \\
    &=\E\big(\hat{H}_Y(X,Z,W;\tilde{p}_z)\ \{Y-\mu(x_{0}, Z, W)\}^2\big)
    \\
    &\hspace{0.5cm}+\E\big(\hat{H}_Y(X,Z,W;\tilde{p}_z)\{\mu(x_{0}, Z, W)-\tilde{\mu}\left(x_{0}, Z, W\right)\}^2\big)
    \\
    &\hspace{0.5cm}+\underbrace{\E\big(2\hat{H}_Y(X,Z,W;\tilde{p}_z)\{Y-\mu(x_{0}, Z, W)\}\ \{\mu(x_{0}, Z, W)-\tilde{\mu}\left(x_{0}, Z, W\right)\}\big)}_{= \ 0}.
\end{align*}
Therefore, $\E(L_Y(\tilde{\mu};\hat{\pi}, \hat{f}Z))$ is minimized when $\mu(x_{0}, Z, W) = \tilde{\mu}(x_{0}, Z, W)$, which makes the second term in the last equality above equal to zero.

\textit{Proof of (C3):}
\begin{align*}
    \frac{\partial}{\partial\varepsilon_Y}L_Y(\hat{\mu}(\varepsilon_Y);\hat{\pi}, \hat{f}_Z)\Big|_{\varepsilon_Y=0}&=2\hat{H}_Y(X,Z,W;\tilde{p}_z)\ \{Y_{i}-\hat{\mu}\left(x_{0}, Z, W\right)\}\propto\Phi_{Y,x_0}(\hat{Q};\tilde{p}_z) \ .
\end{align*}

\underline{Loss function and submodel combination for targeting estimate of $\pi$}

\vspace{-1.5cm}
{\small\begin{align*}
    L_X(\tilde{\pi};\hat{f}_Z) &= - \{\tilde{p}(Z)/\hat{f}_Z(Z \mid W)\} \, \mathrm{log} \ \tilde{\pi}(X\mid Z,W) \\
    \hat{\pi}(\varepsilon_X; \hat{\mu}, \hat{\pi}, \hat{f}_Z) &= \mathrm{expit}\Big\{\mathrm{logit} \ \hat{\pi}(x_0\mid Z,W) + \varepsilon_X \hat{H}_X(Z,W;\tilde{p}_z)\Big\} \ ,
\end{align*}
}

\vspace{-0.8cm}
with $\hat{H}_X(Z,W;\tilde{p}_z) =  \hat{\mu}(x_0, Z, W) - \psi^{\text{pi}}_{x_0}(\hat{Q}; Z)$. 

\textit{Proof of (C1):} $\hat{\pi}(\varepsilon_X=0; \hat{\mu}, \hat{\pi}, \hat{f}_Z)=\mathrm{expit}\{\mathrm{logit} \ \hat{\pi}(x_0\mid Z,W)\} = \hat{\pi}(x_0\mid Z,W)$.

\textit{Proof of (C2):} 

\vspace{-2cm}
\begin{align*}
    \E(L_X(\tilde{\pi};\hat{f}_Z))&=-\int\frac{\tilde{p}(z)}{\hat{f}_Z(z \mid w)} \, \sum_{x\in\{0,1\}}\pi(x\mid z,w)\ \mathrm{log} \ \tilde{\pi}(x\mid Z,W)\ dP(z,w)
    \\
    &=-\int\frac{\tilde{p}(z)}{\hat{f}_Z(z \mid w)} \, \big\{\underbrace{\sum_{x\in\{0,1\}}\pi(x\mid Z,W)\ \mathrm{log} \ \frac{\tilde{\pi}(x\mid z,w)}{\pi(x\mid z,w)}}_{\text{Kullback-Leibler (KL) divergence}} \\ 
    &\hspace{4.5cm}+\sum_{x\in\{0,1\}}\pi(x\mid Z,W)\ \mathrm{log} \ \pi(x\mid z,w)\big\}\ dP(z,w) \ ,
\end{align*}
where for any $z\in\mathcal{Z},\ w\in\mathcal{W}$, the KL divergence from $\pi(x\, | \, z,w)$ to $\tilde{\pi}(x\, | \, z,w)$ is minimized when $\pi(x\, | \, z,w) = \tilde{\pi}(x\, | \, z,w)$. Therefore, $\E(L_X(\tilde{\pi}; \hat{f}_Z))$ is minimized under the same condition.

\textit{Proof of (C3):} 

\vspace{-1.5cm}
\begin{align*}
    &\frac{\partial}{\partial\varepsilon_X}L_X(\hat{\pi}(\varepsilon_X; \hat{\mu}, \hat{\pi}, \hat{f}_Z);\hat{f}_Z)
    \\
    &\hspace{0.5cm}=- \frac{\tilde{p}(Z)}{\hat{f}_Z(Z \mid W)}\frac{(\I(X=x_0)-\I(X=1-x_0))\ \hat{\pi}(x_0\mid Z,W)(1-\hat{\pi}(x_0\mid Z,W))\hat{H}_X(Z,W;\tilde{p}_z)}{\hat{\pi}(X\mid Z,W)}
    \\
    &\hspace{0.5cm}=- \frac{\tilde{p}(Z)}{\hat{f}_Z(Z \mid W)}\ \hat{H}_X(Z,W;\tilde{p}_z)\{\I(X=x_0)- \hat{\pi}(x_0\mid Z,W)\} 
    \propto \Phi_{X,x_0}(\hat{Q};\tilde{p}_z) \ .
\end{align*}

\subsection{The TMLE algorithm}\label{app:tmle_algo}

\newpage 
\begin{singlespacing}
\begin{algorithm}[H]
	\caption{\textproc{TMLE under continuous $Z$ at a pre-specified $\tilde{p}(Z)$ $(\psi_{x_0}(\hat{Q}^*;\tilde{p}_z)$)}}  
	\label{appalg:contiZ}
	
	\begin{algorithmic}[1] 
		
		\State \textbf{Obtain initial nuisance estimates}: $\hat{\mu}$, $\hat{\pi}$, $\hat{f}_Z$, and $\hat{p}_W$.
		
		
		\noindent {\small Estimate of $\pi$ at the $t$-th within-nuisance iterationis denoted by $\hat{\pi}^{(t)}$, with $\hat{\pi}^{(0)} \coloneqq \hat{\pi}$.} 
		
		\vspace{0.1cm}
		\State \textbf{Define loss functions \& submodels} indexed by $\varepsilon_Y,\ \varepsilon_X\in \mathbb{R}$. 
		\par\vspace{0.5em}
		{\small 
			\begin{itemize}
				\item Given $\hat{Q}=\{\hat{\mu},\hat{\pi},\hat{f}_Z,\hat{p}_W\}$, define the loss function \& submodels for $\mu$, indexed by $\varepsilon_Y\in\R$:
				\par\vspace{-2em}
				\begin{align*}
					&L_Y(\tilde{\mu};\hat{\pi}, \hat{f}_Z)=\hat{H}_Y(X,Z,W;\tilde{p}_z)\left\{Y_{i}-\tilde{\mu}\left(x_{0}, Z, W\right)\right\}^2 \ ,  \\
					&\hat{\mu}(\varepsilon_Y) = \hat{\mu}(x_{0}, z^{*}, W) + \varepsilon_Y \ ,
				\end{align*} 
				\par\vspace{-0.75em}
				where $\hat{H}_Y(X,Z,W;\tilde{p}_z)=\{\mathbb{I}(X=x_{0}) \, \tilde{p}(Z)\}/ \hat{f}_{Z}\left(Z \mid W\right)\}$.
				
				\par\vspace{0.5em}
				\item Define the loss function \& submodels for $\pi$, indexed by $\varepsilon_X\in\R$ as follows:
				\par\vspace{-1.5em}
				\begin{align*} 
					&L_X(\tilde{\pi};\hat{f}_Z) = - \{\tilde{p}(Z)/\hat{f}_Z(Z \mid W)\} \, \mathrm{log} \ \tilde{\pi}(X\mid Z,W) \ ,  \\
					&\hat{\pi}(\varepsilon_X; \hat{\mu}, \hat{\pi}, \hat{f}_Z) = \mathrm{expit}\big\{\mathrm{logit} \ \hat{\pi}(x_0\mid Z,W) + \varepsilon_X \hat{H}_X(Z,W;\tilde{p}_z)\big\} \ , 
				\end{align*}
                \par\vspace{-0.75em}
                where $\hat{H}_X(Z,W;\tilde{p}_z) = \{ \hat{\mu}(x_0, Z, W) - \psi^{\text{pi}}_{x_0}(\hat{Q}; \tilde{p}_z) \} $. 
			\end{itemize}        
		}

		\vspace{0.3cm}
		\State: \noindent\textbf{Update $\hat{\mu}$ in one step}. 
		
		\vspace{0.15cm}
		{\small
			\begin{itemize}
				\item Given $\hat{Q}$, fit the following weighted regression: 
				\par\vspace{-0.5em}
				\[
				Y \sim \mathrm{offset}(\hat{\mu}^{(t)}(x_{0}, Z, W))+1, \text{with weight }=\hat{H}_Y(Z,W;\tilde{p}_z) \ .
				\]
				
				\item[] The intercept is the minimizer $\hat{\varepsilon}_Y$. Update: $\hat{\mu}^*(x_{0}, Z, W)=\hat{\mu}(x_{0}, Z, W) + \hat{\varepsilon}_Y.$
				\par\vspace{0.5em}
				\item Update $\hat{Q}=\{\hat{\mu}^*, \hat{\pi}, \hat{f}_Z, \hat{p}_{W}\}$.
			\end{itemize}
		}
		
        \vspace{0.25cm}
		\State \textbf{Update $\hat{\pi}^{(t)}$ iteratively.} {\small At the $t$-th iteration: $\hat{Q}^{(t)} = \{\hat{\mu}^*,\hat{\pi}^{(t)},\hat{f}_Z,\hat{p}_W\}$ }
		
		\vspace{0.15cm}
		{\small 
			\begin{itemize}
				\item Fit the weighted logistic regression without an intercept:
				\par\vspace{-1.5em}
				\[
				\I(X=x_0) \sim \mathrm{offset}(\hat{\pi}^{(t)}(x_0\mid Z,W)) + \hat{H}^{(t)}_X(Z,W), \text{ with weight=}{\tilde{p}(Z)}/{\hat{f}_Z(Z \mid W)} \ . 
				\]
				\par\vspace{-0.5em}
				The coefficient in front of $\hat{H}^{(t)}_X$ is the minimizer $\hat{\varepsilon}^{(t)}_X$. Update $\hat{\pi}^{(t)}$ to $\hat{\pi}^{(t+1)}$:
				\par\vspace{-0.5em}
				\[\hat{\pi}^{(t+1)}(\hat{\varepsilon}^{(t+1)}_X; \hat{\mu}^{(t)} ,\hat{f}_Z) = \mathrm{expit}\{\mathrm{logit} \hat{\pi}^{(t)}(x_0\mid z^*,W) + \hat{\varepsilon}^{(t+1)}_X \ \hat{H}^{(t)}_{X}(Z,W)\} \ .\]
				\item Let $\hat{Q}^{(t+1)} = \{\hat{\mu}^{(t)},\hat{\pi}^{(t+1)},\hat{f}_Z,\hat{p}_W\}$. 
				\vspace{0.15cm}
				\item[] Iteratively update the current estimate of $\pi$ until, for $t=t_X^*$, $|P_n \Phi_{X,x_0}(\hat{Q}^{(t^*_X)})| \leq C_{\text{stop}}$ for pre-specified $C_{\text{stop}}= o_P(n^{-1/2})$. Let $\hat{\pi}^{(t+1)}=\hat{\pi}^*$ and  $\hat{Q}^* = \{\hat{\mu}^*, \hat{\pi}^*, \hat{f}_Z, \hat{p}_{W}\}$.
			\end{itemize}
		}
		
		
		\vspace{0.3cm}
		\State \textbf{Return} The TMLE $\psi_{x_0}(\hat{Q}^*;\tilde{p}_z)$ as
		\par\vspace{-1.5em}
		{\small\[
		\psi_{x_0}(\hat{Q}^*;\tilde{p}(Z)) = \sum_{i=1}^n\int {\displaystyle \hat{\mu}^*(x_0, z, W_i) \ \hat{\pi}^*(x_0 \mid z, W_i) } \tilde{p}(z)\ dz\Big/{\sum_{i=1}^n\int \displaystyle \hat{\pi}^*(x_0 \mid z, W_i)} \ \tilde{p}(z) \ dz \ .
		\]}
	\end{algorithmic}
\end{algorithm}
\end{singlespacing}

\clearpage
\subsection{Extensions to binary outcomes}\label{app:tmle_binary}

Under continuous $Z$, the loss function and submodel for targeting $\hat{\mu}$ are given by 

\vspace{-1.5cm}
{\small\begin{align*}
   L_Y(\tilde{\mu})&=- Y\ \mathrm{log} \ \tilde{\mu}(x_0, Z,W)-(1-Y)\ \mathrm{log} \ (1-\tilde{\mu}(x_0, Z,W)) \ , \\
   \hat{\mu}(\varepsilon_Y;\hat{\pi},\hat{f}_{Z})&=\mathrm{expit}\Big\{\mathrm{logit} \ \hat{\mu}(x_0,Z,W) + \varepsilon_Y \hat{H}_Y(Z,W;\tilde{p}_z)\Big\} \ ,
\end{align*}
}%

\vspace{-0.75cm}
where $\hat{H}_Y(X,Z,W;\tilde{p}_z)=\{\mathbb{I}(X=x_{0}) \, \tilde{p}(Z)\}/\hat{f}_{Z}(Z\, | \, W)$. The validity of this loss function and submodel follows the same reasoning as the proof for targeting $\hat{\pi}$ in Appendix~\ref{app:tmle_validity} and is therefore omitted for brevity.

Accordingly, step (T1) in Section~\ref{subsec:tmle} of the main manuscript is modified as follows:\\
\textit{(T1): One step risk minimization for $\mu$.} 
$\hat{\mu}$ is updated by finding $\hat{\varepsilon}_Y$ that minimizes the loss function $L_Y$. The optimization problem can be solved via a logistic regression without intercept: $Y \sim \mathrm{offset}(\hat{\mu}(x_{0}, Z, W))+\hat{H}_Y(Z,W;\tilde{p}_z)$. The coefficient of $\hat{H}_Y$ gives $\hat{\varepsilon}_Y$, satisfying: $ \hat{\varepsilon}_Y=\underset{\varepsilon_Y \in \mathbb{R}}{\arg \min } \ P_n \Phi_{Y,x_0}(\hat{Q};\tilde{p}_z)$. 

The following updates are then made: $\hat{\mu}^*(x_{0}, Z, W)=\mathrm{expit}\{\mathrm{logit}\hat{\mu}(x_{0}, Z, W) + \hat{\varepsilon}_Y\hat{H}_Y(Z,W;\tilde{p}_z)\}$ and $ \hat{Q}=\{\hat{\mu}^*, \hat{\pi}, \hat{f}_Z, \hat{p}_{W}\}$.

\section{Additional details on simulations} 
\label{app:sim} 

\subsection{Details for Simulation 1}\label{appsubsec:dgp_sim1}
We considered sample sizes of $250, 500, 1000, 2000$ and $4000$. For each sample size, 1000 simulation replicates were conducted to empirically construct confidence intervals for the evaluation metrics, namely $\sqrt{n}$-Bias and $n$-Variance. The true parameter and its variance were obtained using the true nuisance functionals under a large sample of size $10^7$. This same procedure was adopted for computing the true parameter in all subsequent simulations.

We generated data from a DGP that ensures the identification functional for $\E(Y^{x_0})$ is invariant to the choice of $z^*$ while maintaining a simple parametric form that allows theoretical computation of both the true ATE and the nonparametric EIF. The DGP satisfying these conditions is presented below, where “(binary)” refers to the setup with binary $Z$, and “(continuous)” refers to the setup with continuous $Z$.
\begin{align*}
    &W \sim \bin(0.5) 
    \\
    &\text{(binary) }Z \sim \bin(\expit(-1+W)),\quad \text{(continuous) } Z \sim \uniform(0.1, 0.25(1+W))
    \\
    &X \sim \bin(\frac{1}{4}(2-W+ZW))
    \\
    &Y \sim 4+X+\frac{1}{2}Z-\frac{1}{2}ZW-\frac{3}{2}W + (1-W)(1-X)(1-Z)+\N(0,0.1).
\end{align*}

We prove the following expression is irrelevant of $z$.
\begin{align*}
    &\frac{\sum_w \mu\left( x, z, w\right) \ \pi\left(x \mid z, w\right) \ p(w) }{\sum_w \pi\left(x \mid z, w\right) \ p(w)}
    \\
    &=\mu\left( x, z, w=1\right)+\frac{\mu\left( x, z, w=0\right)-\mu\left( x, z, w=1\right)}{\frac{\pi\left(x \mid z, w=1\right)}{\pi\left(x \mid z, w=0\right)}+1}
    \\
    &=\left[4+X-\frac{3}{2}\right]+1 .
\end{align*}

The key idea is to construct $\mu(x, z, w=1, c)$ to be independent of $z$ and to impose the condition $\mu(x, z, w=0) - \mu(x, z, w=1) = \alpha\left(\frac{\pi(x \mid z, w=1)}{\pi(x \mid z, w=0)} + 1\right)$, for some constant $\alpha$.

\subsection{Details for Simulation 2}\label{appsubsec:dgp_sim2}
To induce model misspecification, we estimated the nuisance functionals using incorrect model specifications as follows: \\
\noindent
\begin{minipage}[t]{0.48\textwidth}
\textbf{Univariate binary $Z$ setting}
{\small\begin{itemize}
  \item \textit{Misspecified models for} $\mu$ \textit{and} $\pi$:
    \begin{itemize}
      \item $\mu$: $Y \sim 1 + X + Z + W$
      \item $\pi$: $X \sim 1 + W + Z$ with a logit link
    \end{itemize}
  \item \textit{Misspecified model for} $f_Z$:
    \begin{itemize}
      \item $f_Z$: $Z \sim 1$ with a logit link
    \end{itemize}
  \item \textit{All models misspecified}:
    \begin{itemize}
      \item $\mu$: $Y \sim 1 + X + Z + W$
      \item $\pi$: $X \sim 1 + W + Z$ with a logit link
      \item $f_Z$: $Z \sim 1$ with a logit link.
    \end{itemize}
\end{itemize}}
\end{minipage}
\hfill
\begin{minipage}[t]{0.48\textwidth}
\textbf{Univariate continuous $Z$ setting}
{\small\begin{itemize}
  \item \textit{Misspecified models for} $\mu$:
    \begin{itemize}
      \item $\mu$: $Y \sim 1 + X + Z + W$
    \end{itemize}
  \item \textit{Misspecified model for} $f_Z$:
    \begin{itemize}
      \item $f_Z$: assume a conditional Normal distribution, with conditional mean estimated via a linear regression of the form $Z \sim 1 + W$
    \end{itemize}
  \item \textit{All models misspecified}:
    \begin{itemize}
      \item $\mu$: $Y \sim 1 + X + Z + W$
      \item $\pi$: $X \sim 1 + W + Z$ with a logit link
      \item $f_Z$: assume a conditional Normal distribution, with conditional mean estimated via a linear regression of the form $Z \sim 1$.
    \end{itemize}
\end{itemize}}
\end{minipage}

\subsection{Details for Simulation 3}\label{appsubsec:dgp_sim3}
The DGP for Simulation~3 is given by:
\begin{align*}
    &W \sim \uniform(-2.5,\, 3.5),\quad \text{(binary) }Z \sim \bin(\expit(-\frac{5}{6}+\frac{5}{3} \ W)),
    \\
    &X \sim \bin(\frac{1}{10}(5.5+W-2.75\ Z-0.5\ ZW)),\quad Y \sim 1+XZ+ WX - XZW+\N(0,1).
\end{align*}

One can prove the following expression is irrelevant of $z$:
\begin{align*}
    &\frac{\int \mu(x, z, w) \pi(x \mid z, w) \ p(w) \ dw}{\int  \pi(x \mid z, w) p(w)\ dw}
    \\
    &=\frac{\int (1+xz) \ \pi(x \mid z, w) \ p(w)\ dw + x\ (1-z)\int w \ \pi(x \mid z, w)\ p(w)\ dw}{\int  \pi(x \mid z, w)\ p(w) \ dw}.
\end{align*}
The key idea is to specify the distribution of $W$ and the propensity score $\pi$ such that when $x=1$, we have $\textstyle \int w\ \pi(x \mid z, w)\ p(w)\ dw=\int \pi(x \mid z, w)\ p(w)\ dw$. It is straightforward to show that this is true under the specified distributions:
\begin{align*}
    &\text{When } x=1
    \\
    &\int w\ \pi(1 \mid z, w)\ p(w)\ dw = \int w\ \frac{1}{10}(5.5+w-2.75\ z-0.5\ zw) \ p(w) \ dw 
    \\
    &= \frac{1}{10} (5.5*0.5+3.25-2.75*0.5\ z-0.5*3.25\ z)
    \\
    &\int \pi(1 \mid z, w)\ p(w)\ dw = \int \frac{1}{10}(5.5+w-2.75\ z-0.5\ zw) \ p(w) \ dw 
    \\
    &= \frac{1}{10} (5.5+0.5-2.75\ z-0.5*0.5\ z).
\end{align*}

\subsection{Details for Simulation 4}\label{appsubsec:dgp_sim4}
Under continuous $Z$, the DGP is identical to that in Simulation 1. Under binary $Z$, the DGP parallels that of Simulation 3, except that the conditional distribution of $Z \mid W$ is modified to include interaction and piecewise higher-order terms, specified as
\begin{align*}
    &\text{(binary) } Z\sim \bin(\expit(-1+1 \ W+0.4\ \I(W<0.3)\ W^2)).
\end{align*}

When fitting the nuisance models using generalized linear regressions, key interaction and higher-order terms were intentionally omitted to induce model misspecification. Under continuous $Z$, the nuisance models were fitted following the same approach as in Simulation~2, where all models were misspecified. Under binary $Z$, the nuisance models were specified analogously, except that $f_Z$ was estimated using a logistic regression of the form $Z \sim 1 + W$ with a logit link. The Super Learner incorporated a diverse library of algorithms, including generalized linear models (\texttt{SL.glm}), Bayesian generalized linear models (\texttt{SL.bayesglm}), generalized additive models (\texttt{SL.gam}), multivariate adaptive regression splines (\texttt{SL.earth}), random forests (\texttt{SL.ranger}), and the mean predictor (\texttt{SL.mean}).

\subsection{Details for Simulation 5}\label{appsubsec:dgp_sim5}
The DGP under continuous $Z$ is provided in display \eqref{appeq:sim5_dgp}, where $U_1$ and $U_2$ denote two collections of unmeasured confounders, between $W,\, X$ and $W,\, Y$ respectively.
{\small\begin{equation}\label{appeq:sim5_dgp}
\begin{aligned}
    &C_i \sim \mathrm{Uniform}(0,1), \ i\in\{1,\cdots,10\}, \quad C=[C_1,\cdots, C_{10}]
    \\
    &U_{1,i} \sim \N(0,1),\ i\in\{1,\cdots,10\}, \quad U_1=[U_{2,1},\cdots,U_{2,10}]
    \\
    &U_{2,i} \sim \mathrm{Uniform}(0,1), \ i\in\{1,\cdots,10\}, \quad U_2=[U_{1,1},\cdots,U_{1,10}]
    \\
    &W \sim \operatorname{Binomial}(\expit(C\ V^{1-10}_W+U_2\ V^{11-20}_W+U_1\ V^{21-30}_W+V^{31}_W\sum_{i=1}^{10} (C_i \ U_{2,i}+C_i \ U_{1,i} + U_{2,i} \ U_{1,i} + C_i\ U_{2,i} \ U_{1,i}))
    \\
    &\mathrm{(continuous)}\, Z\sim \N(V^1_Z + V^2_Z W + C\ V^{3-12}_Z,0.5)
    \\
    &X\sim \mathrm{Binomial}(\mathrm{expit}(V^1_X + V^2_X \ Z + C \ V^{3-12}_X + U_1 \ V^{13-22}_X +V^{23}_X \ \sum_{i=1}^{10}(Z\ C_i + Z\ U_{1,i} + C_i\ U_{1,i})))
    \\
    &Y \sim \N(V^1_Y + X \ V^2_Y  + C \ V^{3-12}_Y + U_2\ V^{13-22}_Y + V^{23}_Y \sum_{i=1}^{10}(X\ C_i + X\ U_{2,i} + C_i\ U_{2,i}),0.5),\ \mathrm{where}
    \\
    &V^{T}_M=[0.5, -0.1, -0.4, -0.4, -0.3,  0.3, -0.2,  0.5, -0.3,  0.0, -0.3, -0.3,  0.3 ,-0.4 , 0.0, -0.4 , 0.1 ,-0.5 ,
    \\
    &\hspace{1.2cm} 0.5, -0.2 , 0.1 ,-0.2 , 0.5 , 0.4 , 0.5 ,-0.4 , 0.1 , 0.0 , 0.5 ,-0.1,1.0]
    \\
    &V^{T}_Z=[0.2, 0.5 , 0.4 ,-0.5 ,-0.6 ,-0.6, -0.2 , 0.7 , 0.0 , 0.6 , 0.7, -0.1]
    \\
    &V^{T}_X=[-0.2 , 0.5 , 0.3 ,-0.1 , 0.3 ,-0.1 , 0.4, -0.2 ,-0.4 , 0.3 , 0.4 , 0.5 , 0.1 , 0.2 , 0.3, 0.1 , 0.2 ,-0.1 ,
    \\
    &\hspace{1.2cm} -0.3 ,-0.3 ,-0.1 ,-0.2 , 0.3]
    \\
    &V^{T}_Y=[0.1 , 0.6, -0.3 , 0.0 , 0.3 , 0.2 , 0.0 , 0.2 , 0.0 , 0.0 ,-0.2 ,-0.1 ,-0.2 , 0.1,  0.3,
    \\
    &\hspace{1.2cm} 0.1 ,-0.4 , 0.1 ,-0.1 ,-0.1 , 0.1 , 0.4 , 0.3].
\end{aligned}
\end{equation}}
Here, we use numeric superscripts to denote element selection. For example, $V^1_Z$ refers to the first element of the vector $V_Z$, while $V^{3-12}_Z$ denotes the subvector consisting of the 3rd through 12th elements of $V_Z$, inclusive. We use superscript $T$ to indicate the transpose operator, so that all the coefficient vectors, $V_M$, $V_Z$, $V_X$, and $V_Y$, are column vectors.

Calculating the true ATE via correct nuisance model specification is challenging under this DGP, as the closed-form expressions for the nuisance functionals are complex. Therefore, instead of relying on the identification functional defined by true nuisance functionals, we approximate the true ATE by intervening on $X$, setting it to 1 and 0, respectively, and computing the sample means of the resulting $Y$ values to estimate $\E(Y^1)$ and $\E(Y^0)$. The difference between these two estimates defines the ATE.

\subsection{Additional results on simulations}\label{appsubsec:sim_results}
\begin{figure}
    \centering
    \includegraphics[width=1\linewidth, clip, trim=0 5 0 5]{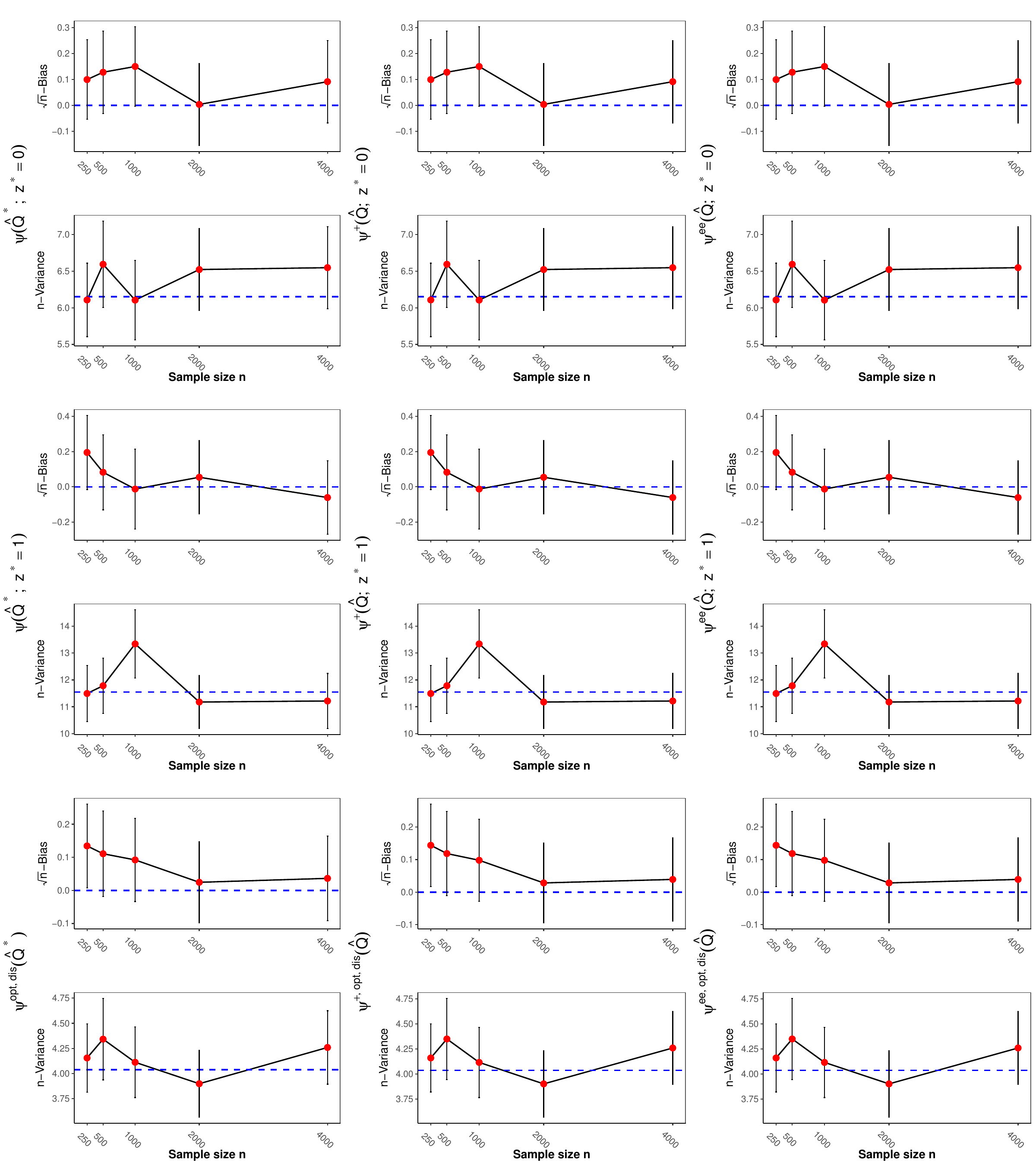}
    \caption{Simulation results demonstrating asymptotic linearity under univariate binary $Z$. The left column is for TMLE, the middle column is for the one-step estimators, and the right column is for the estimating equation estimators.}
    \label{appfig:sim1-binaryZ}
\end{figure}

\begin{figure}
    \centering
    \includegraphics[width=1\linewidth, clip, trim=0 5 0 5]{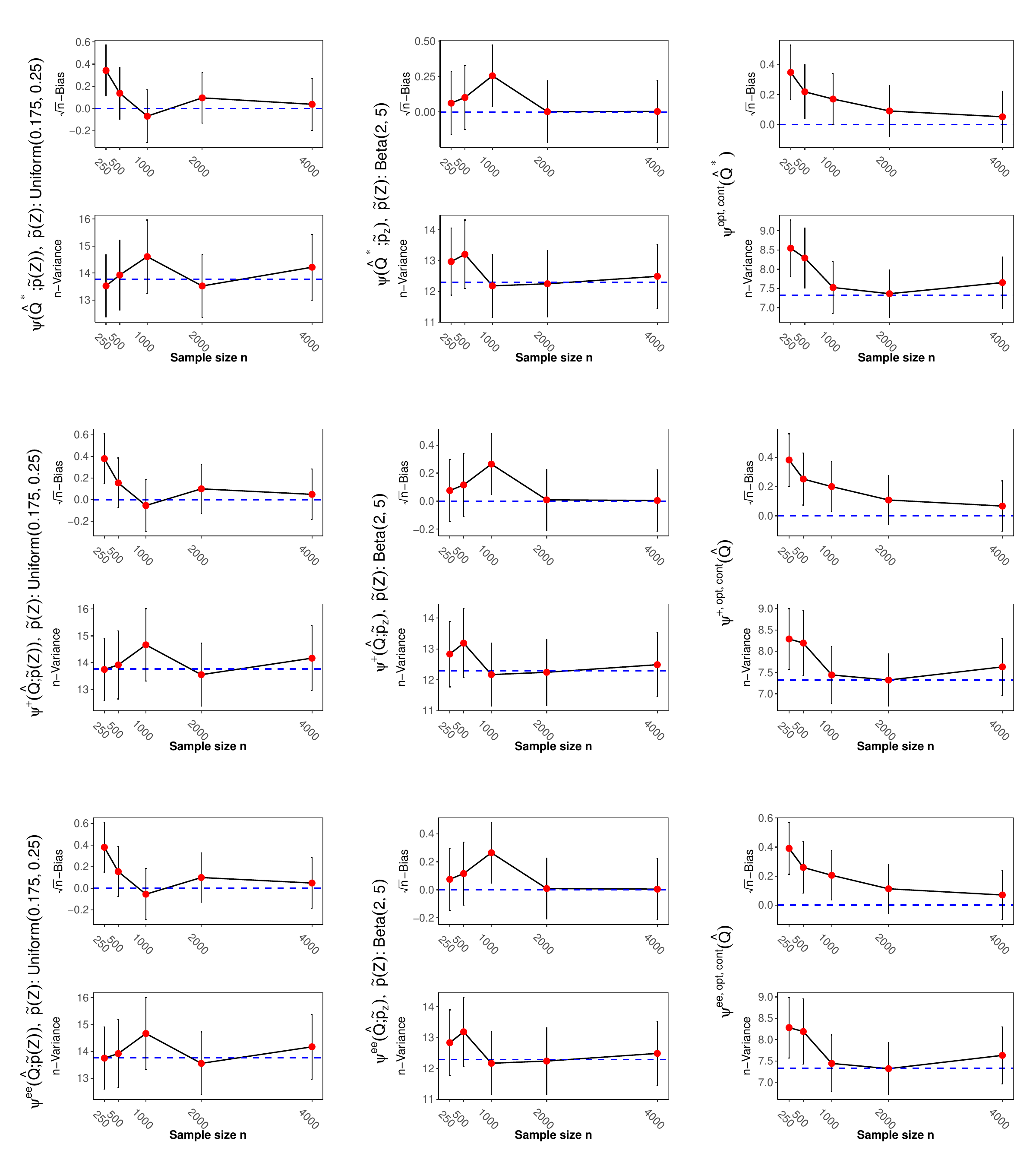}
    \caption{Simulation results demonstrating asymptotic linearity under univariate continuous $Z$. The first, second, and third rows correspond to TMLEs, one-step estimators, and estimating equation estimators, respectively. The first column reports results with $\tilde{p}(Z)$ specified as $\operatorname{Uniform}(0.175,0.25)$, the second column reports results with $\tilde{p}(Z)$ specified as a rescaled $\operatorname{Beta}(2,5)$ distribution on $(0.1,0.25)$, and the third column reports results from estimators formed as optimally weighted linear combinations of the two basis specifications.}
    \label{appfig:sim1-contiZ}
\end{figure}

\begin{figure}
    \centering
    \includegraphics[width=1\linewidth, clip, trim=0 5 0 5]{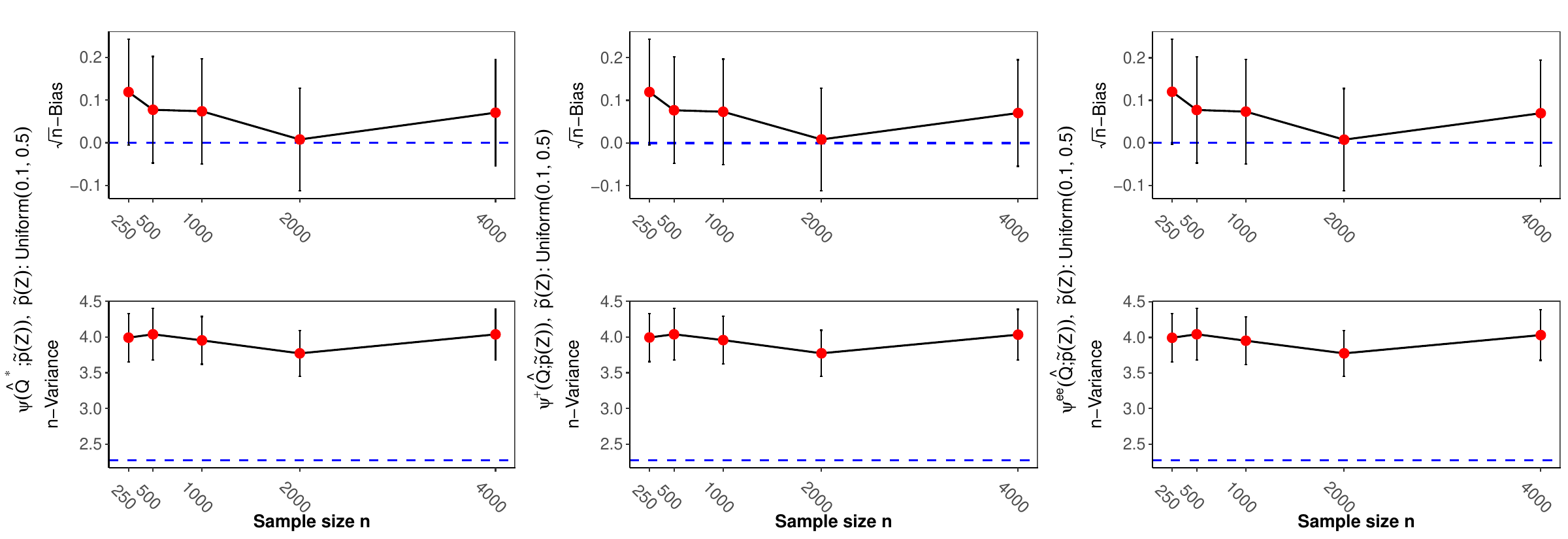}
    \caption{Simulation results illustrating the asymptotic behavior of IF-based estimators constructed with the invalid weighting function $\tilde{p}(Z)=\operatorname{Uniform}(0.1,0.5)$ under univariate continuous $Z$. The first, second, and third columns correspond to TMLEs, one-step estimators, and estimating equation estimators, respectively.}
    \label{appfig:sim1-contiZ-unif05}
\end{figure}

This subsection presents additional results from the simulation studies. Figures~\ref{appfig:sim1-binaryZ} and \ref{appfig:sim1-contiZ} illustrate the asymptotic convergence behavior of the TMLEs, one-step, and estimating equation estimators when all nuisance models are correctly specified, under univariate binary and continuous $Z$, respectively. Figure~\ref{appfig:sim1-contiZ-unif05} illustrates the asymptotic convergence behavior of the three IF-based estimators constructed using an invalid weighting function. Tables~\ref{table:sim2_binary}, \ref{table:sim4_binary}, and \ref{table:sim5_binary} summarize the results for Simulations 2, 4, and 5, respectively, under the binary $Z$ setting.

  \providecommand{\huxb}[2]{\arrayrulecolor[RGB]{#1}\global\arrayrulewidth=#2pt}
  \providecommand{\huxvb}[2]{\color[RGB]{#1}\vrule width #2pt}
  \providecommand{\huxtpad}[1]{\rule{0pt}{#1}}
  \providecommand{\huxbpad}[1]{\rule[-#1]{0pt}{#1}}

\begin{sidewaystable}[ht]
\centering
\captionsetup{justification=raggedright,singlelinecheck=off}
\caption{Simulation results validating the double robustness property of the proposed estimators when Z is binary. Results from the estimating equation estimator coincide with those of the one-step estimator up to three decimal places and are therefore omitted.}
\resizebox{1\textwidth}{!}{
\renewcommand{\arraystretch}{0.5}
 \setlength{\extrarowheight}{0pt}%
 \setlength{\lineskip}{0pt}\setlength{\lineskiplimit}{0pt}%
 \setlength{\tabcolsep}{0pt}
 \setlength{\arrayrulewidth}{0.5pt}
\begin{tabular}{l l l l l l l l l l l l l l l l l l l l}

\thickline

\multicolumn{1}{!{\huxvb{0, 0, 0}{0}}l!{\huxvb{0, 0, 0}{0}}}{\huxtpad{-1pt + 1em}\raggedright \hspace{-1pt} \textbf{{\fontsize{6pt}{7.2pt}\selectfont }} \hspace{-1pt}\huxbpad{-1pt}} &
\multicolumn{1}{c!{\huxvb{0, 0, 0}{0}}}{\huxtpad{-1pt + 1em}\centering \hspace{-1pt} \textbf{{\fontsize{6pt}{7.2pt}\selectfont }} \hspace{-1pt}\huxbpad{-1pt}} &
\multicolumn{9}{c!{\huxvb{0, 0, 0}{0}}}{\huxtpad{-1pt + 1em}\centering \hspace{-1pt} \textbf{{\fontsize{6pt}{7.2pt}\selectfont TMLEs}} \hspace{-1pt}\huxbpad{-1pt}} &
\multicolumn{9}{c!{\huxvb{0, 0, 0}{0}}}{\huxtpad{-1pt + 1em}\centering \hspace{-1pt} \textbf{{\fontsize{6pt}{7.2pt}\selectfont One-step estimators}} \hspace{-1pt}\huxbpad{-1pt}} \tabularnewline
\end{tabular}}\label{table:sim2_binary}
\end{sidewaystable}

  \providecommand{\huxb}[2]{\arrayrulecolor[RGB]{#1}\global\arrayrulewidth=#2pt}
  \providecommand{\huxvb}[2]{\color[RGB]{#1}\vrule width #2pt}
  \providecommand{\huxtpad}[1]{\rule{0pt}{#1}}
  \providecommand{\huxbpad}[1]{\rule[-#1]{0pt}{#1}}

\begin{sidewaystable}[t]
\centering
\captionsetup{justification=raggedright,singlelinecheck=off}
\caption{Comparative analysis of TMLEs, one-step, and estimating equation estimators under model misspecifications when Z is binary. 
  Linear refers to generalized linear regression including only main effects, RF refers to random forest with 500 trees and a minimum node size of 5 for a continuous variable and 1 for binary, 
  and CF denotes random forest with cross fitting using 10 folds. Results from the estimating equation estimator coincide with those of the one-step estimator up to three decimal places and are therefore omitted.}
 \setlength{\tabcolsep}{0pt}
\resizebox{1\textwidth}{!}{
\renewcommand{\arraystretch}{0.5}
 \setlength{\extrarowheight}{0pt}%
 \setlength{\lineskip}{0pt}\setlength{\lineskiplimit}{0pt}%
 \setlength{\tabcolsep}{0pt}
 \setlength{\arrayrulewidth}{0.5pt}
\begin{tabular}{l l l l l l l l l l l l l l l l l l l l}

\thickline

\multicolumn{1}{!{\huxvb{0, 0, 0}{0}}l!{\huxvb{0, 0, 0}{0}}}{\huxtpad{-1pt + 1em}\raggedright \hspace{-1pt} \textbf{{\fontsize{6pt}{7.2pt}\selectfont }} \hspace{-1pt}\huxbpad{-1pt}} &
\multicolumn{1}{c!{\huxvb{0, 0, 0}{0}}}{\huxtpad{-1pt + 1em}\centering \hspace{-1pt} \textbf{{\fontsize{6pt}{7.2pt}\selectfont }} \hspace{-1pt}\huxbpad{-1pt}} &
\multicolumn{9}{c!{\huxvb{0, 0, 0}{0}}}{\huxtpad{-1pt + 1em}\centering \hspace{-1pt} \textbf{{\fontsize{6pt}{7.2pt}\selectfont TMLEs \(\psi(\hat{Q}^*;z^*)\)}} \hspace{-1pt}\huxbpad{-1pt}} &
\multicolumn{9}{c!{\huxvb{0, 0, 0}{0}}}{\huxtpad{-1pt + 1em}\centering \hspace{-1pt} \textbf{{\fontsize{6pt}{7.2pt}\selectfont One-step estimators \(\psi^+(\hat{Q};z^*)\)}} \hspace{-1pt}\huxbpad{-1pt}} \tabularnewline
\end{tabular}
}\label{table:sim4_binary}

\end{sidewaystable}

 \providecommand{\huxb}[2]{\arrayrulecolor[RGB]{#1}\global\arrayrulewidth=#2pt}
  \providecommand{\huxvb}[2]{\color[RGB]{#1}\vrule width #2pt}
  \providecommand{\huxtpad}[1]{\rule{0pt}{#1}}
  \providecommand{\huxbpad}[1]{\rule[-#1]{0pt}{#1}}

\begin{table}[ht]
\centering
\captionsetup{justification=raggedright,singlelinecheck=off}
\caption{Comparative analysis for the impact of cross-fitting on TMLEs, one-step, and estimating equation estimators in conjunction with the use of
random forests. RF refers to random forest with 500 trees and a minimum node size of 5 for a continuous variable and 1 for
binary, and CF denotes random forest with cross fitting using 10 folds. Results from the estimating equation estimator coincide with those of the one-step estimator up to three decimal places and are therefore omitted.}
\resizebox{1\textwidth}{!}{
\renewcommand{\arraystretch}{0.5}
 \setlength{\extrarowheight}{0pt}%
 \setlength{\lineskip}{0pt}\setlength{\lineskiplimit}{0pt}%
 \setlength{\tabcolsep}{0pt}
 \setlength{\arrayrulewidth}{0.5pt}
\begin{tabular}{l l l l l l l l l l l l l l}

\thickline

\multicolumn{1}{!{\huxvb{0, 0, 0}{0}}l!{\huxvb{0, 0, 0}{0}}}{\huxtpad{-1pt + 1em}\raggedright \hspace{-1pt} \textbf{{\fontsize{6pt}{7.2pt}\selectfont }} \hspace{-1pt}\huxbpad{-1pt}} &
\multicolumn{1}{c!{\huxvb{0, 0, 0}{0}}}{\huxtpad{-1pt + 1em}\centering \hspace{-1pt} \textbf{{\fontsize{6pt}{7.2pt}\selectfont }} \hspace{-1pt}\huxbpad{-1pt}} &
\multicolumn{6}{c!{\huxvb{0, 0, 0}{0}}}{\huxtpad{-1pt + 1em}\centering \hspace{-1pt} \textbf{{\fontsize{6pt}{7.2pt}\selectfont TMLEs \(\psi(\hat{Q}^*;z^*)\)}} \hspace{-1pt}\huxbpad{-1pt}} &
\multicolumn{6}{c!{\huxvb{0, 0, 0}{0}}}{\huxtpad{-1pt + 1em}\centering \hspace{-1pt} \textbf{{\fontsize{6pt}{7.2pt}\selectfont One-step estimators \(\psi^+(\hat{Q};z^*)\)}} \hspace{-1pt}\huxbpad{-1pt}} \tabularnewline[-0.5pt]

\cline{2-14}

\multicolumn{1}{!{\huxvb{0, 0, 0}{0}}l!{\huxvb{0, 0, 0}{0}}}{\huxtpad{-1pt + 1em}\raggedright \hspace{-1pt} \textit{\textbf{{\fontsize{6pt}{7.2pt}\selectfont }}} \hspace{-1pt}\huxbpad{-1pt}} &
\multicolumn{1}{c!{\huxvb{0, 0, 0}{0}}}{\huxtpad{-1pt + 1em}\centering \hspace{-1pt} \textit{{\fontsize{6pt}{7.2pt}\selectfont }} \hspace{-1pt}\huxbpad{-1pt}} &
\multicolumn{3}{c!{\huxvb{0, 0, 0}{0}}}{\huxtpad{-1pt + 1em}\centering \hspace{-1pt} \textit{{\fontsize{6pt}{7.2pt}\selectfont RF}} \hspace{-1pt}\huxbpad{-1pt}} &
\multicolumn{3}{c!{\huxvb{0, 0, 0}{0}}}{\huxtpad{-1pt + 1em}\centering \hspace{-1pt} \textit{{\fontsize{6pt}{7.2pt}\selectfont CF}} \hspace{-1pt}\huxbpad{-1pt}} &
\multicolumn{3}{c!{\huxvb{0, 0, 0}{0}}}{\huxtpad{-1pt + 1em}\centering \hspace{-1pt} \textit{{\fontsize{6pt}{7.2pt}\selectfont RF}} \hspace{-1pt}\huxbpad{-1pt}} &
\multicolumn{3}{c!{\huxvb{0, 0, 0}{0}}}{\huxtpad{-1pt + 1em}\centering \hspace{-1pt} \textit{{\fontsize{6pt}{7.2pt}\selectfont CF}} \hspace{-1pt}\huxbpad{-1pt}} \tabularnewline
\end{tabular}}\label{table:sim5_binary}

\end{table}

\clearpage
\section{Additional details on real data application} 
\label{app:realdata} 

To model the relationships among variables flexibly, we estimated the conditional densities $p(Z\mid W)$ or $p(Z \mid W, S, G)$ via semiparametric kernel methods, implemented in the \texttt{np} R package \citep{hayfield2008nonparametric}, and estimated other nuisance parameters using Super Learner with a library including generalized linear models (\texttt{SL.glm}), generalized additive models (\texttt{SL.gam}), multivariate adaptive regression splines (\texttt{SL.earth}), random forests (\texttt{SL.ranger}), and intercept-only models (\texttt{SL.mean}). We specified $\tilde{p}(Z)$ as a uniform distribution, with the range varying according to the population under analysis, to ensure that $\tilde{p}(Z)$ remained valid.

\end{document}